\documentclass[letterpaper,11pt, notoc]{JHEP3}

\usepackage{bbm} % for open 1, JHEP preferred
\usepackage{amsmath, amssymb}
\usepackage{subfigure, graphicx}
\usepackage{braket} 

\allowdisplaybreaks[3]
\newcommand\oldappendix{This will give an error if \backspace oldappendix
   already exists.}
\let\oldappendix\appendix
\renewcommand\appendix{\oldappendix%
   \renewcommand\theequation{\thesection.\arabic{equation}}
   \renewcommand\thesubsection{\thesection.\arabic{subsection}}}
\newcommand\be{\begin{equation}}
\newcommand\bea{\begin{eqnarray}}
\newcommand\ee{\end{equation}}
\newcommand\eea{\end{eqnarray}}
\newcommand\h{\frac{1}{2}}

\newcommand{\bdm}{\begin{displaymath}}
\newcommand{\edm}{\end{displaymath}}

\newcommand{\f}[2]{\frac{#1}{#2}}
\newcommand{\bref}[1]{(\ref{#1})}
\newcommand\p{\partial}

\renewcommand\t{\tilde}

\renewcommand{\choose}[2]{\binom{#1}{#2}}

\newcommand\drm{\mathrm{d}}
\newcommand{\der}[2]{\frac{\drm #1}{\drm #2}}

\newcommand{\pd}{\partial}
\newcommand{\pdb}{\bar{\partial}}
\newcommand{\ints}{\mathbb{Z}}
\DeclareMathOperator{\sech}{\text{\sech}}

\newcommand{\vev}[1]{\langle {#1}\rangle}
\newcommand{\Vev}[1]{\left\langle {#1} \right\rangle}

\newcommand{\com}[2]{[#1,\, #2]}
\newcommand{\ac}[2]{\{#1,\,#2\}}
\newcommand{\vac}{\ket{\varnothing}}
\newcommand{\bvac}{\bra{\varnothing}}
\newcommand{\dg}{\dagger}
\newcommand{\id}{\mathbbm{1}}

\newcommand{\Nsc}{\mathcal{N}}

\newcommand{\ibar}{{\bar{\imath}}}
\newcommand{\jbar}{{\bar{\jmath}}}
\newcommand\normalTag{\addtocounter{equation}{1}\tag{\theequation}}
\newcommand{\bj}{\bar{\jmath}}

\makeatletter 

\newenvironment{calc}{%\allowdisplaybreaks
	\start@align\@ne\st@rredtrue\m@ne}
		     {\normalTag\endalign}
\makeatother

\title{Emission from the D1D5 CFT}
\author{Steven G. Avery \\ 
Department of Physics, The Ohio State University,\\ 
191 West Woodruff Avenue, Columbus, Ohio 43210-1117 U.S.A.\\
E-mail: \email{avery@mps.ohio-state.edu}}

\author{Borun D. Chowdhury\\
Department of Theoretical Physics, Tata Institute of Fundamental Research,\\ 
Homi Bhabha Road, Mumbai 400005, India\\
E-mail: \email{borundev@mps.ohio-state.edu}}

\author{Samir D. Mathur \\ 
Department of Physics, The Ohio State University,\\ 
191 West Woodruff Avenue, Columbus, Ohio 43210-1117 U.S.A.\\
E-mail: \email{mathur@mps.ohio-state.edu}}
    
\abstract{
It is believed that the D1D5 brane system is described by an `orbifold
CFT' at a special point in moduli space. We first develop a general
formulation relating amplitudes in a $d$-dimensional CFT to
absorption/emission of quanta from flat infinity. We then construct
the D1D5 vertex operators for minimally coupled scalars in
supergravity, and use these to compute the CFT amplitude for emission
from a state carrying a single excitation. Using spectral flow we
relate this process to one where we have emission from a highly
excited initial state. In each case the radiation rate is found to
agree with the radiation found in the gravity dual.
}

\keywords{Black Holes in String Theory, AdS--CFT Correspondence, Conformal Field Models in String Theory}

\begin{document}

%\maketitle
%\tableofcontents

\renewcommand{\labelenumi}{(\roman{enumi})} 
%\renewcommand{}
% for lower-case roman numeral enumerate environment
   % place command here so as not to mess up table of contents

\section{Introduction}

The D1D5 system has been very useful in the study of black holes
\cite{stromvafa,radiation-1,radiation-2,radiation-3,radiation-4,
  radiation-5}. This system arises from a bound state of D1 branes and
  D5 branes. The near horizon geometry of these branes is dual to a
  1+1 dimensional CFT.  While the 3+1 dimensional $\Nsc = 4$
  Yang--Mills CFT (arising from D3 branes) has been extensively
  studied~\cite{gkp, witten, Aharony}, there has been much less work
  with the 1+1 dimensional CFT~\cite{maldacena}.

The goal of this paper is to set up a formalism for computing
amplitudes in the orbifold CFT and relating them to
absorption/emission of quanta from the gravitational solution
describing the D-branes. This requires two main steps. For the first,
note that the CFT describes only the physics in the `near-horizon
region' of the branes; vertex operators in the CFT create excitations
that must travel through the `neck' of the D-brane geometry and then
escape to infinity as traveling waves. Thus we set up a general
formalism that relates CFT amplitudes to absorption/emission rates
observed from infinity\footnote{In~\cite{davidmandalwadia}, the
connection between waves coming from infinity and operators in the CFT
was established for $l=0$; however, the effect of the `neck' was
ignored since it is irrelevant for minimal scalars.}	. For the second
step, note that early computations of
radiation~\cite{stromvafa,radiation-1,radiation-2,radiation-3,radiation-4,radiation-5}
used the somewhat heuristic picture of an `effective string' to
describe the D1D5 bound state. We construct states and vertex
operators in the orbifold CFT, setting up notation and tools that
allow us to compute amplitudes with ease.  We apply these steps to
compute the emission rate of supergravity scalars from particular D1D5
states.  In particular we can compute emissions in cases where it was
unclear how to proceed with the effective string model. The CFT
amplitudes, converted to radiation rates by our general formalism,
show exact agreement with the emission rates in the dual gravitational
geometry.

Specifically, we perform the following steps:
\begin{enumerate}
\item Traditionally, one uses AdS/CFT to compute correlation functions
  in the CFT and compares them to quantities computed in the AdS
  geometry, but we are interested in finding the interactions of the
  brane system with quanta coming in from or leaving to \emph{flat}
  infinity.  Thus we must consider the full metric of the branes,
  where at large $r$ the $AdS$ region changes to a `neck' and finally
  to flat space.  We write a general expression for $\Gamma$, the rate
  of radiation to infinity, in terms of the CFT amplitude for the
  decay process.

\item As an example of our CFT techniques we consider  minimal scalars 
  in the D1D5 geometry. An example of such scalars is given by the
  graviton with indices along the compact torus directions. We
  construct the correctly normalized vertex operators for these
  scalars, which are obtained by starting with a twist operator in the
  CFT and dressing it with appropriate modes of the chiral algebra.

\item We use the notion of `spectral flow' to map states from the
  Ramond sector (which describes the D1D5 bound state) to the NS
  sector (which has the vacuum $\vac_{NS}$). This map helps us in two
  ways. First, it is not clear which states in the Ramond sector
  correspond to supergravity excitations (as opposed to string
  excitations). In the NS sector there \emph{is} a simple map: the
  supergravity excitations are given by chiral primaries and their
  descendants under the anomaly-free subalgebra of the chiral algebra.
  Thus the spectral flow map allows us to find the initial and final
  states of our emission process, given the quantum numbers of these
  states. Second, for the process of interest the final state turns
  out to be the \emph{vacuum} in the NS description; thus we do not
  need an explicit vertex operator insertion to create this state when
  computing amplitudes after spectral flowing to the NS sector.

\item We consider a simple decay process where an excited state of the
  CFT decays to the ground state and emits a supergravity quantum. We
  compute the CFT amplitude for this emission process. As mentioned
  above, the final state in the amplitude is nontrivial in the Ramond
  sector, but maps to the NS vacuum $\vac_{NS}$ under spectral flow.
  This converts a 3-point correlator in the Ramond sector to a 2-point
  correlator in the NS sector.  With this amplitude, we use the result
  in (i) to compute $\Gamma$, the rate of emission to flat space.

\item In the above computations we take the initial state to
  contain a few excitations above the ground state, and we compute the
  decay rate for these excitations. Alternatively, we can choose to
  start with the initial state having \emph{no} excitations in the NS
  sector, and then perform a spectral flow on this state. This
  spectral flow adds a large amount of energy to the state, giving a
  configuration which is described in the dual gravity by a
  \emph{nonextremal} classical metric~\cite{ross}. This metric is
  known to emit radiation by ergoregion emission~\cite{myers}, and
  in~\cite{cm1,cm2,cm3} it was shown that for a subset of possible
  supergravity emissions the CFT rate agreed exactly with the gravity
  emission rate.  We can now extend this agreement to all allowed
  emissions of supergravity quanta by using the orbifold CFT
  computations developed in the present paper. It turns out that the
  simple decay process computed in steps~(iii) and (iv) can be used to
  give the emission rate from this highly excited initial state. This
  is because using spectral flow the initial state can be mapped to
  the vacuum, and in fact the entire amplitude maps under spectral
  flow to a time reversed version of the decay amplitude computed
  above.  We find exact agreement with the radiation rate in the dual
  gravity description.
\end{enumerate}

\section{Coupling flat space fields to the CFT}
\label{sec:coupling}

\begin{figure}[ht]
\begin{center}
\subfigure[]
{\label{fig:throats-a}
	\includegraphics[width=6.3cm]{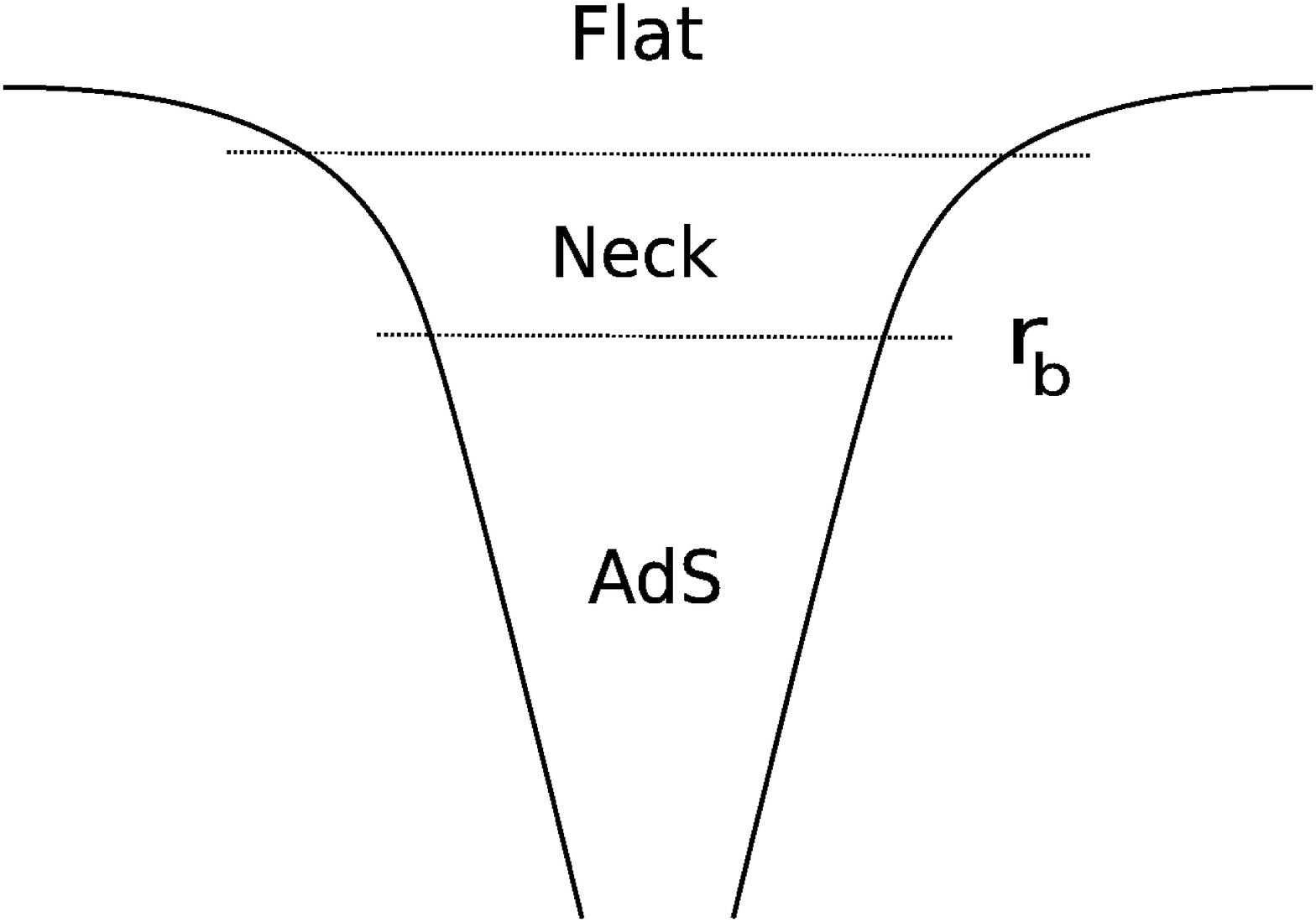}}
\hspace{15pt}
\subfigure[]{\label{fig:throats-b}
	\includegraphics[width=6.3cm]{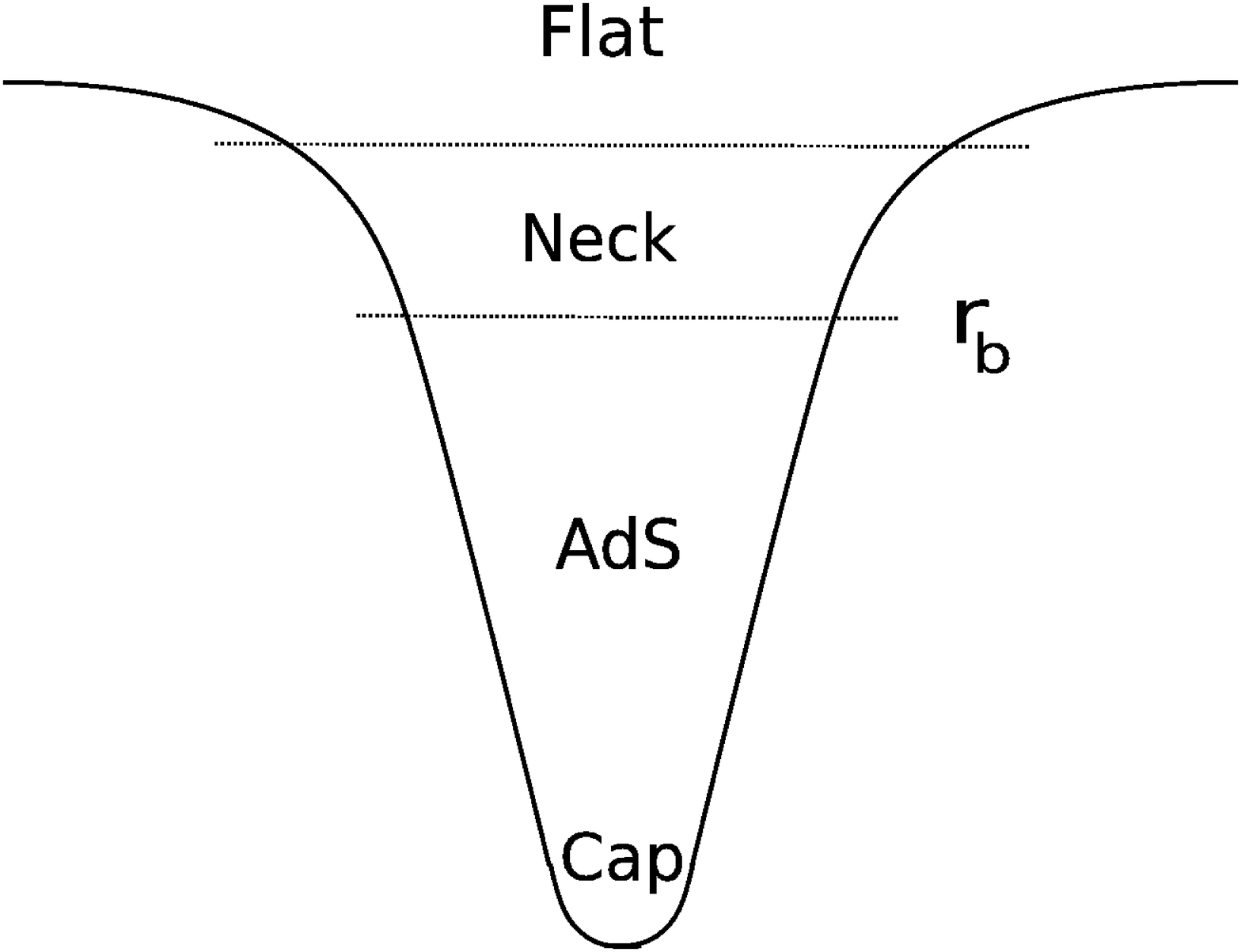}}
\caption{(a) The geometry of branes is flat at infinity, then we have a `neck', and further in the geometry takes the form $AdS_{d+1}\times S^p$ (b) Still further in, the geometry ends in a `fuzzball cap' whose structure is determined by the choice of microstate. For the simple state that we will choose for the D1D5 system, the Cap+AdS region is just a part of global $AdS$. \label{fig:throats}}
\end{center}
\end{figure}

Consider the geometry traditionally written down for branes that have
a near-horizon $AdS_{d+1}\times S^p$ region. This geometry has the
form
\be
ds^2=H^{-{2\over d}} \left[-dt^2+\sum_{i=1}^{d-1} dy_idy_i\right] +
H^{2\over p-1}\left[dr^2+r^2d\Omega_{p}^2\right].
\label{fullmetric}
\ee
If there is only one kind of brane producing the metric (and hence only one length scale) the function $H$ is given by
\be
H=1+ \f{Q}{r^{p-1}}.
\label{caseone}
\ee
The BPS black holes studied in string theory are constructed from $\mathcal B$ sets of mutually BPS branes.  In these cases $H$ is given by
\be
H= \prod_{i=1}^\mathcal B \left( 1+ \f{Q_i}{r^{p-1}} \right)^\f{1}{\mathcal B}.
\ee
(This reduces to (\ref{caseone}) for $\mathcal B=1$.) Let $Q_\text{max}$ be the
largest of the $Q_i$ and $Q_\text{min}$ the smallest.  It is
convenient to define the length scale
\be
R_s = \left( \prod_{i=1}^\mathcal B Q_i \right)^\f{1}{\mathcal B(p-1)}.
\ee
For small $r$ the angular directions give a sphere with radius $R_s$. 

We picture such a geometry in Fig.~\ref{fig:throats-a}. The geometry
has three regions:
\begin{enumerate}
\item {\it The outer region:}\quad For large $r$,
\be
r\gg Q_{\text{max}}^{1\over p-1},
\ee
we have essentially flat space.

\item {\it The intermediate region:}\quad For smaller $r$ we find a `neck', which we write as
\be
 C Q_{\text{min}}^{1\over p-1}<r<D Q_{\text{max}}^{1\over p-1}; ~~~(C\ll1, ~~D\gg 1). \label{neck}
 \ee
 
\item\label{ads-region} {\it The inner region:}\quad For
\be
 r<C Q_{\text{min}}^{1\over p-1},
\ee
we can replace the harmonic function by a power:
\be
 H \rightarrow \left( \f{R_s}{r} \right)^{p-1}.
\ee
The directions $y_i$ join up with $r$ to make an $AdS$ space, and the angular directions become a sphere of constant radius:
\be
ds^2\approx  \left[\left({r \over R_s}\right)^{2 (p-1) \over d}
     \left(-dt^2+\sum_{i=1}^{d-1} dy_idy_i\right)
     +R_s^2 {dr^2\over r^2}\right]+R_s^2 d\Omega_p^2,
\label{metricads}
\ee
which is $AdS_{d+1}\times S^p$. To see this we define the new radial coordinate $\tilde r$ by
\be
\left( \f{\tilde r}{R_s} \right) = \left( \f{r}{R_s} \right)^\f{p-1}{d}. \label{scaledRadialCoordinate}
\ee
In terms of this new radial coordinate one has the inner region metric,
\be
ds^2 \approx \left[ \left( \f{\tilde r}{R_s} \right)^2 \left( -dt^2 + \sum_{i=1}^{d-1} dy_i dy_i \right) + \left( R_s \f{d}{p-1} \right)^2 \f{d {\tilde r}^2}{{\tilde r}^2} \right]+ R_s^2 d\Omega_p^2. \label{metricadsScaled}
\ee
We can scale $t, y_i$ to put the $AdS$ into standard form, but it is
more convenient to leave it as above, since the coordinates $t,
y_i$ are natural coordinates at infinity and we need to relate
the $AdS$ physics to physics at infinity.  The radius of $AdS_{d+1}$
and the sphere $S^p$ are given by
\be
R_{\text{AdS}_{d+1}}=R_s \f{d}{p-1}, \qquad R_{S^p} =R_s.
\ee
\end{enumerate}

In region~(iii), we can put a boundary at
\be
\tilde r={\tilde r}_b,
\ee
which we regard as the boundary of the $AdS$ space. We can then
replace the space at $\tilde r<{\tilde r}_b$ by a dual CFT. Thus,
traditional AdS/CFT calculations are carried out only with the region
$\tilde r<{\tilde r}_b$. Our interest, however, is in the emission and
absorption of quanta between the $AdS$ region and asymptotic
infinity. Thus we need a formalism to couple quanta in region~(i) to
the CFT.

Next, we note that this `traditional' $AdS$ geometry cannot be
completely right. In the case of the D1D5 system, we know that the
ground state has a large degeneracy $\sim
\exp[2\sqrt{2}\pi\sqrt{N_1N_5}]$. At sufficiently large $r$, all these
states have the form \eqref{fullmetric}. But at smaller $r$ these
states differ from each other. None of the states has a horizon;
instead each ends in a different `fuzzball
cap'~\cite{ross,dasmandal,BalaDeBoer,lm4,lm5,lmm,mss,gms1,gms2,st-1,st-2,st-3,st-4,bena-6,bena-7,bena-8,bena-9,gimon-1,gimon-2,gimon-3,gimon-4,BDSMB,BSMB,fuzzballs2-1,fuzzballs2-2,fuzzballs2-3,fuzzballs2-4, lunin, giusto-4,giusto-5}.
We can choose special states where this cap is given by a classical
geometry. In the simplest case the cap is such that the entire region
$\tilde{r}<\tilde{r}_b$ has the geometry of \emph{global} $AdS_3\times
S^3$. We picture the full geometry for this state in
Fig.~\ref{fig:throats-b}.

The geometry without the cap (Fig.~\ref{fig:throats-a}) would have
needed to be supplemented with a boundary condition for the gravity
fields at $\tilde r=0$.  But since the actual states of the system (e.g.
Fig.~\ref{fig:throats-b}) have `caps', we do have a well defined
duality between the physics of the region $\tilde{r}<\tilde{r}_b$ and
the CFT at $\tilde{r}_b$.

\subsection{The coupling between gravity fields and CFT operators}

Let us start with region~(iii), where we have the $AdS$/CFT duality
map. This map says that the partition function of the CFT, computed
with sources $\phi_b$, equals the partition function for gravity in
the AdS, with field values at the boundary equal to
$\phi_b$~\cite{witten}:
\be\label{eq:prelim-coupling}
\int D[X]\, e^{-S_\text{CFT}[X]+\mu\int \drm^d y \sqrt{g_d} \phi_b(y) \mathcal{V}(y)}
                    = \int D[\phi]\, e^{-S_\text{AdS}[\phi]}\bigg|_{\phi({\tilde r}_b)=\phi_b}.
\ee
Here we have rotated to Euclidean spacetime, setting $t=y_{d}$.  The
symbol $\phi$ denotes fields in gravity, and $\hat{\mathcal{V}}$ are
CFT operators (depending on $X$) coupling to the gravity fields. Our
first task is to determine the coupling constant $\mu$, for the
normalizations of $\phi_b$ and $\hat{\mathcal{V}}$ that we choose.

Let us discuss the angular dependence in more detail. On $S^p$ the
fields $\phi(\tilde{r}, y, \Omega)$ can be decomposed into spherical harmonics,
\begin{equation}
\phi(\tilde{r}, y, \Omega) = \sum_{l, \vec{m}}\tilde{\phi}_{l, \vec{m}}(\tilde{r}, y) 
	Y_{l,\vec{m}}(\Omega).
\end{equation}
For the remainder of this discussion we consider a fixed $l$ mode.
Note that the reality of $\phi$ imposes the condition
\begin{equation}
\tilde{\phi}_{l, \vec{m}}^* = \tilde{\phi}_{l, -\vec{m}}.
\end{equation}
Thus if we expand field in the usual spherical harmonics then we should write Equation~\eqref{eq:prelim-coupling} as
\begin{equation}\label{eq:ang-dep-adscft}
\int D[X]\exp\left(-S_\text{CFT}[X] 
	+ \mu \sum_{\vec{m}}\int\drm^d y\sqrt{g_d}\tilde{\phi}_{b,\vec{m}}\mathcal{V}_{\vec{m}}\right)
	= \int D[\phi]
	e^{-S_\text{AdS}}
		\bigg|_{\tilde{\phi}_{\vec{m}}(\tilde{r}_b, y)\equiv\tilde{\phi}_{b,\vec{m}}(y)}.
\end{equation}
Reality of the action requires
\begin{equation}
\hat{\mathcal{V}}_{\vec{m}}^\dg(y) = \hat{\mathcal{V}}_{-\vec{m}}(y).
\end{equation}

At leading order we can replace the gravity path integral by the
classical action evaluated with the given boundary values of the
gravity fields,
\be
\int D[\phi] e^{-S_\text{AdS}[\phi]}\bigg|_{\phi(r_b)=\phi_b}=e^{-S_\text{AdS}[\phi^\text{cl.}]}.
\ee
Then the 2-point function in the CFT is given by
\be
\vev{\hat{\mathcal{V}}_{\vec{m}}(y_1)\hat{\mathcal{V}}_{\vec{m}'}(y_2)}
   ={1\over \mu^2}{\delta\over \delta \tilde{\phi}_{b,\vec{m}}(y_1)}
	{\delta\over \delta \phi_{b,\vec{m}'}(y_2)}
		\big(-S_\text{AdS}\big).
\label{final}
\ee

Let us now define the normalizations of $\phi$ and
$\hat{\mathcal{V}}$. We consider a minimal scalar field for
concreteness, though our computations should be extendable to other
supergravity fields with no difficulty. The gravity action is
\be
S_\text{AdS} = {1\over 16\pi G_D} \int \drm^Dx \sqrt{g} \left({1\over 2}\p\phi\p\phi\right),
\ee
where $D=d+1+p$ is the dimension of the spacetime ($D=10$ for string
theory, $D=11$ for M theory, and we have $D=6$ for the D1D5 system
after we reduce on a compact $T^4$). In region (iii), the spacetime
has the form
\be
AdS_{d+1}\times S^p\times {\cal M}.
\ee
For cases like the D1D5 system we have in addition a compact 4-manifold ${\cal M}=T^4$ or
${\cal M}=K3$.  We take $\phi$ to be a zero mode on ${\cal M}$ and
dimensionally reduce on ${\cal M}$, so that we again have a space
$AdS_{3}\times S^3$, with $G$ now the 6-d Newton's constant.

 If the
spherical harmonics are normalized such that,
\be
\int \drm\Omega\, |Y_{l, \vec{m}}(\Omega)|^2=1,
\ee
then dimensionally reducing on $S^p$ yields
\be
S_\text{AdS}={R_s^p\over 16\pi G_D}\sum_{\vec{m}}\int \drm^{d+1} y \sqrt{g_{d+1}}
   \left[\frac{1}{2} |\p\t\phi_{\vec{m}}|^2 +{1\over 2} {m^2}|\t\phi_{\vec{m}}|^2\right].
\ee
The $l$-dependent mass, $m$, comes from
\be\label{mass}
m^2={\Lambda\over R_s^2}, \qquad \triangle_p Y(\Omega) = -\Lambda Y(\Omega), \qquad
	\Lambda =l(l+p-1).
\ee

The CFT lives on the surface $\tilde r={\tilde r}_b$. The metric on this surface is
(from (\ref{metricadsScaled}))
\be
ds^2=\left({{\tilde r}_b\over R_s}\right)^2\sum _{i=1}^{d}dy_idy_i.
\label{metricb}
\ee
We choose the normalization of the operators $\hat{\mathcal{V}}$ by
requiring the 2-point function to have the short distance expansion
$\sim {1\over (\text{distance})^{2\Delta}}$:
\be
\vev{\hat{\mathcal{V}}_{\vec{m}} (y_1) \hat{\mathcal{V}}_{\vec{m}'}(y_2)}
	= \frac{\delta_{\vec{m} + \vec{m}',0}}
		{\left[({{\tilde r}_b\over R_s})|y_1-y_2|\right]^{2\Delta}}.
\label{first}
\ee
Following~\cite{witten}, we define the boundary-to-bulk propagator
which gives the value of $\phi$ in the $AdS$ region given its value on
the boundary at $r_b$:
\be
\tilde{\phi}_{\vec{m}}(\tilde r, y) = \int K(\tilde r,y; y')
		\tilde{\phi}_{b,\vec{m}}(y') \sqrt{g_d} \drm^d y',
\ee
where $\sqrt{g_d}d^d y'$ is the volume element on the metric
(\ref{metricb}) on the boundary. We have
\be
K(\tilde r,y;y') =  \f{R_\text{AdS}^{2 \Delta-d}}{{\tilde r}_b^{\Delta} \pi^\f{d}{2} } \f{\Gamma(\Delta) }{\Gamma(\Delta- \f{d}{2})}  \left[ \f{\tilde r}{R_\text{AdS}^2 + \f{{\tilde r}^2}{R_s^2}  |y-y'|^2} \right]^\Delta,
\label{kernel}
\ee
where
\be
\Delta=\h(d+\sqrt{d^2+4 m^2 R_\text{AdS}^2}) = (l+p-1) \f{d}{p-1}. \label{delta}
\ee
We then have
\be
{\delta\over \delta \tilde{\phi}_{b,\vec{m}}(y_1)}
	{\delta\over \delta\phi_{b,\vec{m}'}(y_2)}\big(-S_\text{AdS}\big)
  =-\frac{ R_s^p}{16\pi G_D}\delta_{\vec{m}+\vec{m}',0}
	\left({2\Delta-d\over \Delta}\right) 
	\p_{\tilde r} K(\tilde r, y_1; y_2)\Big(\sqrt{g^{\tilde r\tilde r}}
		\Big|_{\tilde r=\tilde r_b}\Big ),
\label{second}
\ee
where the extra factor ${2\Delta-d\over \Delta}$ comes from taking
care with the limit $\tilde{r}_b\rightarrow\infty$ when using the
kernel~\eqref{kernel}~\cite{mathurfreedman}.

Putting (\ref{first}) and (\ref{second}) into (\ref{final}) we get
\be
\mu=\Bigg [ {R_s^{2\Delta-(d+1)+p} \left( \f{d}{p-1}\right)^{2\Delta-(d+1)} \over 16\pi G_D}{(2\Delta-d)\over \pi^{d\over 2}}{\Gamma(\Delta)\over \Gamma(\Delta-{d\over 2})}\Bigg ] ^{1\over 2}.
\ee

\subsection{The outer region}

The wave equation for the minimal scalar is
\be
\square \phi=0.
\ee
We write
\be
\phi=h(r)Y(\Omega)e^{-iE t}e^{i\vec \lambda \cdot \vec y}, \label{WaveAnsatz}
\ee
getting the solution
\be
h={1\over r^{p-1\over 2}}\left[C_1\, J_{l+{p-1\over 2}}\left(\sqrt{E^2-\lambda^2} \,r\right)
     +C_2\, J_{-l-{p-1\over 2}}\left(\sqrt{E^2-\lambda^2} \,r\right)\right].
\ee
Note that we can define a CFT in the $AdS$ region only if the
excitations in the $AdS$ region decouple to leading order from the
flat space part of the geometry \cite{maldacena}. Such an approximate
decoupling happens for quanta with energies $E\ll 1/R_s$: waves
incident from infinity with wavelengths much longer than the $AdS$
curvature scale almost completely reflect off the 'neck' region and
there is only a small probability of absorption into the $AdS$ part of
the geometry. Correspondingly, waves with such energies trapped in the
$AdS$ region have only a small rate of leakage to flat space. Thus we
work throughout this paper with the assumption
\be
E\ll \frac{1}{R_s}.
\ee
With this, we find that to leading order the wave in the outer region
has the $C_2\approx 0$ \cite{hottube}:
\be
h\approx
C_1{1\over \Gamma(l+{p+1\over 2})}
   \left[{\sqrt{E^2-\lambda^2}\over 2}\right]^{l+{p-1\over 2}}r^l.
\ee

A general wave is a superposition of different modes of the
form~\eqref{WaveAnsatz}. We wish to extract a given spherical harmonic
from this wave, so that we can couple it to the appropriate vertex
operator of the CFT. Define
\begin{equation}\label{eq:diff-op}
\big[\p^l \phi\big]^{l,\vec{m}} 
= Y^{k_1k_2\dots k_l}_{l,\vec{m}}
	\p_{k_1}\p_{k_2}\dots\p_{k_l}\phi,
\end{equation}
where the above differential operator is normalized such
that\footnote{See Appendix~\ref{ap:spherical} for more details.}
\begin{equation}\label{eq:diff-op-norm}
Y^{k_1k_2\cdots k_l}_{l,\vec{m}}
	\p_{k_1}\p_{k_2}\dots\p_{k_l}\big[r^{l'} Y_{l', \vec{m}'}(\Omega)\big]
= \delta_{ll'}\delta_{\vec{m}, \vec{m}'}.
\end{equation}
Thus the required angular component of $\phi$ at small $r$ satisfies
\be
\phi\approx [\p^l\phi]\Big|^{l,\vec{m}}_{r\rightarrow 0}~r^l Y_{l,\vec{m}}(\Omega).
\ee

\subsection{The intermediate region}

In the `neck' region we can set $E, \lambda$ to zero in solving the
wave equation, since we assume that we are at low energies and momenta
so the wavelength is large compared to the size of the intermediate
region. Thus the $E, \lambda$ terms do not induce oscillations of the
waveform in the limited domain of the intermediate region.

With this approximation we now have to solve the wave equation in the
intermediate region. From this solution, we need the following
information to construct our full solution. Suppose in the outer part
of the intermediate region $r \sim Q_{\text{max}}^{1\over p-1}$ we
have the solution
\be
\phi\approx r^l.
\ee
Evolved to the inner part of the intermediate region $r\ll Q^{1\over
p-1}$, we have a form given by $AdS$ physics:
\be
\phi\approx b_l {\tilde r}^{\Delta-d}.
\ee
These two numbers, $b_l, \Delta$, give the information we need about
the effect of the intermediate region on the wavefunction. $\Delta$
is known from the CFT, while the number $b_l$ appears in our
final expression for the emission amplitude as representing the
physics of the intermediate region which connects the $AdS$ region to
flat infinity.

We now note that for the case that we work with, the minimally
coupled scalar, we can in fact write down the values of $b_l,
\Delta$. The wave equation for a minimally coupled scalar in the
background \bref{fullmetric} with the ansatz \bref{WaveAnsatz} takes
the form
\be
H^\f{2(d+p-1)}{d(p-1)} (E^2 - \lambda^2) r^2 h(r) + \f{1}{r^{p-2}} \partial_r( r^p \partial_r h(r)) - l(l+p-1) h(r)=0.
\ee
The term
\be
H^\f{2(d+p-1)}{d(p-1)} r^2
\ee
is bounded in the `neck' region \bref{neck}. Therefore, assuming small
$E, \lambda$, the wave equation in the neck is
\be
\f{1}{r^{p-2}} \partial_r( r^p \partial_r h(r)) - l(l+p-1) h(r)=0.
\ee
This has the solution
\be
h(r) = A r^l + B r^{-l-p+1}.
\ee 
Thus we see that if we use the coordinate $r$ throughout the
intermediate region, then there is no change in the functional form of
$\phi$ as we pass through the intermediate region. We are interested
in the $r^l$ solution, so we set $B=0$. We must now write this in
terms of the coordinate $\tilde r$ appropriate for the $AdS$
region. First consider the case $d=p-1$ which holds for the D3 and
D1D5 cases. Then we see that
\be
\tilde r=r, ~~~b_l=1.
\ee
Now consider $d\ne p-1$. 
From \bref{scaledRadialCoordinate} and \bref{delta} we get
\be
b_l=R_s^{l(1- \f{d}{p-1})}.
\ee

In general the scalars are not `minimal'; i.e., they can have
couplings to the gauge fields present in the geometry. Then $b_l$
needs to be computed by looking at the appropriate wave equation.
Examples of such non-minimal scalars are `fixed' scalars discussed
in~\cite{Callan:1996tv}.

To summarize, we find that the change of the waveform through the
intermediate region is given by $b_l, \Delta$. The wave at the
boundary $\tilde{r}=\tilde{r}_b$ is then
\be
\tilde{\phi}_{b,\vec{m}}(y)=b_l ~\tilde{r}_b^{\Delta-d}~ [\p^l\phi(y)]\Big|^{l,\vec{m}}_{r=0}.
\ee

\subsection{The interaction}

We can break the action of the full problem into three parts
\begin{equation}
S_\text{total} = S_\text{CFT} + S_\text{outer} + S_\text{int},
\end{equation}
where the contribution of the interaction between the CFT and the
outer asymptotically flat region, $S_\text{int}$, vanishes in the
strict decoupling limit. We work in the limit where the interaction is
small but nonvanishing, to first order in the interaction.

The coupling of the external wave \emph{at $r_b$} to the
CFT is given by the interaction
\be
S^l_\text{int}= -\mu \sum_{\vec{m}} 
	\int \sqrt{g_d} ~\drm^dy\,  \tilde{\phi}_{b,\vec{m}}(y) \hat{\mathcal{V}}_{l,\vec{m}}(y).
\ee
If we want to \emph{directly} couple to the modes in the \emph{outer
region}, then we can incorporate the intermediate region physics into
$S_\text{int}$ by writing
\be\label{eq:general-S-int}
S^l_\text{int} = -c_l\sum_{\vec{m}}\int \sqrt{g_d} ~\drm^dy\,  [\p^l\phi(y)]\Big|^{l,\vec{m}}_{r=0}~
   \hat{\mathcal{V}}_{l,\vec{m}}(y),
\ee
where
\be
c_l=\mu~ b_l~{\tilde r}_b^{\Delta-d}.
\label{coupling}
\ee
This is the general action connecting modes in the AdS/CFT with the
modes in the asymptotically flat space. We focus specifically on
(first-order) emission processes, but one can consider more general
interactions (absorption, scattering, etc.).

\subsection{Emission}

Suppose we have an excited state in the `cap' region, $\ket{i}$, and
the vacuum in the outer region, $\vac_{\text{outer}}$.  Because of the
coupling~\eqref{coupling}, a particle can be emitted and escape to
infinity, changing the state in the cap to a lower-energy state
$\ket{f}$, and leaving a 1-particle state in the outer region,
$\ket{E,l,\vec{m}, \vec{\lambda}}_\text{outer}$.  We wish to compute the
rate for this emission, $\Gamma$. We can write the total amplitude for
this process as
\begin{equation}
\mathcal{A} = \bigg({}_\text{outer}\bra{E, l, \vec{m}, \vec{\lambda}}
     \bra{f}\bigg)\; iS_\text{int}\;\bigg(\ket{i}\vac_\text{outer}\bigg).
\end{equation}

We quantize the field $\phi$ in the outer region as
\begin{equation}\begin{split}
\hat{\phi} = \sqrt{\frac{16\pi G_D}{2V_y} }
  \sum_{\vec \lambda, l,\vec{m}}\int_0^\infty\drm E\,
  \tfrac{J_{l+{p+1\over 2}}(\sqrt{E^2-\lambda^2}~r)}{r^{{p-1\over 2}}}
  \bigg[&\hat{a}_{E,l, \vec{m},\lambda}
    Y_{l,\vec{m}}e^{i(\vec \lambda\cdot \vec  y - E t)}\\
  &+(\hat{a}_{E,l, \vec{m},\lambda})^\dg 
    Y_{l,\vec{m}}^*e^{-i(\vec \lambda \cdot\vec y - E t)}\bigg],
\end{split}\end{equation}
where
\begin{equation}
\com{\hat{a}_{E,l,\vec{m},\lambda}}
{(\hat{a}_{E',l',\vec{m}', \lambda'})^\dg}
= 
  \delta_{ll'}\delta_{\vec{m},\vec{m}'}
  \delta_{\lambda\lambda'}\delta(E-E').
\end{equation}

Using the asymptotic behavior,
\be
J_\nu(z)\approx {1\over \Gamma(\nu+1)}\Big ( {z\over 2}\Big)^\nu,
\ee
we find that
\begin{multline}
[\p^l\hat{\phi}]\Big|^{\vec{m}}_{r=0} = \sqrt{\frac{16\pi G_D}{2V_y}}
\sum_{\vec{\lambda}}\int_0^\infty\drm E\,
{({\sqrt{E^2-\lambda^2}\over 2})^{l+{p-1\over 2}}\over \Gamma(l+{p+1\over 2})}
    \bigg[\hat{a}_{E,l, \vec{m},\lambda}
    e^{i(\vec \lambda\cdot \vec  y - E t)}\\
  +(\hat{a}_{E,l, -\vec{m},\lambda})^\dg 
    e^{-i(\vec \lambda \cdot\vec y - E t)}\bigg].
\end{multline}
Using the coupling $c_l$ from (\ref{coupling}) we get the interaction
Lagrangian
\begin{multline}
S_\text{int} =  -\mu b_l {\tilde r}_b^{\Delta-d}\sqrt{\frac{16\pi G_D}{2V_y} }
\sum_{\vec \lambda, \vec{m}}\int\sqrt{g_d} \drm^d y\int_0^\infty\drm E\\    
{({\sqrt{E^2-\lambda^2}\over 2})^{l+{p-1\over 2}}\over \Gamma(l+{p+1\over 2})}
    \bigg[\hat{a}_{E,l, \vec{m},\lambda}e^{i(\vec \lambda\cdot \vec  y - E t)}
	  +(\hat{a}_{E,l,-\vec{m},\lambda})^\dg 
    e^{-i(\vec \lambda \cdot\vec y - E t)}\bigg]~\hat{\mathcal{V}}_{l,\vec{m}}(t,\vec y).
\end{multline}

We can pull out the $(t,\vec{y})$ dependence of the CFT part of the
amplitude by writing
\be 
\bra{f} \hat{\mathcal{V}}(t, \vec{y})\ket{i}=
     e^{-iE_0 t + i \vec \lambda_0\cdot \vec y}
         \bra{f}\hat{\mathcal{V}}(t=0, \vec y=0)\ket{i},
\ee
where $E_0$ and $\lambda_0$ can be determined from the initial and
final states of the CFT, $\ket{i}$ and $\ket{f}$. We also work in the
case where the initial and final CFT states select out a single $l,m$
mode in the interaction, whose indices have been suppressed. The CFT
lives on a space of (coordinate) volume $V_y$. When computing CFT
correlators we work in a `unit-sized' space with volume
$(2\pi)^{d-1}$. Scaling the operator $\hat{\mathcal{V}}$ we have
\be
\bra{f} \hat{\mathcal{V}}(t=0, \vec y=0)\ket{i}
     = \left[{(2\pi)^{d-1}\over ({{\tilde r}_b\over R_s})^{d-1} V_y}\right]^{\Delta\over d-1}
       \bra{f}\hat{\mathcal{V}}(t=0, \vec y=0)\ket{i}_\text{unit}.
\ee
Rotating back to Lorentzian signature, the amplitude for emission of a
quantum from an excited state of the CFT is
\begin{multline}
\mathcal{A}=-i \mu b_l {\tilde r}_b^{\Delta-d}\sqrt{\frac{16\pi G_D}{2V_y}}
{({\sqrt{E^2-\lambda^2}\over 2})^{l+{p-1\over 2}}\over \Gamma(l+{p+1\over 2})}
\left[ \frac{(2\pi)^{d-1}}{\big(\frac{{\tilde r}_b}{R_s}\big)^{d-1} V_y}\right]^{\Delta\over d-1}
   \bra{f} \hat{\mathcal{V}}_{l,\vec{-m}}(t=0, \vec y=0)\ket{i}_\text{unit}
  \\
 \times
\int_0^T \drm t \int \big(\tfrac{{\tilde r}_b}{R_s}\big)^d 
   \drm^{d-1} y\, e^{-i(E_0-E)t}e^{i(\vec{\lambda}_0-\vec{\lambda})\cdot \vec y}.
\end{multline}
The amplitude gives the emission rate in a straightforward calculation:
\begin{calc}\label{eq:general-decay-rate}
\der{\Gamma}{E} &= \lim_{T\to\infty}\frac{|\mathcal{A}|^2}{T}\\
       &=(2\pi)^{2\Delta+1}{R_s^{4\Delta-3d+p-1} 
         \left( \f{d}{p-1}\right)^{2 \Delta -(d+1)} \over V_y^{{2\Delta-d+1\over d-1}}}
\left[ {(2\Delta-d)\over 2\pi^{d\over 2}}{\Gamma(\Delta)\over \Gamma(\Delta-{d\over 2})}\right]
  |b_l|^2\left[{1\over \Gamma(l+{p+1\over 2})}\right]^2\\
  &\qquad\times\left({{E^2-\lambda^2}\over 4}\right)^{l+{p-1\over 2}}
   \Big|\bra{0}\hat{\mathcal{V}}_{l,-\vec{m}}(0)\ket{1}_\text{unit}\Big|^2\,
   \delta_{\vec{\lambda},\vec{\lambda}_0}\delta(E- E_0).
\end{calc}
From this expression, we see that in the strict decoupling limit where $E R_s\rightarrow \infty$ this rate vanishes as expected.

We have derived the above result for a general CFT and its corresponding brane geometry.  In the remainder of the
paper we work with the D1D5 system. For a minimal scalar in the D1D5 geometry we have
\begin{equation}
d = 2, \qquad p = 3\qquad \Delta_\text{tot.} = l + 2\qquad b_l =1\qquad
R_s = (Q_1Q_5)^\frac{1}{4}\qquad V_y = 2\pi R.
\end{equation}
Plugging in, we reduce the decay rate formula to the form
\begin{equation}\label{eq:D1D5-decay-rate}
\der{\Gamma}{E} = 
	\frac{2\pi}{2^{2l+1}\,l!^2}\frac{(Q_1Q_5)^{l+1}}{R^{2l+3}}(E^2-\lambda^2)^{l+1}\,
	|\bra{0}\hat{\mathcal{V}}\ket{1}_\text{unit}|^2\,
	\delta_{\lambda,\lambda_0}\delta(E - E_0).
\end{equation}

\section{The D1D5 CFT at the orbifold point}

\subsection{The CFT}

Consider type IIB string theory, compactified as
\be
M_{9,1}\rightarrow M_{4,1}\times S^1\times T^4.
\label{compact}
\ee
Wrap $N_1$ D1 branes on $S^1$, and $N_5$ D5 branes on $S^1\times
T^4$. The bound state of these branes is described by a field
theory. We think of the $S^1$ as being large compared to the $T^4$, so
that at low energies we look for excitations only in the direction
$S^1$.  This low energy limit gives a conformal field theory (CFT) on
the circle $S^1$.

We can vary the moduli of string theory (the string coupling $g$, the
shape and size of the torus, the values of flat connections for gauge
fields etc.). These changes move us to different points in the moduli
space of the CFT. It has been conjectured that we can move to a point
called the `orbifold point' where the CFT is particularly simple
\cite{swD1D5, fmD1D5,deBoerD1D5, dijkgraafD1D5,frolov-1,frolov-2, Jevicki,David}. At this orbifold point the CFT is
a 1+1 dimensional sigma model. The 1+1 dimensional base space is
spanned by $(y, t)$, where
\be
0\le y<2\pi R
\ee
is a coordinate along the $S^1$, and $t$ is the time of the 10-d
string theory. For our CFT computations, we rotate time to Euclidean
time, and also use scaled coordinates $(\sigma, \tau)$ where the space
direction of the CFT has length $2\pi$:
\begin{equation}
\tau = i\frac{t}{R} \qquad \sigma = \frac{y}{R}.
\end{equation}
Moreover, we find it convenient to work in the complex plane with
coordinates $(z,\bar{z})$ defined by the exponential map,
\begin{equation}
z = e^{\tau + i\sigma}\qquad \bar{z} = e^{\tau - i\sigma}.
\end{equation}
We continue back to Lorentzian signature at the end.

The target space of the sigma model is the `symmetrized product' of
$N_1N_5$ copies of $T^4$,
\be
(T_4)^{N_1N_5}/S_{N_1N_5},
\ee
with each copy of $T^4$ giving 4 bosonic excitations $X^1, X^2, X^3,
X^4$. It also gives 4 fermionic excitations, which we call $\psi^1,
\psi^2, \psi^3, \psi^4$ for the left movers, and $\bar\psi^1,
\bar\psi^2,\bar\psi^3,\bar\psi^4$ for the right movers. The fermions can be
antiperiodic or periodic around the $\sigma$ circle. If they are
antiperiodic on the $S^1$ we are in the Neveu-Schwarz (NS) sector, and
if they are periodic on the $S^1$ we are in the Ramond (R)
sector\footnote{The periodicities flip when mapping to the complex
plane because of a Jacobian factor.}.

\subsubsection{Twist operators}

\begin{figure}[ht]
\begin{center}
\subfigure[~Untwisted component strings]{\label{fig:twist1}
	\includegraphics[width=6.3cm]{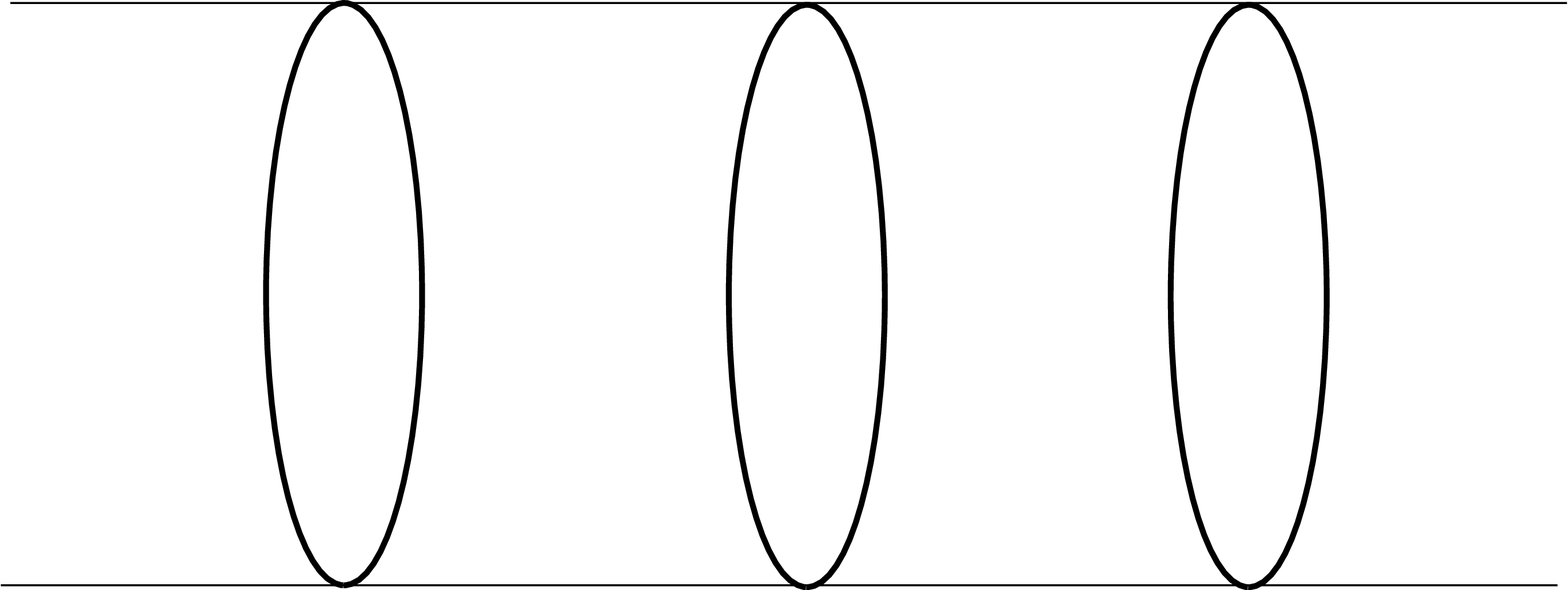}}
\raisebox{30pt}{$\xrightarrow{\hspace{5pt}{\displaystyle \sigma_3}\hspace{3pt}}$}
\subfigure[~The twisted component string]{\label{fig:twist2}
	\includegraphics[width=6.3cm]{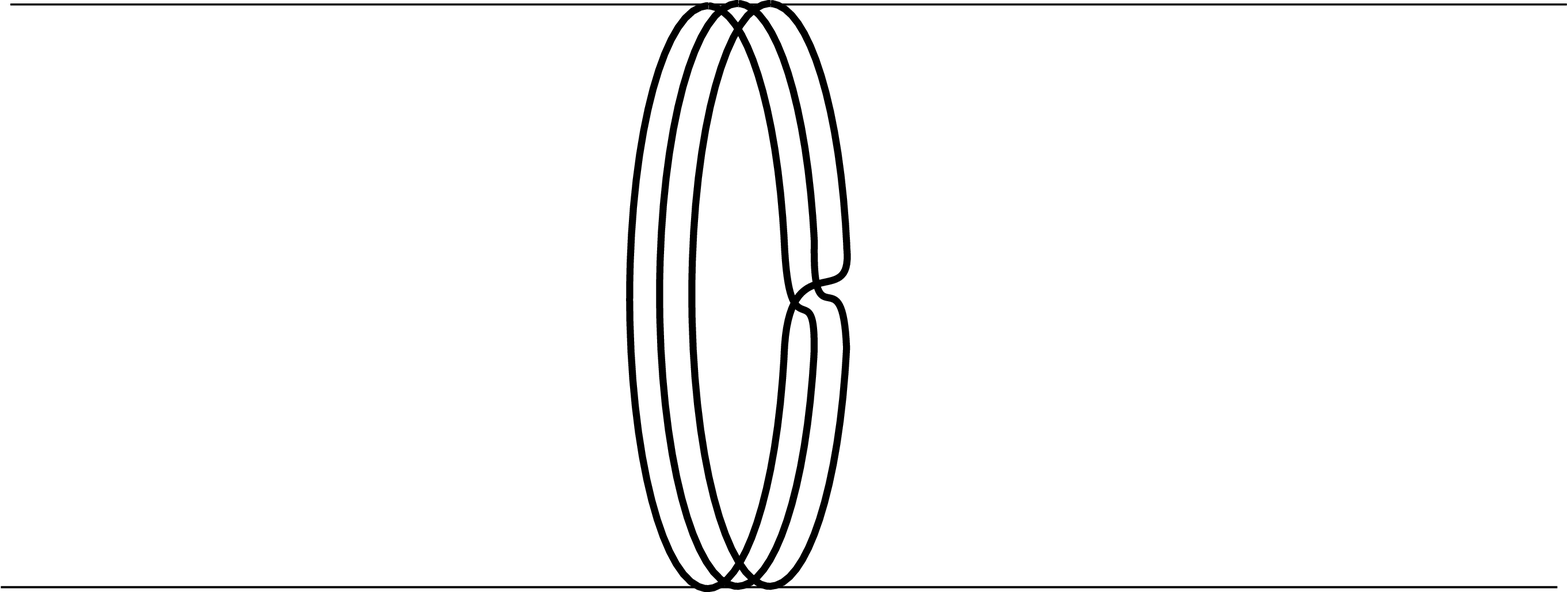}}
      \caption{The twist operator $\sigma_3$.   Each loop represents a
        `copy' of the CFT wrapping the $S^1$. The twist operator joins these copies into one single copy of the CFT living on a circle of three times the length of the original circle. \label{fig:twist}}
\end{center}
\end{figure}

Since we orbifold by the symmetric group $S_{N_1N_5}$, we generate
`twist sectors', which can be obtained by acting with `twist
operators' $\sigma_n$ on an untwisted state. Suppose we insert a
twist operator at a point $z$ in the base space. As we circle the
point $z$, different copies of $T^4$ get mapped into each other. Let
us denote the copy number by a subscript $a=1, 2, \dots n$. The twist
operator is labeled by the permutation it generates. For instance,
every time one circles the twist operator
\be
\sigma_{(123\dots n)},
\label{qone}
\ee 
the fields $X^i_{(a)}$ get mapped as
\begin{equation}
X^i_{(1)} \rightarrow
X^2_{(2)} \rightarrow
\cdots
\rightarrow
X^i_{(n)} \rightarrow X^i_{(1)},
\end{equation}
and the other copies of $X^i_{(a)}$ are unchanged. We have a similar
action on the fermionic fields. We depict this `twisting' in
Fig.~\ref{fig:twist}. Each set of linked copies of the CFT is called
one `component string'.

We often abbreviate a twist operator like the one in
Equation~\eqref{qone} with $\sigma_n$ for simplicity (we have to give
the indices involved in the permutation explicitly when we use
$\sigma_n$ in a correlator). We call these operators, $\sigma_n$,
`bare twists' to distinguish them from the more relevant operators for
our purpose, which have additional `charges' added to the `bare twist'
forming operators that are chiral primaries for the supersymmetric
CFT.

\subsubsection{Symmetries of the CFT}

The D1D5 CFT has $(4,4)$ supersymmetry, which means that we have
$\mathcal{N}=4$ supersymmetry in both the left and right moving
sectors. This leads to a superconformal ${\cal N}=4$ symmetry in both
the left and right sectors, generated by operators $L_{n}, G^\pm_{r},
J^a_n$ for the left movers and $\bar L_{n}, \bar G^\pm_{r}, \bar
J^a_n$ for the right movers. The expressions for the left generators
in terms of the $X^i, \psi^i$ are given in
Equation~\eqref{eq:currents-def}. The OPEs are given in
Equation~\eqref{eq:currents-OPE}, and the (anti-)commutation relations
between modes is given in Equation~\eqref{eq:mode-algebra}.

Each $\Nsc = 4$ algebra has an internal R symmetry group
SU(2),\footnote{In fact, the full R symmetry group of the
$\mathcal{N}=4$ algebra is $SO(4)$; however, the other $SU(2)$ does
not have a current associated with it within the algebra.} so there is
a global symmetry group $SU(2)_L\times SU(2)_R$.  We denote the
quantum numbers in these two $SU(2)$ groups as
\be
SU(2)_L: ~(j, m);~~~~~~~SU(2)_R: ~ (\bj, \bar m).
\ee
In the geometrical setting of the CFT, this symmetry arises from the
rotational symmetry in the 4 space directions of $M_{4,1}$ in
Equation~\eqref{compact},
\be
SO(4)_E\simeq SU(2)_L\times SU(2)_R.
\ee
Here the subscript $E$ stands for `external', which denotes that these
rotations are in the noncompact directions. These quantum numbers
therefore give the angular momentum of quanta in the gravity
description.  We have another $SO(4)$ symmetry in the four directions
of the $T^4$. This symmetry we call $SO(4)_I$ (where $I$ stands for
`internal'). This symmetry is broken by the compactification of the
torus, but at the orbifold point it still provides a useful organizing
principle. We write
\be
SO(4)_I\simeq SU(2)_1\times SU(2)_2.
\ee
We use spinor indices $\alpha, \dot\alpha$ for $SU(2)_L$ and $SU(2)_R$
respectively. We use spinor indices $A, \dot A$ for $SU(2)_1$ and
$SU(2)_2$ respectively.

The 4 real fermions of the left sector can be grouped into complex
fermions $\psi^{\alpha A}$ with the reality constraint
\begin{equation}
\big(\psi^{\alpha\dot{A}}\big)^\dg 
	=
	-\epsilon_{\alpha\beta}\epsilon_{\dot{A}\dot{B}}\psi^{\beta\dot{B}}
	= - \psi_{\alpha\dot{A}}.
\end{equation}
The right fermions have indices $\bar{\psi}^{\dot\alpha \dot A}$ with
a similar reality constraint. The bosons $X^i$ are a vector in the
$T^4$. Thus they have no charge under $SU(2)_L$ or $SU(2)_R$ and are
given by
\begin{equation}
[X]^{\dot{A}A} = X^i(\sigma^i)^{\dot{A}A}.
\end{equation}
where $\sigma^i, i=1, \dots 4$ are the three Pauli matrices and the identity. (The
notations described here are explained in full detail in
Appendix~\ref{ap:CFT-notation}.)

\subsection{States of the CFT}

The CFT arising from the D1D5 brane bound state is in the Ramond (R)
sector. One can understand this because the periodicities of the
fermions around the $S^1$ are inherited from the behavior of fermionic
supergravity fields around the $S^1$ in \eqref{compact}. These
supergravity fields must be taken to be periodic, since otherwise we
would generate a nonzero vacuum energy in our spacetime and the metric
far from the branes would not be flat.

Even though the physical CFT problem is in the R sector, we find it
convenient to map our R sector states to the NS sector using
\emph{spectral flow}, which simplifies the calculation. Using
spectral flow we can relate one calculation to a whole family of
related processes. Thus let us first look at states in the NS sector.

\subsection{States in the NS sector}

There are two general pieces of information needed to describe the
states of the orbifold CFT. First we have to look at the `twist
sector'; i.e., note which copies of the CFT are linked to which copies
as we go around the $S^1$.  The second thing we have to look at are
the bosonic and fermionic excitations in the given twist sector.

\begin{figure}[ht]
\begin{center}
\subfigure[~The NS vacuum state]{\label{fig:nsvacuumCFT}
	\raisebox{40pt}{\includegraphics[width=6.3cm]{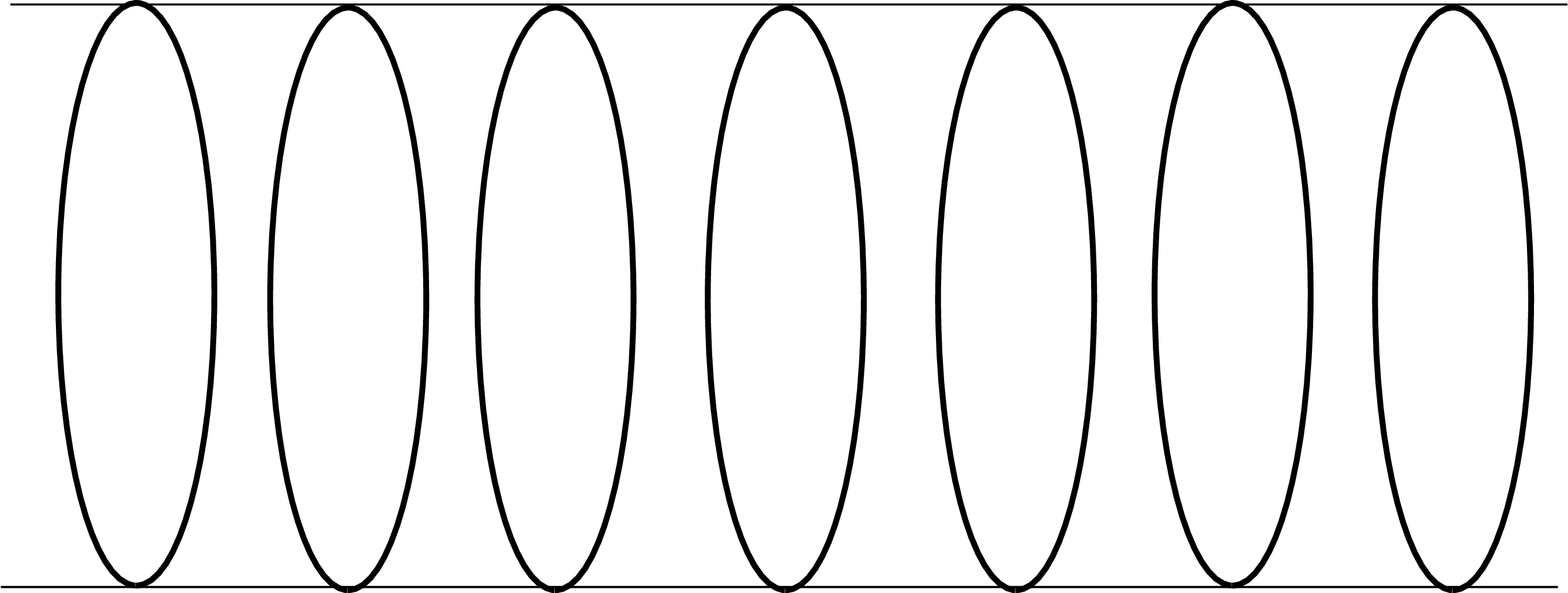}}}
\hspace{15pt}
\subfigure[~The gravity dual of the NS vacuum state]{\label{fig:nsvacuumGrav}
	\includegraphics[width=6.3cm]{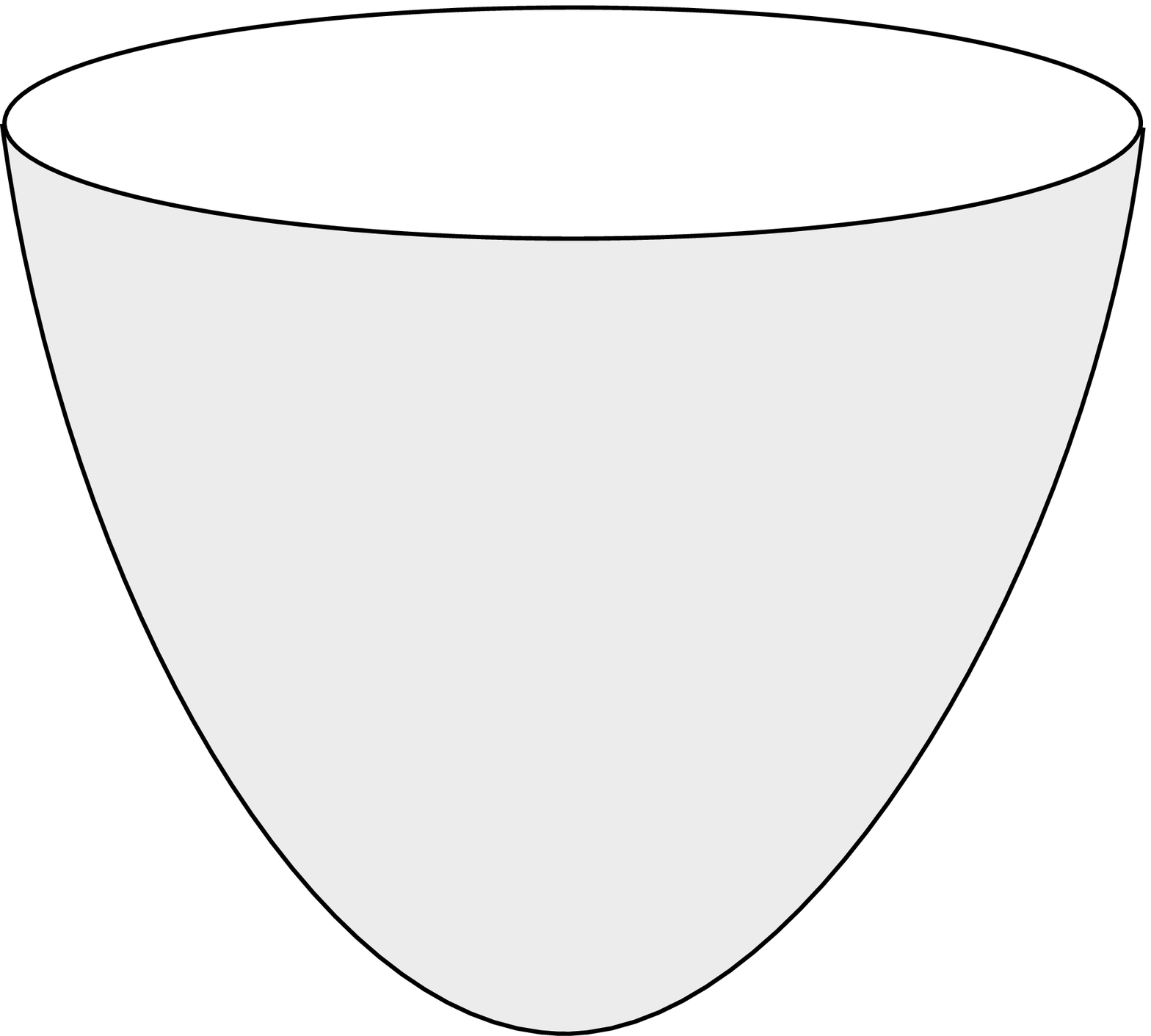}}
      \caption{(a) The NS vacuum state in the CFT and (b) its gravity dual, which is global $AdS$.
        The NS vacuum is the simplest possible state
        having no twists, no excitations, and no base
        spin.\label{fig:nsvacuum}}
\end{center}
\end{figure}

The simplest sector is the `untwisted sector', where the copies are
all delinked from each other. Let us take the state in this sector
with \emph{no} excitations. This state is depicted in
Fig.~\ref{fig:nsvacuumCFT}. It is the NS vacuum, and has quantum numbers
\be
\vac_{NS}:~~~~h=\bar h=0; ~~~j=m=\bj=\bar m=0.
\ee
The gravity dual of this state is `global AdS', depicted in
Fig.~\ref{fig:nsvacuumGrav}.

Now consider excited states. The simplest states are `chiral
primaries', which have `dimension=charge':
\be
h=m, ~~~\bar h=\bar m.
\ee
We can start with a chiral primary and act with the `anomaly-free
subalgebra' of the chiral algebra to make descendants. This subalgebra
is spanned by
\be
\big\{L_{-1}, L_0, L_1, G^\pm _{\pm {1\over 2}}, J^a_0\big\}.
\label{small}
\ee
All states obtained by starting with a chiral primary and applying
operators (\ref{small}) correspond to supergravity excitations in the
dual gravity theory. Other states correspond to `stringy' states. We
are interested in the emission of a supergravity quantum from the CFT
state, so let us look at such states in more detail.

\subsubsection{The basic chiral primary operators $\sigma^0_{l+1}$}

Let us recall the construction of chiral primary operators introduced
in~\cite{lm2}.

Start with the NS vacuum state $\vac_{NS}$, which is in the completely
untwisted sector, where all the component strings are `singly wound'. The
gravity dual is global AdS.

Now suppose in this gravity dual we want to add one supergravity
quantum carrying angular momentum ${l\over 2}$ in each of the two
factors of $SU(2)_L\times SU(2)_R$ (scalars must have $j=\jbar$).  We
take a set of $l+1$ copies of the CFT and join them by using a twist
operator $\sigma_{l+1}$ into one `multiply wound' component
string. Let us take the lowest energy state in this twist sector. This
state has dimensions~\cite{lm2}
\be
h=\bar h= \frac{c}{24}\left[(l+1) - \frac{1}{(l+1)}\right],
\label{start}
\ee
but it does not yet have any charge, so it is not a chiral
primary; it has more dimension than charge. The current operator $J^+$
carries positive charge, and we can apply contour integrals of $J^+$
to our state to raise its charge. Since all operators in the theory
have $h>|m|$, one might think that we cannot reach a chiral primary with
$h=j=m$ if we start with the state (\ref{start}); however, on the twisted
component string we can apply \emph{fractional} modes $J^a_{k\over
l+1}$ of the current operators, because any contour integral around
the twist operator insertion has to close only after going around the
insertion $l+1$ times. 

Before we apply these fractionally-moded current operators, there is
one more point to note. For the case of $l+1$ even one finds that the
twist operator $\sigma_n$ yields antiperiodic boundary conditions for
the fermion field when we traverse around the twist insertion $l+1$
times. Since we wanted the fermion field to return to itself after
going $l+1$ times around the insertion, we must insert a `spin field'
to change the periodicity of these fermions. The construction of these
spin fields was explained in detail in \cite{lm2}, but for now we just
denote the twist with spin field insertions (for both left and right
fermions) as $(S_{l+1}^+\bar{S}_{l+1}^+ \sigma_{l+1})$. With this
notation we find that the chiral primaries are given
by\footnote{Chiral primary operators can also be written using
alternative descriptions which make use of bosonized fermions (see for
example~\cite{Jevicki,davidmandalwadia}).}
\be
\sigma^{0}_{l+1}=
\begin{cases}
J^+_{-\frac{l-1}{l+1}}J^+_{-\frac{l-3}{l+1}}\cdots J^+_{-\frac{1}{l+1}}
\bar{J}^+_{-\frac{l-1}{l+1}}\bar{J}^+_{-\frac{l-3}{l+1}}\cdots
\bar{J}^+_{-\frac{1}{l+1}} \sigma_{l+1} & \text{$(l+1)$ odd}\\
J^+_{-\frac{l-1}{l+1}}J^+_{-\frac{l-3}{l+1}}\cdots J^+_{-\frac{2}{l+1}}
\bar{J}^+_{-\frac{l-1}{l+1}}\bar{J}^+_{-\frac{l-3}{l+1}}\cdots
\bar{J}^+_{-\frac{2}{l+1}}(S_{l+1}^+\bar{S}_{l+1}^+ \sigma_{l+1})
             & \text{$(l+1)$ even}.
\end{cases}
\label{chiral}
\ee
This construction generates chiral primaries with dimensions and charges
\be
\sigma^{0}_{l+1}:~~~h=m={l\over 2}, ~~~\bar h=\bar{m}={l\over 2}.
\ee
Note that $J^+\sim \psi^{+\dot{1}}\psi^{+\dot{2}}$, and the current
operators in (\ref{chiral}) fill up the left and right moving Fermi
seas up to a `Fermi level' (here we write only the left sector)
\be
\sigma^0_{l+1}\sim 
\begin{cases}
\psi^{+\dot{1}}_{-\frac{l-1}{2(l+1)}}\psi^{+\dot{2}}_{-\frac{l-1}{2(l+1)}}
\psi^{+\dot{1}}_{-\frac{l-3}{2(l+1)}}\psi^{+\dot{2}}_{-\frac{l-3}{2(l+1)}}
\cdots \psi^{+\dot{1}}_{-\frac{1}{2(l+1)}}\psi^{+\dot{2}}_{-\frac{1}{2(l+1)}}
     \sigma_{l+1}
                 & \text{$(l+1)$ odd}\\
   \psi^{+\dot{1}}_{-\frac{l-1}{2(l+1)}}\psi^{+\dot{2}}_{-\frac{l-1}{2(l+1)}} 
\psi^{+\dot{1}}_{-\frac{l-3}{2(l+1)}}\psi^{+\dot{2}}_{-\frac{l-3}{2(l+1)}}
\cdots \psi^{+\dot{1}}_{-\frac{1}{l+1}}\psi^{+\dot{2}}_{-\frac{1}{l+1}}
   (S_{l+1}^+  \sigma_{l+1})
                 & \text{$(l+1)$ even}.
              \end{cases}
\label{fermion}
\ee

\subsubsection{Additional chiral primaries}

Above we described the construction of the simplest chiral primary
$\sigma^0_{l+1}$. We can make additional chiral primaries as follows:
\begin{enumerate}
\item The next available fermion level for the fermion $\psi^{+\dot 1}$
  in~\eqref{fermion} is $\psi^{+\dot 1}_{-{1\over 2}}$. If we fill this level,
  we raise both dimension and charge by ${1\over 2}$, so we get
  another chiral primary.
\item We can do the same with the fermion  $\psi^{+\dot 2}$.
\item We can add both fermions $\psi^{+\dot 1},\psi^{+\dot 2}$, which is equivalent to an application of $J^+_{-1}$.
\end{enumerate}
Of course we can make analogous excitations to the right sector as
well. This gives a total of $4\times 4=16$ chiral primaries in a given
twist sector. This exhausts all the possible chiral primaries for this
system.

\subsubsection{Anti-chiral primaries}

We define anti-chiral primaries as states with
\be
h=-m, ~~~\bar h=-\bar m.
\ee
To construct these states, we again start with a twist operator
$\sigma_{l+1}$ and the apply modes of $J^- $ instead of
$J^+$. Proceeding in the same way as for chiral primaries, we get the
anti-chiral primary (denoted with a tilde over the $\sigma$) as
\be
\tilde\sigma^0_{l+1}=
\begin{cases}
J^-_{-\frac{l-1}{l+1}}J^-_{-\frac{l-3}{l+1}}\cdots J^-_{-\frac{1}{l+1}}
\bar{J}^-_{-\frac{l-1}{l+1}}\bar{J}^-_{-\frac{l-3}{l+1}}\cdots
\bar{J}^-_{-\frac{1}{l+1}} \sigma_{l+1} 
       & \text{$(l+1)$ odd}\\
       J^-_{-\frac{l-1}{l+1}}J^-_{-\frac{l-3}{l+1}}\cdots J^-_{-\frac{2}{l+1}}
\bar{J}^-_{-\frac{l-1}{l+1}}\bar{J}^-_{-\frac{l-3}{l+1}}\cdots
\bar{J}^-_{-\frac{2}{l+1}}(S_{l+1}^-\bar{S}_{l+1}^- \sigma_{l+1}) 
       & \text{$(l+1)$ even}.
\end{cases}
\ee 
We can construct additional anti-chiral primaries just as in the case
of chiral primaries. A chiral primary has a nonvanishing 2-point
function with its corresponding anti-chiral primary. The chiral and
anti-chiral twist operators are normalized such that the 2-point
function is unity at unit separation. We use this fact later in the
paper.

\subsubsection{Ramond sector states}

To get states in the Ramond sector we have to change the boundary
conditions on fermions, making them periodic around the $\sigma$
circle. This requires the insertion of a `spin field'. While this is
not hard to do, we perform our computations by mapping the Ramond
sector states to the NS sector by spectral flow. Thus we do not give
the explicit structure of the Ramond sector states in this paper. In
Appendix~\ref{ap:CFT-notation} we give a brief description of the
Ramond vacua.

\section{The initial state, the final state, and the vertex operator}\label{sec:states-and-op}

Our main computation addresses the following physical process.
Consider a bound state of $N_1$ D1 branes and $N_5$ D5 branes, sitting
at the origin of asymptotically flat space.  As mentioned above, the
CFT describing this bound state is in the Ramond sector, which has a
number of degenerate ground states. We pick a particular Ramond ground
state. Instead of describing this choice directly in the Ramond
sector, we note that all Ramond sector ground states are obtained by
one unit of spectral flow from chiral primary states of the NS sector.
We pick the Ramond ground state that arises from the simplest chiral
primary: the NS vacuum $\vac_{NS}$. The gravity dual of this state can
be described as follows~\cite{bal, mm, lm3, lm4}: there is flat space
at infinity, then a `neck', then an $AdS$ region, and then a `cap', as
pictured in Fig.~\ref{fig:throats-b}. While the structure of the `cap'
depends on the choice of Ramond ground state, in the present case the
structure is particularly simple: below $r_b$ in
Fig.~\ref{fig:throats-b} the geometry is a part of global $AdS_3\times
S^3$.

By itself such a D1D5 state is stable, and does not radiate energy. We
therefore add an excitation to the D1D5 brane state. In the
supergravity dual, the excitation we choose corresponds to adding a
supergravity quantum sitting in the `cap'. The supergravity field we
choose is a scalar $\phi_{ij}$, where $i,j=1, \dots 4$ are vector
indices valued in the $T^4$ in (\ref{compact}). These scalars arise
from the following fields:
\begin{enumerate}
\item A symmetric traceless matrix $h_{ij}$ with $i, j=1,\dots 4$
  giving the transverse traceless gravitons with indices in the $T^4$.

\item An antisymmetric matrix $B^{RR}_{ij}$ giving the components of
  the Ramond-Ramond $B$ field with indices in the torus.

\item The dilaton, which is a scalar in the full 10-dimensional theory.
\end{enumerate}
We can put all these scalars together into a $4\times 4$ matrix
$\phi_{ij}$, with the symmetric traceless part coming from $h_{ij}$,
the antisymmetric part from $B^{RR}_{ij}$ and the trace from the
dilaton. (Such a description was used for example in
\cite{radiation-3, lm4}. But we may need  to scale the above fields
by some function.  For instance, it is not the graviton, $h_{ij}$, but
$(H_5/H_1)^\frac{1}{4} h_{ij}$ which behaves as a minimal scalar in
the 6-d space obtained by dimensional reduction on $T^4$.

The supergravity particle is described by its angular momenta in the
$S^3$ directions given by $SU(2)_L\times SU(2)_R$ quantum numbers
$({l\over 2}, m), ({l\over 2},\bar{m})$; and a `radial quantum
number', $N$, where $N=0$ gives the lowest energy state with the given
angular momentum, and $N=1, 2, \dots$ give successively higher energy
states.

Adding this quantum to the $r<r_b$ region of the geometry corresponds
to making an excitation of the D1D5 CFT. Since we compute all
processes after spectral flowing to the NS sector, we should describe
this excitation in the NS sector. In the NS sector, the excitation is
a supersymmetry descendant of a chiral primary state, which is acted
on $N$ times with $L_{-1}$ to further raise the energy. We describe
the construction of this initial state in more detail below.

The process of interest is the emission of this supergravity particle
from the cap out to infinity. The final state is thus simple: in the
 Ramond sector description we return to the Ramond
ground state that we started with. In the spectral flowed NS sector
description that we compute with, the final state is just the NS
vacuum $\vac_{NS}$.

The emission is caused by the interaction Lagrangian in
Equation~\eqref{eq:general-S-int} which couples excitations in the CFT
to modes at infinity; the general structure of this coupling was
discussed in Section~\ref{sec:coupling}. We write down the vertex
operator $\hat{\mathcal{V}}$ which leads to the emission of the quanta
$\phi_{ij}$, and compute the emission amplitude
$\bra{f}\hat{\mathcal{V}}\ket{i}$.
  
We now describe in detail the initial state, the final state, and the
vertex operator.
 
\subsection{The initial state in the NS sector}

Let us first write the state, and then explain its structure. The left
and right parts of the state have similar forms, so we only write the
left part (indicated by the subscript $L$):
\begin{equation}
\ket{\phi_{N+1}^{\frac{l}{2},\frac{l}{2}-k}}_L^{A\dot{A}} 
      = \mathcal{C}_L
        L_{-1}^N(J_0^-)^k G^{-A}_{-\frac{1}{2}}
	\psi_{-\frac{1}{2}}^{+\dot{A}}\sigma^0_{l+1}\vac_{NS}.
\end{equation}

Let us describe the structure of this state starting with the elements on the rightmost end:

\begin{enumerate}
\item We start with the NS vacuum $\vac_{NS}$. In this state each copy
  of the CFT is `singly wound', and each copy is unexcited. In the
  supergravity dual, we have global AdS space with no particles in it.

\item We apply the chiral primary $\sigma^0_{l+1}$, thereby twisting
  together $l+1$ copies of the CFT into one `multiply wound' component
  string. It also adds charge, so that we get a state with 
\be
  h=m=\bar h=\bar m= \frac{l}{2}.  
\ee 
In the gravity dual, we have one supergravity quantum, with angular
momenta $(m, \bar m)$.

\item We act with $\psi_{-\frac{1}{2}}^{+\dot{A}}$. This increases
  both $h$ and $m$ by ${1\over 2}$, and so gives another chiral
  primary. We do the same for the right movers, so that overall the
  new state created is again bosonic. In the supergravity dual, it
  corresponds to a different bosonic quantum in the AdS.

\item We act with elements of the `anomaly-free subalgebra' of the
  chiral algebra:
\begin{enumerate}
\item The $G^{-A}_{-\frac{1}{2}}$ changes the chiral primary to a
  supersymmetry descendant of the chiral primary, corresponding to a
  different supergravity particle in the gravity dual. Again, because
  we apply this supersymmetry operator on both left and right movers,
  the new supergravity quantum is bosonic. We now find that the
  indices carried by this quantum are those corresponding to a minimal
  scalar with both indices along the $T^4$ in the gravity description.
  Thus we have finally arrived at the supergravity quantum that we
  wanted to consider.

\item The $(J_0^-)^k$ rotate the quantum in the $S^3$ directions.
  Before this rotation, the quantum numbers $(m, \bar m)$ were the
  highest allowed for the given supergravity particle state. The
  application of the $(J_0^-)^k, (\bar{J}_0^-)^k$ give us other
  members of the $SU(2)_L\times SU(2)_R$ multiplet.

\item The $L_{-1}^N$ move and boost the quantum around in the $AdS$,
  thus increasing its energy and momentum.
\end{enumerate}

\item Finally, we have a normalization constant. Below, we derive this
  in detail since the final expression for the radiation rate involves
  the factors appearing in this normalization.
\end{enumerate}

\subsubsection{Normalizing the initial state}

To find the normalization constant $\mathcal{C}_L$, we take the
Hermitian conjugate to find
\begin{equation}
\mathop{\vphantom{a}}_{A\dot{A}}^L\hspace{-3pt}
\bra{\phi_{N+1}^{\frac{l}{2},\frac{l}{2}-k}}
  = -\mathcal{C}_L^*\, 
  \mathop{\vphantom{a}}_{NS}\hspace{-3pt}\bvac\tilde{\sigma}_{l+1}\epsilon_{\dot{A}\dot{B}}
		\psi_{\frac{1}{2}}^{-\dot{B}}\epsilon_{AB}G^{+B}_\frac{1}{2}(J_0^+)^k L_1^N,
\end{equation}
and then compute the norm,
\begin{calc}
\mathop{\vphantom{a}}_{A\dot{A}}^L \hspace{20pt}&\hspace{-23pt}
\braket{\phi_{N+1}^{\frac{l}{2},\frac{l}{2}-k}|\phi_{N+1}^{\frac{l}{2},\frac{l}{2}-k}}_L^{B\dot{B}}
	\\
  &= -|\mathcal{C}_L|^2 \epsilon_{\dot{A}\dot{C}}\epsilon_{AC}
\,\mathop{\vphantom{a}}_{NS}\hspace{-3pt}\bvac\tilde{\sigma}^0_{l+1}
	\psi_{\frac{1}{2}}^{-\dot{C}} G^{+C}_\frac{1}{2}(J_0^+)^k
			L_1^N L_{-1}^N
	(J_0^-)^k G^{-B}_{-\frac{1}{2}}
	\psi_{-\frac{1}{2}}^{+\dot{B}}\sigma^0_{l+1}\vac_{NS}\\
  &= -|\mathcal{C}_L|^2 \epsilon_{\dot{A}\dot{C}}\epsilon_{AC}
\,\mathop{\vphantom{a}}_{NS}\hspace{-3pt}\bvac\tilde{\sigma}^0_{l+1}
	\psi_{\frac{1}{2}}^{-\dot{C}} G^{+C}_\frac{1}{2}(J_0^+)^k
		\left(\prod_{j=1}^N j(2L_0 + j-1)\right)
	(J_0^-)^k G^{-B}_{-\frac{1}{2}}
	\psi_{-\frac{1}{2}}^{+\dot{B}}\sigma^0_{l+1}\vac_{NS}\\
  &= -\frac{N!(N+l+1)!}{(l+1)!}|\mathcal{C}_L|^2 \epsilon_{\dot{A}\dot{C}}\epsilon_{AC}
\,\mathop{\vphantom{a}}_{NS}\hspace{-3pt}\bvac\tilde{\sigma}^0_{l+1}
	\psi_{\frac{1}{2}}^{-\dot{C}} G^{+C}_\frac{1}{2}(J_0^+)^k
	(J_0^-)^k G^{-B}_{-\frac{1}{2}}
	\psi_{-\frac{1}{2}}^{+\dot{B}}\sigma^0_{l+1}\vac_{NS}\\
  &= -\frac{k!\,l!}{(l-k)!}\frac{N!(N+l+1)!}{(l+1)!}|\mathcal{C}_L|^2 
		\epsilon_{\dot{A}\dot{C}}\epsilon_{AC}
	\,\mathop{\vphantom{a}}_{NS}\hspace{-3pt}\bvac\tilde{\sigma}^0_{l+1}
	\psi_{\frac{1}{2}}^{-\dot{C}} 
		G^{+C}_\frac{1}{2}G^{-B}_{-\frac{1}{2}}
	\psi_{-\frac{1}{2}}^{+\dot{B}}\sigma^0_{l+1}\vac_{NS}\\
  &= -2\delta_A^B\frac{k!\,l!}{(l-k)!}\frac{N!(N+l+1)!}{(l+1)!}|\mathcal{C}_L|^2 
	\epsilon_{\dot{A}\dot{C}}
	\,\mathop{\vphantom{a}}_{NS}\hspace{-3pt}\bvac\tilde{\sigma}^0_{l+1}
	\psi_{\frac{1}{2}}^{-\dot{C}} 
		(L_0 + J_0^3)
	\psi_{-\frac{1}{2}}^{+\dot{B}}\sigma^0_{l+1}\vac_{NS}.
\end{calc}
Proceeding with the calculation, one finds
\begin{calc}
\mathop{\vphantom{a}}_{A\dot{A}}^L\hspace{-3pt}
\braket{\phi_{N+1}^{\frac{l}{2},\frac{l}{2}-k}|\phi_{N+1}^{\frac{l}{2},\frac{l}{2}-k}}_L^{B\dot{B}}
  &= -2(l+1)\delta_A^B\frac{k!\,l!}{(l-k)!}\frac{N!(N+l+1)!}{(l+1)!}|\mathcal{C}_L|^2 
	\epsilon_{\dot{A}\dot{C}}
	\,\mathop{\vphantom{a}}_{NS}\hspace{-3pt}\bvac\tilde{\sigma}^0_{l+1}
	\psi_{\frac{1}{2}}^{-\dot{C}} 
	\psi_{-\frac{1}{2}}^{+\dot{B}}\sigma^0_{l+1}\vac_{NS}\\
  &= -2\delta_A^B\frac{N!(N+l+1)!k!}{(l-k)!}|\mathcal{C}_L|^2 
	\epsilon_{\dot{A}\dot{C}}
	\,\mathop{\vphantom{a}}_{NS}\hspace{-3pt}\bvac\tilde{\sigma}^0_{l+1}
	\psi_{\frac{1}{2}}^{-\dot{C}} 
	\psi_{-\frac{1}{2}}^{+\dot{B}}\sigma^0_{l+1}\vac_{NS}\\
   &= 2\,\delta_A^B\,\delta_{\dot{A}}^{\dot{B}}
      \frac{N!(N+l+1)!k!}{(l-k)!}(l+1)|\mathcal{C}_L|^2,
\end{calc}
where we have used the fact that the chiral primary twist operators
are correctly normalized. The factor of $l+1$ comes from the fermion
anticommutator, since in the twisted sector there are $l+1$ copies of
the fermion field that go into what we call $\psi$.  One can
understand this factor most easily by using
Equation~\eqref{eq:frac-mode-cover}. By demanding that
\begin{equation}
\mathop{\vphantom{a}}_{A\dot{A}}^L\hspace{-3pt}
\braket{\phi_{N+1}^{\frac{l}{2},\frac{l}{2}-k}|\phi_{N+1}^{\frac{l}{2},\frac{l}{2}-k}}_L^{B\dot{B}} = \delta_A^B\delta_{\dot{A}}^{\dot{B}},
\end{equation}
we conclude that the normalized state (with the left \emph{and} right parts)
is
\begin{equation}\begin{split}
\ket{\phi}^{A\dot{A}B\dot{B}} 
      &= \sqrt{\frac{(l-k)!(l-\bar{k})!}{4N!\bar{N}!(N+l+1)!(\bar{N}+l+1)!k!\bar{k}!(l+1)^2}}\\
	&\qquad\times L_{-1}^N(J_0^-)^k  G^{-A}_{-\frac{1}{2}}
	\psi_{-\frac{1}{2}}^{+\dot{A}}\,\,
	\bar{L}_{-1}^{\bar{N}}(\bar{J}_0^-)^{\bar{k}} \bar{G}^{\dot{-}B}_{-\frac{1}{2}}
	\bar{\psi}_{-\frac{1}{2}}^{\dot{+}\dot{B}}\sigma^0_{l+1}\vac_{NS}.
\end{split}\end{equation}

In this computation we use the identity
\begin{equation}
\prod_{j=1}^k \big(j l -j(j-1)\big)
	=  \frac{k!\, l!}{(l-k)!}.
\end{equation}

\subsubsection{The state in $SO(4)_I$ notation}

In the gravity description it is natural to write the quantum as
$\phi_{ij}$, with vector indices $ij$ of the internal symmetry group
$SO(4)_I$ of the $T^4$ directions. For CFT computations it is more
useful to use indices $A\dot A$ for $SU(2)_1\times SU(2)_2$, as we do
above. The conversion is achieved by
\begin{equation}
\ket{\phi_{N+1}^{\frac{l}{2},\frac{l}{2}-k}}_L^i
= \frac{1}{\sqrt{2}}(\sigma^i)^{A\dot{A}}\epsilon_{AB}\epsilon_{\dot{A}\dot{B}}
\ket{\phi_{N+1}^{\frac{l}{2},\frac{l}{2}-k}}_L^{B\dot{B}}. 
\end{equation}
We then have
\begin{equation}
\mathop{\vphantom{E}}_L^i\hspace{-3pt}\braket{\phi_{N+1}^{\frac{l}{2},\frac{l}{2}-k}|\phi_{N+1}^{\frac{l}{2},\frac{l}{2}-k}}_L^j
  = \delta^{ij}.
\end{equation}

Similarly, one typically labels the angular momentum eigenstates in
terms of $(l, m_\psi, m_\phi)$, instead of $(l, m, \bar{m})$. The two
bases are related via
\begin{equation}\begin{split}
m_\psi &= -(m + \bar{m})\\ 
m_\phi &= m - \bar{m},
\end{split}\end{equation}
where the values on the right-hand side are the angular momenta of the
initial state in the NS sector.
 
\subsection{The final state}

In the supergravity description the initial state had one quantum in
it. The emission process of interest leads to the emission of this
quantum. Thus the final state has no quanta, and in the NS description
is just the vacuum
\be
\ket{f}=\vac_{NS}.
\ee

\subsection{The vertex operator}

We need the vertex operator that emits the supergravity quantum
described by the initial state $\ket{i}$. Vertex operators describing
supergravity particles are given by chiral primaries and their
descendants under the anomaly-free part of the chiral algebra.

For the process of interest the emission vertex must have appropriate
charges to couple to the supergravity field under consideration.
Thus, one naturally concludes that the vertex operator has essentially
the same structure as the state $\ket{i}$, with two differences.
First, the operator has charges that are opposite to the charges
carried by the state. (We get a nonvanishing inner product between
$\ket{i}$ and the Hermitian \emph{conjugate} of $\ket{i}$.) Second,
the operator does not have the $L_{-1}$ modes present in the
description of the CFT state. This is because applying an $L_{-1}$
mode is equivalent to translating the location of the vertex
insertion, and we have already chosen the insertion to be the point
$(\sigma, \tau)$. Note that after applying the supercurrent to give
the operator the correct $SO(4)_I$ index structure, one finds that the
operator already has the correct weight to couple to a minimal scalar
in Equation~\eqref{eq:general-S-int} and form a scale invariant
action.

The vertex operator, then, is given by (we drop the hat on the vertex
operator from now on)
\begin{equation}
\widetilde{\mathcal{V}}^{A\dot{A}B\dot{B}}_{l, l-k-\bar{k}, k-\bar{k}}(\sigma, \tau)
	= \frac{1}{2}\sqrt{\frac{(l-k)!(l-\bar{k})!}{(l+1)^2(l+1)!^2\,k!\,\bar{k}!}}
	(J^+_0)^k(\bar{J}^+_0)^{\bar{k}}
	G^{+A}_{-\frac{1}{2}}\psi^{-\dot{A}}_{-\frac{1}{2}}
	\bar{G}^{\dot{+}B}_{-\frac{1}{2}}\bar{\psi}^{\dot{-}\dot{B}}_{-\frac{1}{2}}
	\tilde{\sigma}^0_{l+1}(\sigma, \tau).
\end{equation}
The subscript on the vertex operator $\mathcal{V}_{l,m_\psi, m_\phi}$
are the $SO(4)_E$ angular momenta labels. Again the normalization is a
crucial part of the final amplitude, so we perform it in more detail
below.

Note that for $l=0$, the vertex operator reduces to $[\pd
X]^{A\dot{A}}[\pdb X]^{B\dot{B}}$, the old `effective string' coupling
found by expanding the DBI action~\cite{radiation-3}.

\subsubsection{Mapping to the complex plane}

Before normalizing the vertex operator, we first map the operator from
the cylinder onto the complex plane via $z=e^{\tau + i\sigma}$. The
vertex operator has weight $\frac{l}{2}+1$ on both the left and the
right, so we get
\begin{equation}\begin{split}
\widetilde{\mathcal{V}}^{A\dot{A}B\dot{B}}_{l, l-k-\bar{k}, k-\bar{k}}(\sigma, \tau) 
&= |z|^{l+2}
	\frac{1}{2}\sqrt{\frac{(l-k)!(l-\bar{k})!}{(l+1)^2(l+1)!^2\,k!\,\bar{k}!}}
	\left((J^+_0)^k(\bar{J}^+_0)^{\bar{k}}
	G^{+A}_{-\frac{1}{2}}\psi^{-\dot{A}}_{-\frac{1}{2}}
	\bar{G}^{\dot{+}B}_{-\frac{1}{2}}\bar{\psi}^{\dot{-}\dot{B}}_{-\frac{1}{2}}
	\tilde{\sigma}^0_{l+1}(z,\bar{z})\right)_{z,\bar{z}}\\
&= |z|^{l+2}
	\mathcal{V}^{A\dot{A}B\dot{B}}_{l, l-k-\bar{k},k-\bar{k}}(z, \bar{z}).
\end{split}\end{equation}

\subsubsection{Normalizing the vertex operator}

The left part of the vertex operator is given by
\begin{equation}
\mathcal{V}_{L; l,k}^{A\dot{A}}(z)
	= N_L\left(\big(J_0^+\big)^kG_{-\frac{1}{2}}^{+A}\psi_{-\frac{1}{2}}^{-\dot{A}}
		\tilde{\sigma}^0_{l+1}(z)\right)_z.
\end{equation}
We need to normalize the vertex operator. To that end, we begin by
writing its Hermitian conjugate:
\begin{equation}
{\mathcal{V}^{A\dot{A}}_{L;l,k}}^\dg(z) = -(-1)^{k+1}\epsilon_{AB}\epsilon_{\dot{A}\dot{B}}\,N_L^*\,
	\left(\big(J_0^-\big)^k G_{-\frac{1}{2}}^{-B}\psi_{-\frac{1}{2}}^{+\dot{B}}
			\sigma^0_{l+1}\left(z\right)\right)_{z}
                      = \epsilon_{AB}\epsilon_{\dot{A}\dot{B}}\mathcal{V}_{l,2l-k}^{B\dot{B}}(z),
\end{equation}
where the second equality is the condition needed to ensure the total
interaction action is Hermitian.

The factor of $(-1)^{k+1}$ comes from the $G$ and the $J_0$'s. We
illustrate below with $J_0^+$:
\begin{calc}
\left[\big(J_0^+\big)_z\right]^\dg &= \left[\oint_z\frac{\drm z'}{2\pi i}J^+(z')\right]^\dg\\
  &= -\oint_{\bar{z}}\frac{\drm {\bar{z}}'}{2\pi i}J^-\left(\tfrac{1}{{\bar{z}}'}\right)
          \frac{1}{{\bar{z}}'^2}\\
  &= -\oint_{\frac{1}{\bar{z}}}\frac{\drm \xi}{2\pi i}J^-(\xi)\\
  &= -\big(J_0^-\big)_{\frac{1}{\bar{z}}},
\end{calc}
where when making the change of variables $\xi = 1/\bar{z}'$ there are
two minus signs. One comes from the Jacobian $\drm\bar{z}' = - 1/\xi^2
\drm \xi$, and the other comes from making the contour
counter-clockwise. The $G_{-\frac{1}{2}}$ behaves in the same way;
however, the $\psi_{-\frac{1}{2}}$ is different:
\begin{calc}
\left[\big(\psi^{-\dot{A}}_{-\frac{1}{2}}\big)_z\right]^\dg 
  &= \left[\oint_z\frac{\drm z'}{2\pi i}\frac{\psi^{-\dot{A}}(z')}{z'-z}\right]^\dg\\
  &= -(-\epsilon_{-+}\epsilon_{\dot{A}\dot{B}})
       \oint_{\bar{z}}\frac{\drm \bar{z}'}{2\pi i}\,\psi^{+\dot{B}}\big(\tfrac{1}{\bar{z}'}\big)\,
       \frac{1}{\bar{z}'(\bar{z}'-\bar{z})}\\
  &= -\epsilon_{\dot{A}\dot{B}}\oint_{\frac{1}{\bar{z}}}\frac{\drm \xi}{2\pi i}\psi^{+\dot{B}}(\xi)
        \frac{1}{\xi\left(\frac{1}{\xi} - \bar{z}\right)}\\
  &= \frac{\epsilon_{\dot{A}\dot{B}}}{\bar{z}}
      \oint_\frac{1}{\bar{z}}\frac{\drm \xi}{2\pi i}
         \frac{\psi^{+\dot{B}}(\xi)}{\xi - \frac{1}{\bar{z}}}\\
  &= \frac{\epsilon_{\dot{A}\dot{B}}}{\bar{z}}
     \big(\psi^{+\dot{B}}_{-\frac{1}{2}}\big)_{\frac{1}{\bar{z}}};
\end{calc}
it receives an extra minus sign from the integrand.

We use the notation
\begin{equation}
\vev{\cdot} = {}_{NS}\bvac\cdot\vac_{NS}
\end{equation}
for the NS-vacuum expectation value.  Proceeding with the
normalization, the 2-point function is given by
\begin{calc}
\Vev{{\mathcal{V}_{L;l,k}^{A\dot{A}}}^\dg(z)\mathcal{V}_{L;l,k}^{B\dot{B}}(0)}
	&= (-1)^{k+2}|N_L|^2\epsilon_{AC}\epsilon_{\dot{A}\dot{C}}
		\Vev{\left(\big(J_0^-\big)^k G_{-\frac{1}{2}}^{-C}\psi_{-\frac{1}{2}}^{+\dot{C}}
			\sigma^0_{l+1}\left(z\right)\right)_{z}
		\left(\big(J_0^+\big)^kG_{-\frac{1}{2}}^{+B}\psi_{-\frac{1}{2}}^{-\dot{B}}
		\tilde{\sigma}^0_{l+1}(0)\right)_0}\\
	&= -|N_L|^2\epsilon_{AC}\epsilon_{\dot{A}\dot{C}}
		\Vev{\left(\psi_{-\frac{1}{2}}^{+\dot{C}}\sigma^0_{l+1}\left(z\right)\right)_{z}
		\left(G_{-\frac{1}{2}}^{-C}\big(J_0^-\big)^k\big(J_0^+\big)^kG_{-\frac{1}{2}}^{+B}
		\psi_{-\frac{1}{2}}^{-\dot{B}}\tilde{\sigma}^0_{l+1}(0)\right)_0} \\
	&= -|N_L|^2\epsilon_{AC}\epsilon_{\dot{A}\dot{C}}\frac{k!\,l!}{(l-k)!}
		\Vev{\left(\psi_{-\frac{1}{2}}^{+\dot{C}}\sigma^0_{l+1}\left(z\right)\right)_{z}
		\left(G_{-\frac{1}{2}}^{-C}G_{-\frac{1}{2}}^{+B}
		\psi_{-\frac{1}{2}}^{-\dot{B}}\tilde{\sigma}^0_{l+1}(0)\right)_0} \\
	&= 2|N_L|^2\delta_A^B\epsilon_{\dot{A}\dot{C}}\frac{k!\,l!}{(l-k)!}
		\Vev{\left(\psi_{-\frac{1}{2}}^{+\dot{C}}\sigma^0_{l+1}\left(z\right)\right)_{z}
		\left(L_{-1}\psi_{-\frac{1}{2}}^{-\dot{B}}\tilde{\sigma}^0_{l+1}(0)\right)_0} \\
	&= 2|N_L|^2\delta_A^B\epsilon_{\dot{A}\dot{C}}\frac{k!\,l!}{(l-k)!}
		\lim_{v\to 0}\pd_v
		\Vev{\left(\psi_{-\frac{1}{2}}^{+\dot{C}}\sigma^0_{l+1}\left(z\right)\right)_{z}
		\left(\psi_{-\frac{1}{2}}^{-\dot{B}}\tilde{\sigma}^0_{l+1}(v)\right)_v} \\
	&= 2|N_L|^2\delta_A^B\epsilon_{\dot{A}\dot{C}}\epsilon^{\dot{C}\dot{B}}
		\frac{k!\,l!}{(l-k)!}
		\lim_{v\to 0}\pd_v
		\frac{l+1}{(z-v)^{l+1}}\\
	&= 2|N_L|^2\delta_A^B\delta_{\dot{A}}^{\dot{B}}
		\frac{k!\,(l+1)!}{(l-k)!}
		(l+1)\frac{1}{z^{l+2}}.
\end{calc}
The factor of $l+1$ has the same origin as in the normalization of the
initial state.  Using the above, one finds
\begin{equation}
N_L = \frac{1}{\sqrt{2}}\sqrt{\frac{(l-k)!}{k!\,(l+1)!(l+1)}},
\end{equation}
and thus the left part of the vertex operator is
\begin{equation}
\mathcal{V}_{L; l,k}^{A\dot{A}}(z)
	= \frac{1}{\sqrt{2}}\sqrt{\frac{(l-k)!}{k!\,(l+1)!(l+1)}}
  \left(\big(J_0^+\big)^kG_{-\frac{1}{2}}^{+A}\psi_{-\frac{1}{2}}^{-\dot{A}}
		\tilde{\sigma}^0_{l+1}(z)\right)_z.
\end{equation}

The normalization is chosen such that the vertex operator in the
complex plane satisfies
\begin{equation}\begin{split}
\vev{\mathcal{V}^{A\dot{A}B\dot{B}}_{l,-m_\psi, -m_\phi}(z)
\mathcal{V}^{C\dot{C}D\dot{D}}_{l,m_\psi, m_\phi}(0)}
 &= \frac{\epsilon^{AC}\epsilon^{\dot{A}\dot{C}}\epsilon^{BD}\epsilon^{\dot{B}\dot{D}}}
	{|z|^{l+2}}\\
\vev{\mathcal{V}^{ij}_{l,-m_\psi, -m_\phi}(z)
	\mathcal{V}^{kl}_{l,m_\psi, m_\phi}(0)} 
	&= \frac{\delta^{ik}\delta^{jl}}{|z|^{l+2}}.
\end{split}\end{equation}
Note that this is the normalization of the operator corresponding to
one particular way of permuting $l+1$ copies of the CFT. As mentioned
earlier, the actual vertex operator coupling to $\phi_{ij}$ is a
symmetrized sum over all possible ways of permtuing $l+1$ copies from
the $N_1N_5$ available copies. We discuss the combinatorics of this
choice in Section~\ref{sec:combinatorics} below, and at that time note
the extra normalization factor which is needed to agree
with~(\ref{first}).

\section{Using spectral flow}

We wish to relate a CFT amplitude computed in the NS sector,
\begin{equation}
\mathcal{A}' = \bra{f'}\mathcal{V}(z, \bar{z})\ket{i'},
\end{equation}
to an amplitude in the Ramond sector, since the physical D1D5 system
has its fermions periodic around the $y$ circle.  In this section, we
show how to spectral flow~\cite{spectral,spectral-yu,vafa-warner} the
computation in the NS sector to the physical problem in the R sector.
Furthermore, we find that we can relate this NS sector computation to
a whole family of Ramond sector amplitudes, and each member of the
family describes a different physical emission process.

If spectral flowing the states $\ket{i'}$ and $\ket{f'}$ by $\alpha$
units is given by
\begin{equation}
\ket{i'}\mapsto \ket{i} = \mathcal{U}_\alpha\ket{i'}\qquad
\bra{f'}\mapsto \bra{f} = \bra{f'}\mathcal{U}_{-\alpha},
\end{equation}
then we can compute the amplitude in the Ramond sector by using
\begin{equation}
\mathcal{A}_\text{Ramond} = \bra{f} \mathcal{V}(z,\bar{z})\ket{i}
 			  = \big(\bra{f}\mathcal{U}_{\alpha}\big)
			     \big(\mathcal{U}_{-\alpha}\mathcal{V}\mathcal{U}_{\alpha}\big)
			     \big(\mathcal{U}_{-\alpha}\ket{i}\big)
			  = \bra{f'}\mathcal{V}'(z, \bar{z})\ket{i'}.
\end{equation}
Note that one finds $\mathcal{V}'$ by spectral flowing $\mathcal{V}$
by $-\alpha$ units.

We need to determine how the vertex operator transforms under
spectral flow. First, we demonstrate that the $G\psi$ part is
unaffected, since
\begin{calc}
\left(G^{+A}_{-\frac{1}{2}}\psi^{-\dot{A}}_{-\frac{1}{2}}\right)_z
	&= \oint_z\frac{\drm z_1}{2\pi i}\oint_z\frac{\drm z_2}{2\pi i} 
		\frac{G^{+A}(z_1)\psi^{-\dot{A}}(z_2)}{z_2-z}\\
	&= -\oint_z\frac{\drm z_1}{2\pi i}\oint_z\frac{\drm z_2}{2\pi i} 
		\frac{[\pd X(z_2)]^{\dot{A}A}}{(z_2-z)(z_1-z_2)}
\end{calc}
and the bosons are unaffected by spectral flow.

Therefore, we need only spectral flow the $k$ $J_0^+$'s and the chiral
primary. The effect of spectral flow by \emph{negative} $\alpha$ units
on chiral ($h=m$) and anti-chiral primaries ($h=-m$) is very simple:
\begin{equation}
\mathcal{O}'_\text{c.p.}(z) = z^{\alpha m}\mathcal{O}_\text{c.p.}(z)\qquad
\mathcal{O}'_\text{a.c.p.}(z) = z^{\alpha m}\mathcal{O}_\text{a.c.p.}(z).
\end{equation}
One can see this most directly after bosonizing the fermions; see 
Appendix~\ref{sec:spectral-flow} for details.

Under spectral flow by $-\alpha$ units the $J^\pm$ transform as
\begin{equation}
J^\pm(z) \mapsto z^{\pm\alpha}J^\pm(z),
\end{equation}
from which we see that
\begin{equation}
(J^+_0)_z = \oint_z\frac{\drm z'}{2\pi i} J^+(z')
\mapsto \oint_z \frac{\drm z'}{2\pi i} J^+(z') z'^{\alpha}
         = z^{\alpha}(J^+_0)_z +\alpha z^{\alpha-1}(J^+_1)_z + \dots.
\end{equation}
Only the first term contributes since the positive modes annihilate a
chiral primary. Therefore, we conclude that spectral flowing the
vertex operator by $-\alpha$ units has the effect of
\begin{equation}
\mathcal{V}'_{l,l-k-\bar{k}, k-\bar{k}}(z,\bar{z}) = z^{-\alpha(\frac{l}{2}-k)}
	\bar{z}^{-\bar{\alpha}(\frac{l}{2}-\bar{k})}
		\mathcal{V}_{l,l-k-\bar{k}, k-\bar{k}}(z,\bar{z}).
\end{equation}

Thus we observe that we can spectral flow the initial and final
states, keep the vertex operator unchanged, and compute the amplitude
\be 
\mathcal{A}'=\bra{f'}\mathcal{V}(z,\bar{z})\ket{i'}.
\ee 
The result we want, $\mathcal{A}_\text{Ramond}$, is then given by
\begin{equation}\label{eq:unspectral-flowed-V}
\mathcal{A}_\text{Ramond} = z^{-\alpha(\frac{l}{2}-k)}\bar{z}^{-\bar{\alpha}(\frac{l}{2}-\bar{k})}
	\bra{f'}\mathcal{V}(z,\bar{z})\ket{i'}
	= z^{-\alpha(\frac{l}{2}-k)}\bar{z}^{-\bar{\alpha}(\frac{l}{2}-\bar{k})}
		\mathcal{A}'.
\end{equation}
Here $\alpha$ is chosen to have a value that spectral flows from the
NS to the Ramond sector, but this can be achieved by \emph{any} odd
integral value of $\alpha$:
\begin{equation}
\alpha = (2n+1)\qquad \bar{\alpha} = (2\bar{n} + 1) \qquad n,\bar{n}\in\ints.
\end{equation}
For these values of $\alpha$ the initial and final states have weight and charge
\begin{equation}\begin{aligned}
h &= h' + (2n+1) m' + (2n+1)^2\frac{c_\text{tot.}}{24}\\
m &= m' + (2n+1)\frac{c_\text{tot.}}{12},
\end{aligned}\end{equation}
where $c_\text{tot.}$ is $c=6$ times the number of copies being
spectral flowed. A similar relation holds on the right sector.

In our present computation in the NS sector, we have
\begin{equation}\begin{aligned}
h'_i &= \frac{l}{2} + N + 1 \hspace{20pt}& h'_f &= 0\\
m'_i &= \frac{l}{2} - k                  & m'_f &= 0.
\end{aligned}\end{equation}
In the next section we look at the Ramond sector process for
$\alpha=\bar\alpha=1$. In this case the weights and charges of the
Ramond sector states are
\begin{equation}\begin{aligned}
h_i &= \frac{l}{2} + N + 1  + \left(\frac{l}{2} - k\right) + \frac{l+1}{4}
                     \hspace{20pt}& h_f &= (l+1)\frac{1}{4}\\
m_i &= \frac{l}{2} - k + \frac{l+1}{2}     & m_f &= (l+1)\frac{1}{2}.
\end{aligned}\end{equation}
We see that the final state has the weight and charge of the
`spin-up' Ramond vacuum, while the initial state has the correct
weight and charge above the Ramond vacuum. Although the current
calculation is $\alpha = 1$, we leave $\alpha$ as an explicit
parameter in following calculations for later use and illustration.

In section~\ref{sec:nonextremal}, the full $\alpha$ and $\bar{\alpha}$
dependence is of physical interest, since how big $\alpha$ and
$\bar{\alpha}$ are roughly corresponds to how nonextremal the initial
state is.

\section{Evaluating the CFT amplitude}\label{sec:CFT-evaluation}

Let us now compute the amplitude
\be
{\mathcal{A}'}^{A\dot{A}}(\sigma, \tau)=\bra{f'}\widetilde{\mathcal{V}}(\sigma, \tau)\ket{i'}
                                = |z|^{l+2}\bra{f'}\mathcal{V}(z, \bar{z})\ket{i'}.
\ee
We choose the charges of the initial state and the vertex operator so
that we get a nonvanishing amplitude. The nonvanishing amplitude is
\begin{equation}\begin{split}
{\mathcal{A}'}_L^{A\dot{A}} &= \frac{1}{\sqrt{2}}(\sigma^\ibar)_{B\dot{B}}\,
        z^{\frac{l}{2}+1}
	\sqrt{\frac{(l-k)!}{2(l+1)(l+1)!\,k!}}
	\mathop{\vphantom{a}}_{NS}\hspace{-3pt}\bvac\left((J^+_0)^k
	G^{+A}_{-\frac{1}{2}}\psi^{-\dot{A}}_{-\frac{1}{2}}
	\tilde{\sigma}^0_{l+1}\right)_z \\
 &\hspace{20pt}\times \sqrt{\frac{(l-k)!}{2N!(N+l+1)!k!(l+1)}} 
        \left((J_0^-)^k G^{-B}_{-\frac{1}{2}}
	L_{-1}^N \psi_{-\frac{1}{2}}^{+\dot{B}}\sigma_{l+1}^0\right)_0
      \vac_{NS},
\end{split}\end{equation}
where $A,\dot{A}$ and $\ibar$ are free indices. The $\ibar$ is the
index of the initial state excitation and $A,\dot{A}$ are the indices
on the vertex operator. We have commuted the $L_{-1}$'s to the right
for calculational convenience.

To evaluate the correlator, we first note the subscript $z$ on the
first parenthetical expression indicates that the contours for the
modes circle $z$. Since there are no obstructions, we can shift the
contour to orbit the origin instead. We have
\begin{equation}\begin{split}
  (-1)^{k+1}\Vev{\left(\psi^{-\dot{A}}_{-\frac{1}{2}}\tilde{\sigma}^0_{l+1}\right)_z
       \left(G^{+A}_{-\frac{1}{2}}(J^+_0)^k(J_0^-)^k G^{-B}_{-\frac{1}{2}}
	L_{-1}^N \psi_{-\frac{1}{2}}^{+\dot{B}}\sigma_{l+1}^0\right)_0}
    \hspace{-150pt}& \\
 &= (-1)^k2\frac{k!\,l!}{(l-k)!}\epsilon^{AB}\epsilon^{+-}
 \Vev{\left(\psi^{-\dot{A}}_{-\frac{1}{2}}\tilde{\sigma}^0_{l+1}\right)_z
    \left(L_{-1}^{N+1}\psi_{-\frac{1}{2}}^{+\dot{B}}\sigma^0_{l+1}\right)_0}.
\end{split}\end{equation}
For the final step, we should note the action of $L_{-1}$ on a primary
field $\mathcal{O}$ is
\begin{equation}
L_{-1}\mathcal{O}(0) = \oint\frac{\drm z}{2\pi i}T(z)\mathcal{O}(0)
  = \pd\mathcal{O}(0);
\end{equation}
therefore, we may write
\begin{calc}
\Vev{\left(\psi^{-\dot{A}}_{-\frac{1}{2}}\tilde{\sigma}^0_{l+1}\right)_z
     \left(L_{-1}^{N+1}\psi_{-\frac{1}{2}}^{+\dot{B}}\sigma^0_{l+1}\right)_0}
 &= \lim_{v\to 0} \pd_v^{N+1}
 \Vev{\left(\psi^{-\dot{A}}_{-\frac{1}{2}}\tilde{\sigma}^0_{l+1}\right)_z
      \left(\psi_{-\frac{1}{2}}^{+\dot{B}}\sigma^0_{l+1}\right)_v}\\
 &= -\epsilon^{\dot{A}\dot{B}}\lim_{v\to 0}\pd_v^{N+1}\frac{l+1}{(z-v)^{l+1}}\\
 &= -\epsilon^{\dot{A}\dot{B}}\frac{(N+l+1)!(l+1)}{l!}\frac{1}{z^{l+N+2}}.
\end{calc}
Finally, one finds the left amplitude reduces to the simple form
\begin{equation}\label{eq:NS-untwisting-amp}
{\mathcal{A}'}_L^{A\dot{A}} = (-1)^k\frac{1}{\sqrt{2}}(\sigma^\ibar)^{A\dot{A}}
   \frac{1}{z^{\frac{l}{2} + N+1}}
   \sqrt{\choose{N+l+1}{N}}.
\end{equation}

From Equation~\eqref{eq:unspectral-flowed-V}, we find that the left
part of the CFT amplitude in the Ramond sector is given by
\begin{calc}
\mathcal{A}_L^{A\dot{A}}  &= z^{-\alpha(\frac{l}{2} - k)}\mathcal{A}_L'\\
 &= (-1)^{k}\frac{1}{\sqrt{2}}(\sigma^\ibar)^{A\dot{A}}
    \frac{1}{z^{(1+\alpha)\frac{l}{2} -\alpha k +  N+1}}
    \sqrt{\choose{N+l+1}{N}}.
\end{calc}
Finally, converting back to $SO(4)$ indices for the vertex operator,
one gets in the Ramond sector
\begin{calc}\label{eq:final-L-amp}
\mathcal{A}_L^\ibar(z) &= \frac{1}{\sqrt{2}}
   (\sigma^\ibar)_{A\dot{A}}\mathcal{A}_L^{A\dot{A}}\\
   &= (-1)^{k+1}\frac{1}{z^{(1+\alpha)\frac{l}{2} -\alpha k +  N+1}}
    \sqrt{\choose{N+l+1}{N}},
\end{calc}
The free index $\ibar$ and a similar index from the right movers
$\jbar$ correspond to the indices $\phi_{ij}$ for the field coupling
to the emission vertex.

\section{Combinatorics}
\label{sec:combinatorics}

The full CFT has $N_1N_5$ copies of the basic $c=6$ CFT. In the above
section, we took a set of $l+1$ copies twisted together, and look at
an emission process where the emission vertex untwists these copies.
Now, we must put this computation in its full CFT context, by doing the
following:
\begin{enumerate}
\item We must compute the combinatorics of how we pick the particular
  way of twisting $l+1$ copies in the initial state from all $N_1N_5$
  copies.

\item We must similarly consider all the ways that the vertex operator
  can twist copies. This allows us to normalize the vertex operator in
  the full theory so that we reproduce (\ref{first}).

\item We can take the limit $N_1N_5\rightarrow\infty$ to get the
  `classical limit' of the D1D5 system; the result in this limit
  should agree with the computation in the dual supergravity theory.
\end{enumerate}

In fact we start with something a little more general. We assume that
the initial state has $\nu$ quanta of the same kind, and let the
emission process lead to the final state with $\nu-1$ quanta.  We then
observe a Bose enhancement of the emission amplitude by a factor $\sqrt{\nu}$,
which agrees with the enhancement observed in both CFT and dual gravity
computations in~\cite{cm1}.

\subsection{The initial state}

We wish to have $\nu$ excitations, each of which involve twisting
together $l+1$ copies of the $c=6$ CFT. We can pick the needed copies
in any way from the full set of $N_1N_5$ copies, and because of the
orbifold symmetry between these copies the state must be a symmetrized
sum over these possibilities:
\begin{equation}\label{eq:comb-initial-state}
\ket{\Psi_\nu} = \mathcal{C}_\nu\bigg[\ket{\psi_\nu^1} + \ket{\psi_\nu^2} + \dots\bigg],
\end{equation}
where $\mathcal{C}_\nu$ is the overall normalization and each
$\ket{\psi^i_\nu}$ is individually normalized. To understand what we are
doing better, note that the state $\ket{\psi_\nu^1}$ can be written
schematically as
\begin{equation}
\ket{\psi_\nu^1} = \Ket{
	\big[12\cdots(l+1)\big]
	\big[(l+2)\cdots 2(l+1)\big]\cdots
	\big[\big(\nu(l+1)- l\big)\cdots\nu(l+1)\big]},
\end{equation}
where the numbers in the square brackets are indicating particular ways
of twisting individual strands corresponding to particular cycles of
the permutation group. For instance,
\begin{equation}
\ket{[1234]},
\end{equation}
indicates that we twist strand 1 into strand 2 into strand 3 into
strand 4 into strand 1 and leave strands 5 through $N_1N_5$ untwisted.

Our first task is to determine the number of terms in
Equation~\eqref{eq:comb-initial-state} and thereby its normalization
$\mathcal{C}_\nu$. To count the number of states we imagine
constructing one of these states and see how many choices we have
along the way. First, we choose $\nu(l+1)$ of the total $N_1N_5$
strands that are going to be twisted in some way. The remaining
strands are untwisted. Those $\nu(l+1)$ strands must now be broken
into sets of $l+1$. To do this, we first choose $l+1$ of the
$\nu(l+1)$, then the next set of $l+1$ from the remaining
$(\nu-1)(l+1)$, and so on. Note that $\ket{[12][34]}=\ket{[34][12]}$,
and so there is no sense in talking about the `first' set versus the
`second set'. Therefore we should divide by the number of ways to
rearrange the $\nu$ sets of $l+1$. Finally, we should choose a
particular cycle for each set of $(l+1)$; since it does not matter where we start on the final cycle, this gives a factor $l!$ for each twisted cycle. Putting all of these factors
together yields the number of terms in
Equation~\eqref{eq:comb-initial-state},
\begin{calc}
N_\text{terms}&= \choose{N_1N_5}{\nu(l+1)}\times
	\choose{\nu(l+1)}{l+1}\choose{(\nu-1)(l+1)}{l+1}\cdots
	\choose{l+1}{l+1}\times\frac{1}{\nu!}\times (l!)^\nu\\
	&= \frac{(N_1N_5)!}{(l+1)^\nu\,\nu![N_1N_5 - \nu(l+1)]!}.
\end{calc}
Without loss of generality, let us choose $\mathcal{C}_\nu$ to be
real, which gives
\be
\mathcal{C}_\nu=\left[\frac{(N_1N_5)!}{(l+1)^\nu\,\nu![N_1N_5 - \nu(l+1)]!}\right]^{-\frac{1}{2}}.
\ee

\subsection{The final state}

The final state is simply $\ket{\Psi_{\nu-1}}$, with its corresponding
normalization $\mathcal{C}_{\nu-1}$.

\subsection{The vertex operator}

The vertex operator can twist together any $l+1$ copies of the CFT
with any $l+1$-cycle, and it should be written as a symmetrized sum
over these possibilities:
\begin{equation}
\mathcal{V}_\text{sym} = \mathcal{C}\sum_i\mathcal{V}_i.
\label{checktwo}
\end{equation}
Since the joined copies form a single long loop, the order of copies
matters but not which copy is the `first one' in the loop. Thus the
number of terms in the sum is
\begin{equation}
\choose{N_1N_5}{l+1}l! = \frac{(N_1N_5)!}{(l+1)[N_1N_5 - (l+1)]!}.
\end{equation}
This gives the normalization
\be
\mathcal{C}=\left[\frac{(N_1N_5)!}{(l+1)[N_1N_5 - (l+1)]!}\right]^{-\frac{1}{2}}.
\ee

\subsection{The amplitude}

To compute the amplitude
\begin{equation}\label{eq:comb-amp}
\bra{\Psi_{\nu-1}}\mathcal{V}_\text{sym}\ket{\Psi_\nu},
\end{equation}
we have to count the different ways that terms in the initial state
can combine with terms in the vertex operator and terms in the final
state to produce a nonzero amplitude. For a given initial state term
$\ket{\psi_\nu^i}$, there are exactly $\nu$ vertex operators
$\mathcal{V}_i$ that can de-excite it into a final state. There is
only one final state that works, obviously. Thus the number of ways
that we can get a nonzero amplitude is simply
\[
\nu N_\text{terms} = \frac{\nu}{\mathcal{C}_\nu^2}.
\]
Let 
\be
\bra{\psi_{\nu-1}^1}\mathcal{V}_1\ket{\psi_\nu^1}
\ee
be the amplitude obtained by using only one allowed initial state $
\ket{\psi_\nu^1}$ from the set in Equation~\eqref{eq:comb-initial-state} and
one allowed vertex operator $\mathcal{V}_1$ from the set in
Equation~\eqref{checktwo}. Then we have
\begin{calc}
\bra{\Psi_{\nu-1}}\mathcal{V}_\text{sym.}\ket{\Psi_\nu}
	&= \mathcal{C}\mathcal{C}_\nu\mathcal{C}_{\nu-1}\cdot 
        \frac{\nu}{\mathcal{C}_\nu^2}
	   \bra{\psi_{\nu-1}^1}\mathcal{V}_1\ket{\psi_\nu^1}\\
	&= \sqrt{\nu}
        \sqrt{\frac{[N_1N_5-(\nu-1)(l+1)]![N_1N_5-(l+1)]!}
          {[N_1N_5-\nu(l+1)]!(N_1N_5)!}}
		\bra{\psi_{\nu-1}^1}\mathcal{V}_1\ket{\psi_\nu^1}.
\end{calc}

\subsection{The large $N_1N_5$ limit}

We are ultimately interested in the limit of large $N_1N_5$. Then we have
\begin{equation}
\frac{[N_1N_5-(\nu-1)(l+1)]!}{[N_1N_5-\nu(l+1)]!}
	\longrightarrow
	(N_1N_5)^{l+1}
\hspace{20pt}
\frac{[N_1N_5-(l+1)]!}{(N_1N_5)!}
	\longrightarrow
	(N_1N_5)^{-(l+1)},
\end{equation}
which gives
\begin{equation}
\bra{\Psi_{\nu-1}}\mathcal{V}_\text{sym.}\ket{\Psi_\nu}
	\longrightarrow \sqrt{\nu}
	\bra{\psi_{\nu-1}^1}\mathcal{V}_1\ket{\psi_\nu^1}.
	\label{finalq}
\end{equation}
The prefactor $\sqrt{\nu}$ gives a `Bose enhancement' effect which
tells us that if we start with $\nu$ quanta, the amplitude to emit
another quantum is amplified by a factor $\sqrt{\nu}$ (compared to the
case when there was only one quantum). This gives an enhancement $\nu$
in the probability, which just tells us that if we start with $\nu$
quanta in the initial state, then the rate of emission is proportional
to $\nu$.

\section{The Rate of Emission}\label{sec:calc-rate-emission}

We now put together all the computations of the above sections to get
the emission rate for a quantum from the excited CFT state. We need to
do the following:
\begin{enumerate}
\item We use (\ref{finalq}) to relate the decay amplitude for one
  $(l+1)$-permutation to the amplitude with all the required
  symmetrizations put in
\begin{equation}
\bra{\Psi_{0}}\mathcal{V}_\text{sym.}\ket{\Psi_1}
	= \sqrt{\nu}\bra{\psi_{0}^1}\mathcal{V}_1\ket{\psi_1^1}.
\end{equation}

\item From Equation~\eqref{eq:final-L-amp}, we have 
  the decay amplitude for a given $l+1$-permutation (we put the
  right sector back in now):
\begin{calc}\label{eq:final-euc-amp}
\bra{\psi_{0}^1}\mathcal{V}_1\ket{\psi_1^1} &= \mathcal{A}^{\ibar\jbar}(z,\bar{z})\\ 
  &= (-1)^{k+\bar{k}}\sqrt{\choose{N+l+1}{N}\choose{\bar{N} + l+1}{\bar{N}}}
	 z^{-(\alpha +1)\frac{l}{2} +\alpha k - N - 1}
	\bar{z}^{-(\bar{\alpha} +1)\frac{l}{2} +\bar{\alpha} \bar{k} - \bar{N} - 1}.
\end{calc}
We now rotating back to Lorentzian signature and replacing   $\tau, \sigma$ by the physical $(t,y)$ coordinates. Note that we are still working with a CFT with spatial section of `unit size' where the spatial circle has length $(2\pi)$. The  `unit-sized'
amplitude is thus
\begin{equation}\begin{split}\label{eq:physical-CFT-amp}
\mathcal{A}_\text{unit}^{\ibar\jbar}(t,y) &= (-1)^{k+\bar{k}}
	\sqrt{\choose{N+l+1}{N}\choose{\bar{N} + l+1}{\bar{N}}}\\
	 &\hspace{50pt} \times 
	e^{\frac{i}{R}\left(-(\alpha +1)\frac{l}{2} +\alpha k - N - 1\right)(y+t)}
	e^{-\frac{i}{R}\left(-(\bar{\alpha} +1)\frac{l}{2} +\bar{\alpha} \bar{k} 
		- \bar{N} - 1\right)(y-t)}.
\end{split}\end{equation}

\item From Equation~\eqref{eq:physical-CFT-amp} or by comparing the initial
and final states, we can read off
\begin{equation}\begin{split}
E_0 &= \frac{1}{R} \left[(\alpha + \bar{\alpha} + 2)\tfrac{l}{2} - \alpha k - \bar{\alpha}\bar{k}
	+ N + \bar{N} + 2\right]
	= \frac{1}{R}\left[2l - k - \bar{k} + N + \bar{N} + 2\right]\\
\lambda_0 &= \frac{1}{R}\left[-(\alpha -\bar{\alpha})\tfrac{l}{2} + \alpha k - \bar{\alpha}\bar{k}
	- N + \bar{N}\right]
	= \frac{1}{R}\left[k - \bar{k} - N + \bar{N}\right],
\end{split}\end{equation}
where we have set $\alpha = \bar{\alpha} = 1$ for the physical process
of interest. We also can determine the `unit-sized' amplitude with the
position dependence removed,
\begin{equation}
\bra{f}\mathcal{V}\ket{i}_\text{unit} = 
	\mathcal{A}^{\ibar\jbar}_\text{unit}(0,0) = 
	(-1)^{k+1}\sqrt{\nu}\sqrt{\choose{N+l+1}{N}\choose{\bar{N} + l+1}{\bar{N}}}.
\end{equation}
\end{enumerate}
Putting this into Equation~\eqref{eq:D1D5-decay-rate}, one finds the final
emission rate 
\begin{equation}
\der{\Gamma}{E} = \nu\frac{2\pi}{2^{2l+1}l!^2}\frac{(Q_1Q_5)^{l+1}}{R^{2l+3}}(E^2-\lambda^2)^{l+1}
	\choose{N+l+1}{N}\choose{\bar{N} + l+1}{\bar{N}}\delta_{\lambda, \lambda_0}\delta(E-E_0).
\end{equation}
This is the emission rate for one of $\nu$ excitations in the CFT to
de-excite and emit a supergravity particle with energy $E_0$,
$S^1$-momentum $\lambda_0$, and angular momentum
\begin{equation}\begin{split}\label{eq:emitted-ang-mom}
m_\psi &= -(m + \bar{m}) = -l + k + \bar{k}\\
m_\phi &= m - \bar{m} = \bar{k} - k.
\end{split}\end{equation}
The angular momentum can be read off from the angular momentum of the
NS sector initial state, or the difference in angular momentum between the
initial and final physical states.

The expression for the emission rate obtained above matches the one
obtained in \cite{gms2} where it is given in a slightly different
form. There the expression for a minimally coupled scalar to be
absorbed into the geometry and re-emerge is given in Equation~(5.34)
along with the time of travel in Equation~(5.33). The total
probability is the product of the probabilities to be absorbed and to
re-emerge, which are equal. Therefore, the rate of emission is the
square root of the total probability, with the energy and other
quantum numbers taking the corresponding values for excitations in the
background, divided by the time of travel. This expression is seen to
match the emission rate obtained above.

\section{Emission from nonextremal microstate}\label{sec:nonextremal}

From a physics point of view the emission computed above corresponds
to a very simple process. We take an extremal 2-charge D1D5
microstate, excite it by adding a quantum, and compute the rate at
which the state de-excites by emitting this quantum.

But this same computation can be slightly modified to obtain the
emission rate for a more interesting physical process. We start with a
nonextremal D1D5 microstate which has a large energy above
extremality. This particular microstate is obtained by taking an
extremal D1D5 microstate and performing a spectral flow on both the
left and right moving sectors. Such a spectral flow adds fermionic
excitations to \emph{every} component string. Thus we get a large
energy above extremality, not just the energy of one nonextremal
quantum as was the case with our earlier
computations~\cite{gms1,gms2,ross}.

This nonextremal state emits radiation, and we wish to compute the
rate of emission after $\nu$ quanta have been emitted. We again get
the `Bose enhancement' like that in (\ref{finalq}), so that the rate
of emission keeps increasing as more quanta are emitted. In~\cite{cm1}
it was shown that the resulting decay behavior is exactly the Hawking
radiation expected from this particular microstate. But the
computation of~\cite{cm1} was restricted to certain choices of spins
and excitation level $N=0$ for the emitted quantum; now we are able to
get a general expression for all values of spins and $N$.

\subsection{The CFT process}

\begin{figure}[ht]
\begin{center}
\subfigure[]{\label{fig:spectral-a}
	\includegraphics[width=6.3cm]{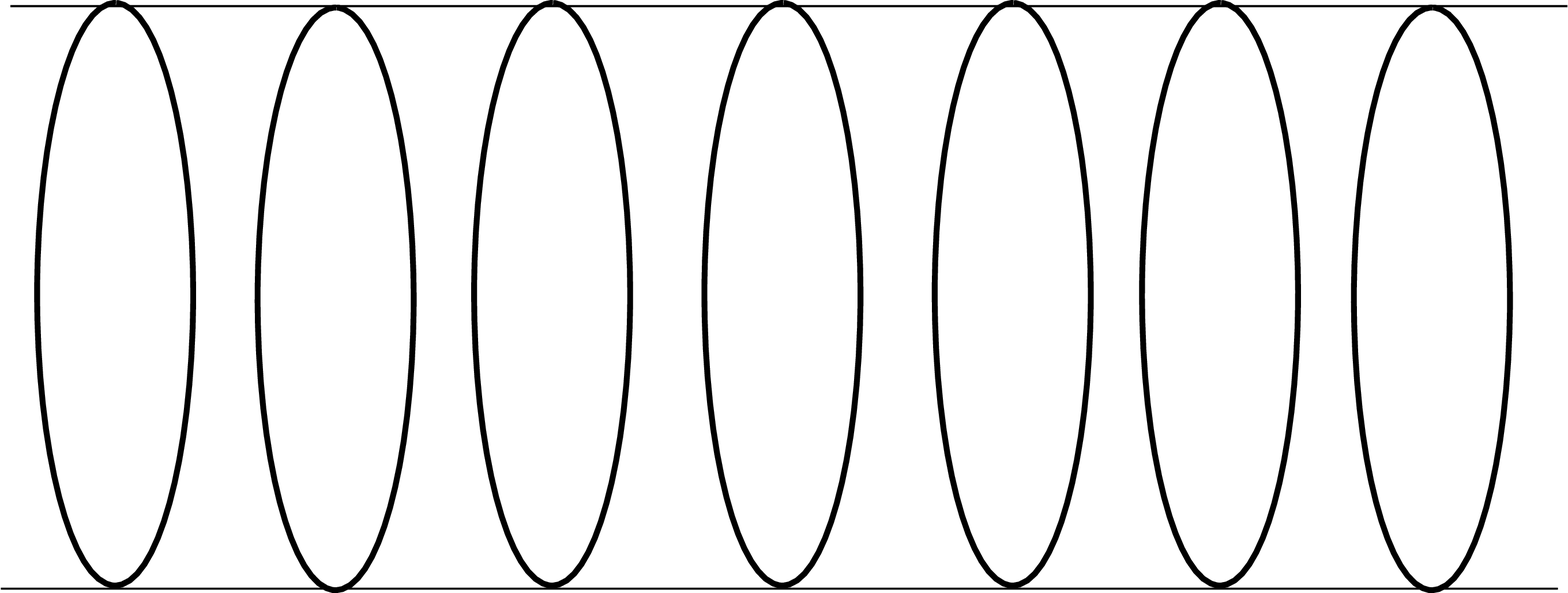}}
\raisebox{28pt}{$\xrightarrow{{\displaystyle \text{spectral flow}}}$}
\subfigure[]{\label{fig:spectral-b}
	\includegraphics[width=6.3cm]{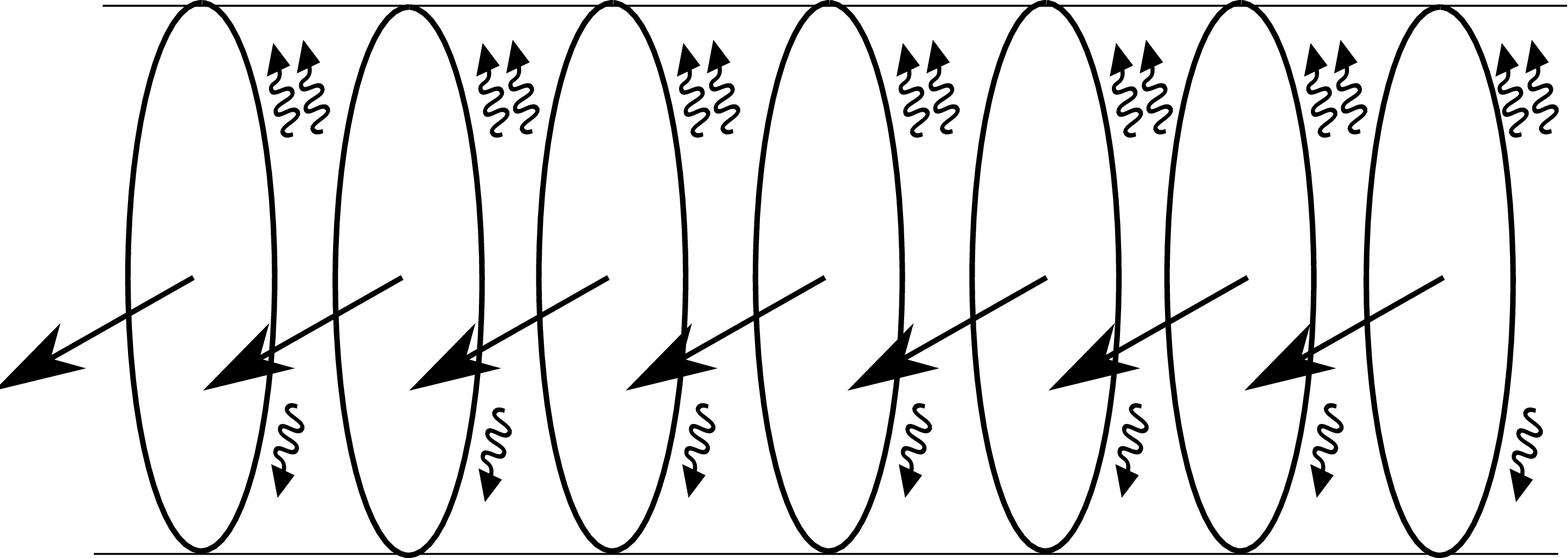}} \caption{(a) The NS
	vacuum state in the CFT and (b) the CFT state after spectral
	flow.  The arrows at the center of the circle indicate the
	`base spin' of component strings in the Ramond sector. The
	wavy arrows on top (bottom) of the strands represent fermionic
	excitations in the left (right)
	sector.\label{fig:SpectralFlow}}
\end{center}
\end{figure}

As discussed earlier, the physical D1D5 system is in the Ramond
sector. We can relate Ramond sector states to NS sector states by
spectral flow. Recall that under spectral flow the dimensions and
charges change as follows:
\begin{equation}\begin{split}
h' &= h + \alpha m+ \frac{c\alpha^2}{24}\\
m' &= m + \frac{c\alpha}{12}.
\end{split}\end{equation}
If we start with the NS vacuum $\vac_{NS}$ and spectra flow by
$\alpha=1$, we reach the Ramond vacuum state with $h={c\over
24}$. But we can also reach a Ramond state by spectral flow by
$\alpha=3, 5, \dots$, which are excited states with energy more than
the energy of the Ramond vacua. Let us take our initial state in the
Ramond sector to be the state obtained by spectral flow of $\vac_{NS}$
by $\alpha=2n+1$ on the left and $\bar{\alpha} = 2\bar{n} + 1$ on the
right. The spectral flow adds fermions to the left and right sectors,
raising the level of the Fermi sea on both these sectors. Thus we get
an excited state of the D1D5 system, which we depict in
Fig.~\ref{fig:spectral-b}.

The vertex operator we have constructed can twist together $l+1$
copies of the CFT. In our earlier computation, we started with a set
of twisted copies, and the vertex operator `untwisted' these, leading
to a final state with no twists. This time the initial state has all
copies of the CFT `untwisted', but these copies are all in an excited
state. The vertex operator can therefore twist together $l+1$ copies,
leading to a twisted component string in the final state. Even though
twisting a set of strings increases the energy, this component string
in the final state can have lower energy than the strings in the
initial state because of the fermionic excitations present on the
initial component strings. The energy difference between the initial
and final states is the energy of the emitted supergravity particle.

Let us now set up the CFT computation needed for this process. We
observe that the amplitude can be obtained in a simple way from the
amplitude that we have already computed.

\subsection{The initial state}

As before, we do all our computations in the NS sector. If we spectral
flow the starting state depicted in Fig.~\ref{fig:spectral-b} by
$-(2n+1)$ units, we arrive at the NS vacuum $\vac_{NS}$ depicted in
Fig.~\ref{fig:spectral-a}. It may appear that this vacuum state cannot
lead to any emission, but recall that we have used spectral flow only
as a technical trick; the actual initial state has a much higher
energy, and indeed leads to emission.

\begin{figure}[ht]
\begin{center}
\subfigure[~The initial state in the NS sector]{\label{fig:RossNSInitial}
	\includegraphics[width=6.3cm]{NSNuEq1}}
\hspace{15pt}
\subfigure[~The final state in the NS sector]{\label{fig:RossNSFinal}
	\includegraphics[width=6.3cm]{NSNuEq2}} \\
\subfigure[~The initial state in the Ramond sector]{\label{fig:RossRRInitial}
	\includegraphics[width=6.3cm]{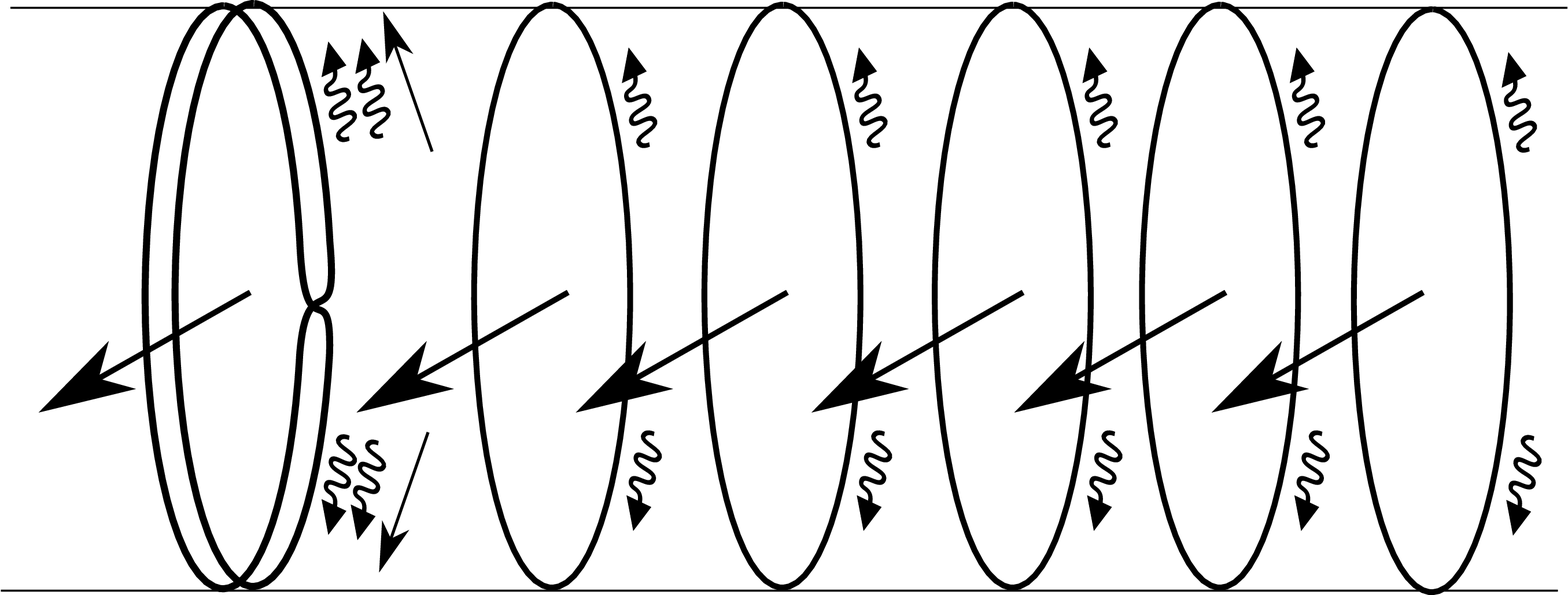}}
\hspace{15pt}
\subfigure[~The final state in the Ramond sector]{\label{fig:RossRRFinal}
	\includegraphics[width=6.3cm]{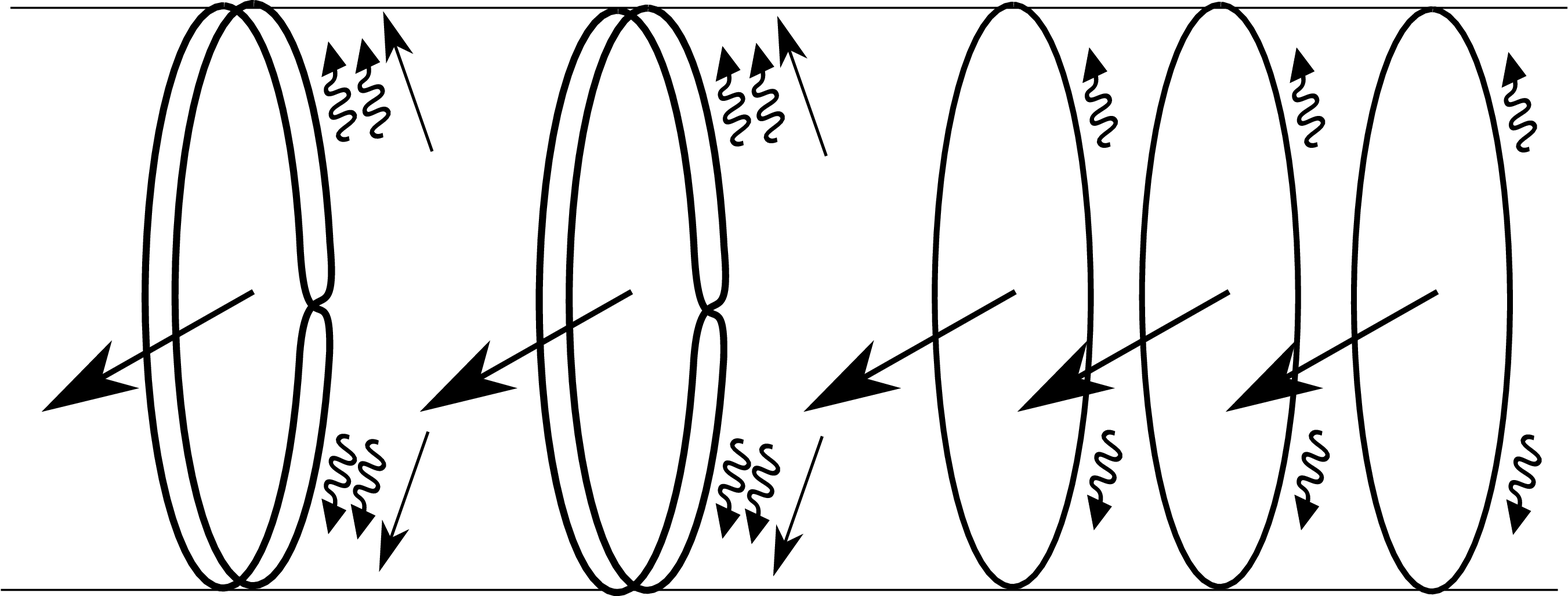}} \caption{The initial
	and final states for the nonextremal emission process
	discussed in Section~\protect\ref{sec:nonextremal}.  The specific case
	depicted is $\nu=2$ and $l=1$.\label{fig:RossDecay}}
\end{center}
\end{figure}

If we wanted to start with this state and proceed with the computation
we would set $\ket{i'}=\vac_{NS}$. But we instead look at a slightly
more general situation where $\nu-1$ quanta have already been emitted.
In this case, the initial state looks like the one depicted in
Fig.~\ref{fig:RossRRInitial}, where $\nu-1$ sets of $l+1$ copies have
already been twisted together.

This may look like a complicated initial state, but we look only at a
specific amplitude: the amplitude for emission of a further quantum of
the same kind as the quanta already present. This process therefore
requires us to take $l+1$ of the \emph{untwisted} copies of the CFT,
and use the vertex operator to twist them together. The other copies
of the CFT are unaffected by the vertex operator. Thus, for the
purposes of computing the amplitude, the initial state of the $l+1$
copies of interest is
\be
\ket{i'}=\vac_{NS}.
\ee

\begin{figure}[ht]
\begin{center}
\subfigure[~The initial state in the NS sector]{\label{fig:alphaNSInitial}
	\includegraphics[width=6.3cm]{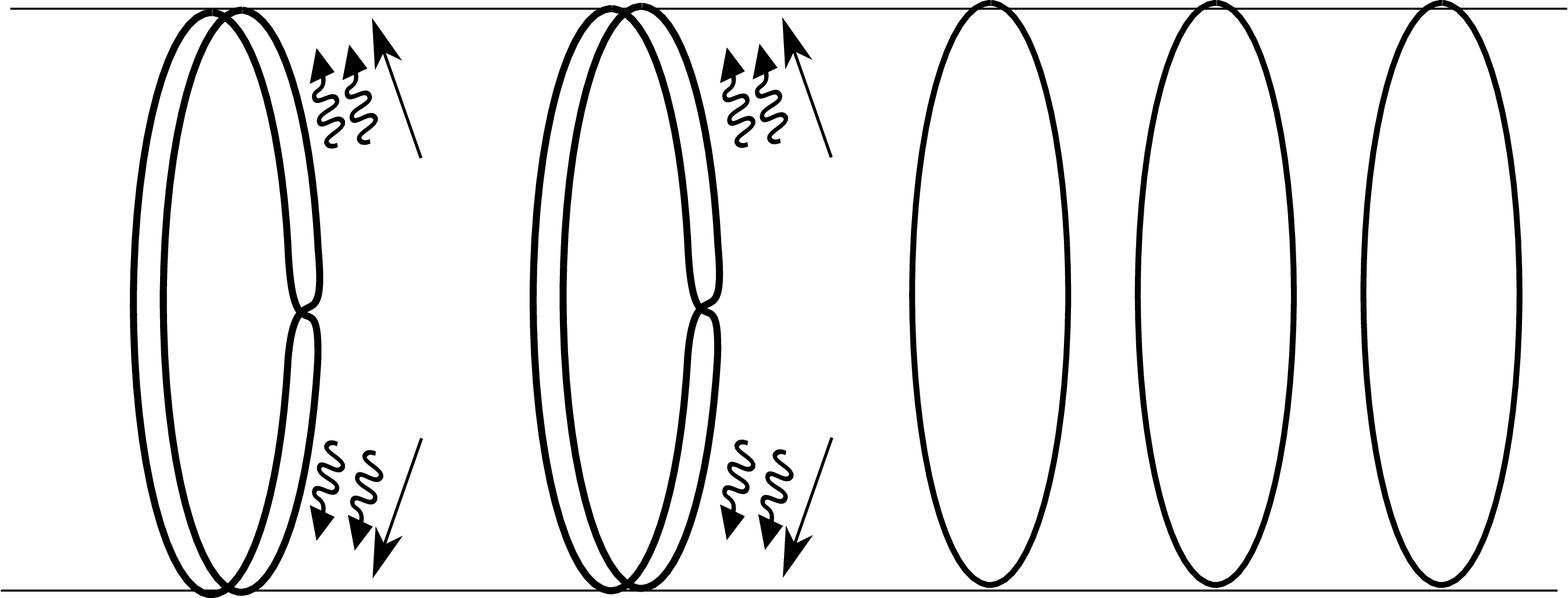}}
\hspace{15pt}
\subfigure[~The final state in the NS sector]{\label{fig:alphaNSFinal}
	\includegraphics[width=6.3cm]{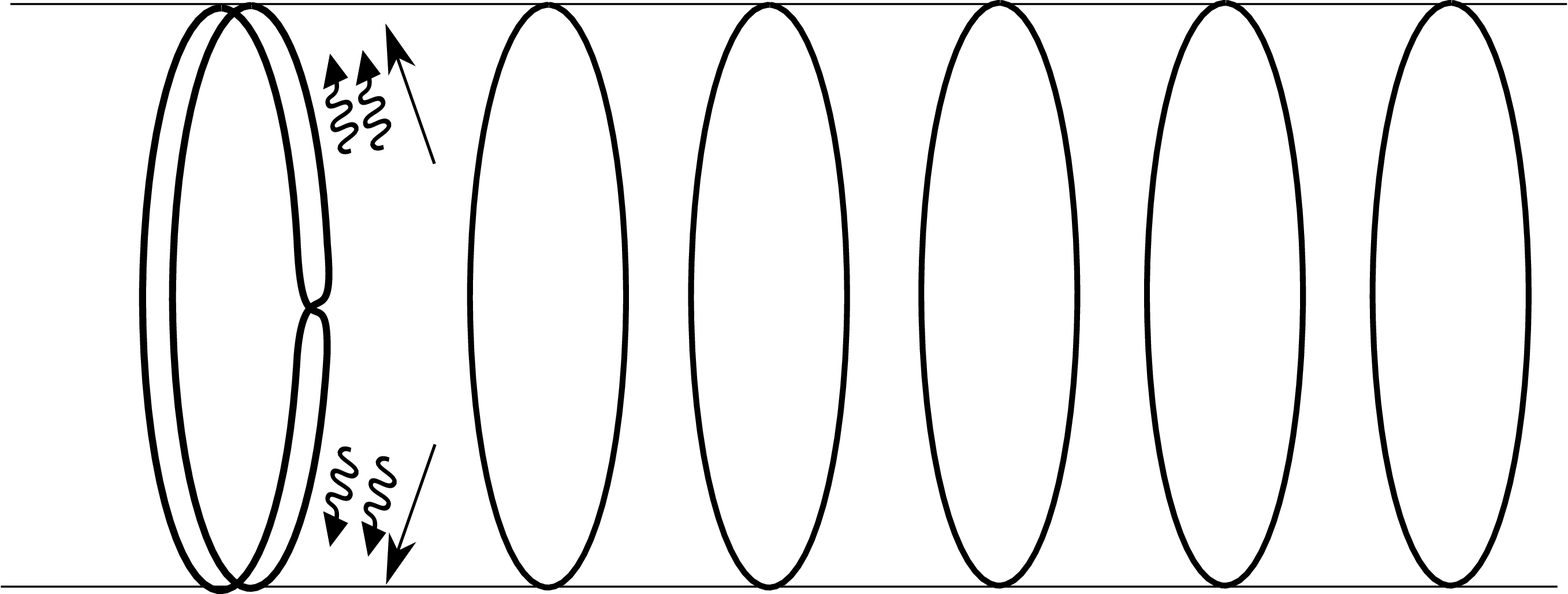}} \\
\subfigure[~The initial state in the Ramond sector]{\label{fig:alphaRRInitial}
	\includegraphics[width=6.3cm]{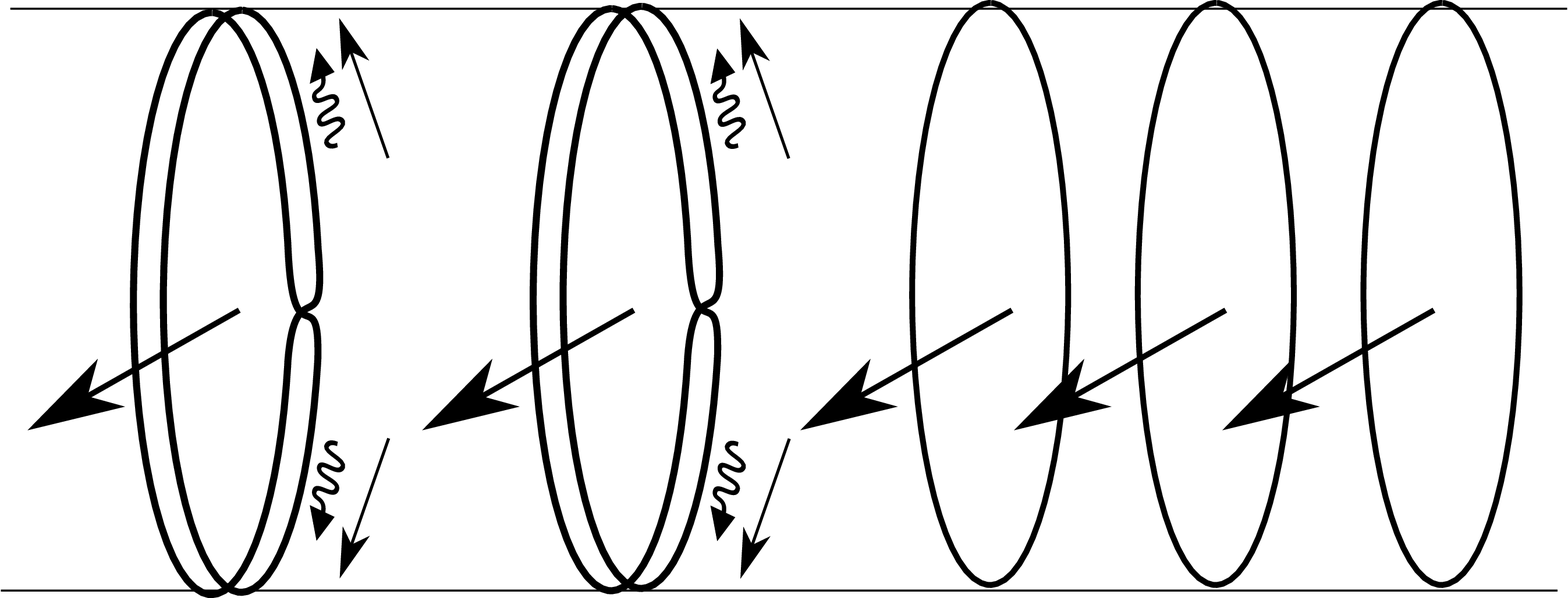}}
\hspace{15pt}
\subfigure[~The final state in the Ramond sector]{\label{fig:alphaRRFinal}
	\includegraphics[width=6.3cm]{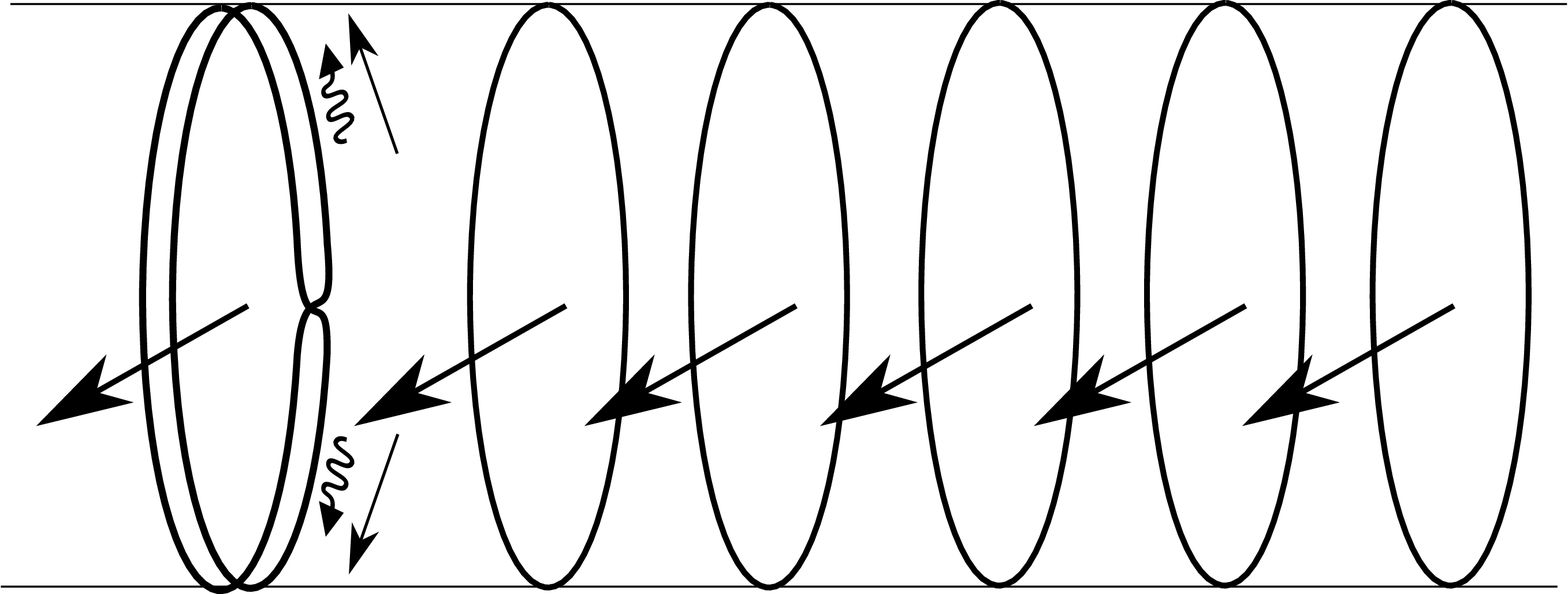}} \caption{The initial
	and final states for the emission process discussed in
Sections~\protect\ref{sec:states-and-op},~\protect\ref{sec:CFT-evaluation},
	and~\protect\ref{sec:calc-rate-emission}. The pictures correspond to
	$\nu=2$ and $l=1$ emission.  The straight arrows pointing up
	(down) on the loops indicate bosonic excitations in the left
	(right) sector.\label{fig:AlphaDecay} }
\end{center}
\end{figure}

\subsection{The final state}

The final state is determined by the fact that we are looking for the
amplitude to transition to a supergravity state, and we have a unique
supergravity excitation with given twist and angular quantum
numbers. Working again in the NS sector, arrived at by spectral flow
by $-(2n+1)$ units, we get
\be
\ket{f'} = \ket{\phi_{N+1}^{\frac{l}{2},\frac{l}{2}-k}};
\ee
the initial state of our previous calculation.

\subsection{The vertex operator}

The vertex operator is independent of the states it acts on. It is
completely determined by the supergravity scalar to which it couples
in Equation~\eqref{eq:general-S-int}.

We now see that the present process is similar to the amplitude we
computed earlier if we reverse the direction of time $\tau$. That is,
the initial state now is untwisted, while in our earlier computation
the \emph{final} state was untwisted. The vertex operator then leads
to a final twisted state, and since there is a unique supergravity
state with given quantum numbers, we can write down this state.

\subsection{Relating the emission amplitude to the earlier computed amplitude}

We term the supergravity excitation emission in the previous sections
`untwisting' emission since a twisted component string in the initial
state `untwists' under the action of the vertex operator and leads to
a final state with no twists. We call the emission of the present
section `twisting' emission since the initial state has no twists, and
the vertex operator leads to a twisted component string in the final
state.

By comparing the initial and final states of the two process, we
immediately see that the current NS sector twisting amplitude is
simply the Hermitian conjugate of the previous NS sector untwisting
amplitude,
\begin{equation}
{\mathcal{A}'}_{l, m_\psi, m_\phi}^\text{twisting}(t,y) 
	= \big[{\mathcal{A}'}_{l, -m_\psi, -m_\phi}^\text{untwisting}(-t,y)\big]^\dg,
\end{equation}
with flipped $SO(4)_E$ charges and reversed time. In the complex
plane, this statement becomes
\begin{equation}\label{eq:herm-conj-amp}
{\mathcal{A}'}^\text{twisting}_{l, m_\psi, m_\phi}(z, \bar{z}) = 
    \big[{\mathcal{A}'}^\text{untwisting}_{l, -m_\psi, -m_\phi}\big(\tfrac{1}{\bar{z}},\tfrac{1}{z}\big)\big]^\dg.
\end{equation}

To see the above relation explicitly, consider the Hermitian conjugate
of the previous, untwisting amplitude:
\begin{calc}
\big[{\mathcal{A}'}_{l,m_\psi, m_\phi}^\text{untwisting}(z,\bar{z})\big]^\dg 
	&= \left[|z|^{l+2}\bra{f'}\mathcal{V}_{l,m_\psi, m_\phi}
                (z, \bar{z})\ket{i'}\right]^\dg\\
	&= |z|^{l+2}\bra{i'}
            \big[\mathcal{V}_{l,m_\psi, m_\phi}(z,\bar{z})\big]^\dg\ket{f'}\\
	&= \frac{1}{|z|^{l+2}}
		\bra{i'}\mathcal{V}_{l,-m_\psi, -m_\phi}
                \big(\tfrac{1}{\bar{z}},\tfrac{1}{z}\big)\ket{f'},	
\end{calc}
where the $i'$ and $f'$ are from the previous calculation. The
amplitude we now wish to compute is (in terms of the previous
calculation's states)
\begin{equation}
{\mathcal{A}'}_{l,m_\psi, m_\phi}^\text{twisting}(z,\bar{z}) = |z|^{l+2}
                       \bra{i'}\mathcal{V}_{l,m_\psi, m_\phi}(z,\bar{z})\ket{f'}
\end{equation}
Comparing these two expressions, one arrives at
Equation~\eqref{eq:herm-conj-amp}.

From Equation~\eqref{eq:final-L-amp}, we have
\begin{equation}
{\mathcal{A}'}^\text{untwisting}(z, \bar{z})
	 = (-1)^{k+\bar{k}}
\sqrt{\choose{N+l+1}{N}\choose{\bar{N} + l+1}{\bar{N}}}
 z^{-\frac{l}{2} - N - 1}
\bar{z}^{-\frac{l}{2} - \bar{N} - 1},
\end{equation}
and so using Equation~\eqref{eq:herm-conj-amp} gives
\begin{equation}
{\mathcal{A}'}^\text{twisting} = (-1)^{k+\bar{k}}
	\sqrt{\choose{N+l+1}{N}\choose{\bar{N}+l+1}{\bar{N}}}
	z^{\frac{l}{2} + N + 1}\bar{z}^{\frac{l}{2} + \bar{N} +1}.
\end{equation}
Note that this amplitude is in the NS sector, and before we can use it
to get emission we have to spectral flow it to the Ramond sector.
Spectral flowing by $\alpha=2n+1$ units to the Ramond sector gives
\begin{calc}\label{eq:cft-twisting-amp}
\mathcal{A}^{(\alpha, \nu)}_\text{twisting} 
	&= z^{-\alpha(\frac{l}{2}-k)}\bar{z}^{-\bar{\alpha}(\frac{l}{2}-\bar{k})}
		\mathcal{A}'^{(\nu)}_\text{twisting}\\
	&= (-1)^{k+\bar{k}}\sqrt{\nu}\,
	\sqrt{\choose{N+l+1}{N}\choose{\bar{N}+l+1}{\bar{N}}}
		z^{\frac{l}{2} + N + 1- \alpha(\frac{l}{2}-k)}
	\bar{z}^{\frac{l}{2} + \bar{N} +1 - \bar{\alpha}(\frac{l}{2}-\bar{k})},
\end{calc}
where we have put the combinatoric factor in as well. Note that the
combinatorics work out the same as before since the combinatorics
cannot be affected by Hermitian conjugation; however, the
interpretation is different. The initial state starts with $\nu-1$
sets of $(l+1)$-twisted component strings, while the final state has
$\nu$ $(l+1)$-twisted component strings. Therefore, if at some initial
time all of the strands were untwisted and each twist corresponds to
an emitted supergravity particle, then the above is the amplitude for
the emission of the $\nu$th particle.

Comparing Equation~\eqref{eq:cft-twisting-amp} with
Equation~\eqref{eq:final-euc-amp}, we see that the amplitudes agree
except for the power of $z$, which is different because of the
different energies of the concerned states in the two processes. Thus
we can immediately write down the emission rate for the $\nu$th
particle from the cap into the flat space
\begin{equation}
\der{\Gamma}{E}
  = 
  \nu\frac{2\pi}{2^{2l+1}l!^2}\frac{(Q_1Q_5)^{l+1}}{R^{2l+3}}(E^2-\lambda^2)^{l+1}
	\choose{N+l+1}{N}\choose{\bar{N} + l+1}{\bar{N}}\delta_{\lambda, \lambda_0}\delta(E-E_0)
\end{equation}
where
\begin{equation}\begin{split}
E_0 &= \frac{1}{R}\left[(\alpha + \bar{\alpha} - 2)\tfrac{l}{2} 
  - \alpha k - \bar{\alpha}\bar{k} - N - \bar{N} - 2\right]\\
\lambda_0 &= \frac{1}{R}\left[-(\alpha - \bar{\alpha})\tfrac{l}{2} + \alpha k 
   - \bar{\alpha}\bar{k} + N - \bar{N}\right].
\end{split}\end{equation}
For sufficiently large $\alpha$ or $\bar{\alpha}$, $E_0$ is positive
and the physical process is emission and not absorption.  In taking the
hermitian conjugate we have flipped the the angular momentum of the
emitted particle from that in Equation~\eqref{eq:emitted-ang-mom};
therefore, the above emission rate is for
\begin{equation}
m_\psi =  l - k - \bar{k}\qquad m_\phi = k - \bar{k}.
\end{equation}

One can check that the emission rate above agrees with the emission
rate from the gravity dual \cite{myers,cm1}. Such a check was carried
out in \cite{cm1} only for states with excitation level $N=0$, because
it was not clear how to construct the initial state for $N>0$ in the
effective string description of the D1D5 bound state. With our present
construction of states and vertex operators in the orbifold CFT, we
can compute amplitudes for emission of supergravity quanta from all
initial states containing supergravity excitations.

\section{Discussion}

The D1D5 system is described by a 1+1 dimensional CFT, just as D3
branes give rise to a 3+1 dimensional Yang-Mills CFT. The Yang-Mills
case has a simple `free' point where the coupling vanishes and
computations are simple. The analog of the `free' point for the 1+1
CFT is expected to be the orbifold CFT.

Even though the orbifold CFT is expected to be `simple', one faces the
complication that the orbifold group is nonabelian (it is the
permutation group $S_{N_1N_5}$). Thus computations of 3-point
functions take more effort than would be required in a free theory or
an orbifold theory with an abelian orbifold group~\cite{lm2,lm1, 
rastelli1,rastelli2}.

Our goal is to relate CFT computations to gravity computations,
particularly those related to black holes. In this paper we have made
definitive progress toward this end. First we set up a general
formalism relating amplitudes computed in the CFT to scattering
amplitudes for quanta incident from flat infinity. This is important
because the quanta coming from flat infinity do not directly reach the
$AdS$ boundary; they have to pass through an `intermediate region',
which introduces a deformation that we must consider in
general. However, for minimal scalars, which we considered in detail,
it turns out that the intermediate region does not introduce a
deformation.

Next, we constructed the vertex operators that describe the emission
and absorption of supergravity particles. While some supergravity
quanta are described by chiral primary operators, others arise from
the descendants of these operators obtained by acting with the
anomaly-free part of the chiral algebra. The minimal scalars that we
considered arise as descendants, and we constructed these operators
and normalized them.

We then computed the amplitude of emission from a simple excited state
of the D1D5 system. This computation is performed most easily by
spectral flowing the system to the NS sector.  We examined how
spectral flow changed the states and vertex operator, we performed the
computation in the NS sector, and then we spectral flowed back to the
Ramond sector where the physical theory actually lives.

To finish this computation we had to take into account combinatoric
factors that count the different ways in which we can permute the
copies of the orbifold theory involved in the interaction. Through
these steps, we obtained the rate of emission of quanta from the
excited D1D5 CFT state. We then compared the rate to the known
result in the dual gravity description, and found exact agreement. It
is true that the CFT and gravity theories hold at different couplings,
but it is well known that simple low energy processes involving
supergravity quanta often agree at leading order between the two
descriptions.

Finally we noted that the CFT amplitude we computed above could be
related to the CFT amplitude for a different process: emission from a
highly excited state obtained by spectral flow from the CFT vacuum
state. Again the emission rate computed from the CFT was found to
agree exactly with the emission rate obtained in the gravity
description~\cite{myers}.

Note that once we normalize the CFT 2-point function to agree with the
gravity 2-point function, we will get an agreement between the CFT and
gravity for all processes that involve only the 2-point function. We
have made such a normalization using the short distance behavior of
the 2-point function in the CFT, and then used this to compute
emission: a process that depends on `long distance physics' because it
depends on the length of the effective string. Our physics goal is to
compare the emission rate found from ergoregion instability
in~\cite{myers} with a CFT computation; which is what we achieve in
the end.

Such comparisons of emission rates between CFT and gravity
computations have been performed many times in the past
\cite{radiation-3, radiation-4, radiation-5, cm1}, but the vertex
leading to emission was previously modeled somewhat heuristically. In
particular, in~\cite{cm1} the heuristic vertex was used to get the
emission for supergravity quanta with $N=0$ (the lowest energy state
in a given angular momentum sector), but it was not clear how to make
the CFT state for higher $N$. In the present paper we have constructed
the CFT state for all $N$, so emissions of all supergravity quanta can
be computed.  It is hoped that these explicit constructions will allow
us to perform a large set of further computations, including those
that lead to deformations away from the orbifold point.

\begin{acknowledgments}

We are grateful for collaboration with J.~Michelson very early on in
the project. Additionally, we thank O.~Lunin, P.~Kraus, E.~Matrinec,
and Y.~Srivastava for insightful comments, conversations, or other
assistance during the project. Furthermore, we thank A.~Taliotis for
typos and ambiguities found while proofreading.

This work was supported in part by DOE grant~DE-FG02-91ER-40690.

\end{acknowledgments}

\appendix

\section{Notation and conventions for the orbifolded CFT} \label{ap:CFT-notation}

Here, we carefully define the notation and conventions used throughout
this paper, and to be used in future work. The system we describe
lives in $M_{4,1}\times T^4\times S^1$. We restrict our attention to
the case where the compact space is a torus, although one may also
consider K3.

The base space of the CFT is the $S^1$ and time, with fields living in
the orbifolded target space: ${(T^4)^{N_1N_5}}/{S_{N_1N_5}}$. The
expressions in this appendix are exclusively given for the complex
plane.

\subsection{Symmetries and indices}

The symmetries of our theory are $SU(2)_L\times SU(2)_R$ and the
$SO(4)_I\simeq SU(2)_1\times SU(2)_2$ rotations of the torus. Indices
correspond to the following representations
\begin{gather*}
\alpha, \beta \qquad \text{doublet of $SU(2)_L$} \hspace{60pt}
\dot{\alpha}, \dot{\beta} \qquad \text{doublet of $SU(2)_R$}\\
A, B \qquad \text{doublet of $SU(2)_1$}\hspace{60pt}
\dot{A}, \dot{B} \qquad \text{doublet of $SU(2)_2$}\\
i, j \qquad \text{vector of $SO(4)$}.
\end{gather*}
One can project vectors of $SO(4)$ into the doublets of $SU(2)_1\times
SU(2)_2$, using the usual Pauli spin matrices and the identity matrix
\[
(\sigma^i)^{\dot{A}A},\qquad \sigma^4 = i\id_2.
\]
The indices are such that, for instance, $(\sigma^2)^{\dot{2}1}= i$. 

We use indices $a,b,c=1,2,3$ for the triplet of any $SU(2)$. Their
occurrence is rare enough that which $SU(2)$ is being referred to
should be unambiguous. Note that the $SU(2)$ generators
${(\sigma^a)_\alpha}^\beta$ naturally come with one index raised and
one index lowered. On the other hand the Clebsch-Gordan coefficients
to project a vector of $SO(4)$ into two $SU(2)$'s naturally come with
both indices raised (or lowered), as above.

We raise and lower all $SU(2)$ doublet indices in the same way so that
\begin{equation}
\epsilon_{\alpha\beta}v^\beta = v_\alpha\qquad v^\alpha = \epsilon^{\alpha\beta}v_\beta,
\end{equation}
where 
\begin{equation}
\epsilon_{12} = -\epsilon_{21} = \epsilon^{21} = -\epsilon^{12} = 1,
\end{equation}
and therefore
\begin{equation}
\epsilon_{\alpha\beta}\epsilon^{\beta\gamma} = \delta_\alpha^\gamma.
\end{equation}

\subsection{Field content}

The bosonic field content of each copy of the CFT consists of a vector
of $SO(4)_I$, $X^i(z,\bar{z})$, giving the position of that effective
string in the torus. The fermions on the left sector have indices in
$SU(2)_L\times SU(2)_2$, while the fermions in the right sector have
indices in $SU(2)_R\times SU(2)_2$:
\begin{equation}
\psi^{\alpha \dot{A}}(z)\qquad \bar{\psi}^{\dot{\alpha}\dot{A}}(\bar{z}).
\end{equation}
These fermions are complex, so there are two complex fermions and
their Hermitian conjugates in the left sector and two complex fermions
and their Hermitian conjugates in the right sector. 

Note that we use the abbreviated notation
\begin{equation}
[X]^{\dot{A}A} = X^i(\sigma^i)^{\dot{A}A}.
\end{equation}

\subsection{Currents}

The holomorphic currents of our theory that form a closed OPE algebra
are an $SU(2)_L$ current, $J^a(z)$; the supersymmetry currents,
$G^{\alpha A}(z)$; and the stress--energy $T(z)$. The right sector has
the corresponding anti-holomorphic currents. Obviously, in this case,
the index $a$ on $J$ transforms in $SU(2)_L$. 

For each copy of the CFT, the currents are realized in terms of the
fields as
\begin{subequations}\label{eq:currents-def}
\begin{align}
J^a(z) &= \frac{1}{4}\epsilon_{\dot{A}\dot{B}}
	\psi^{\alpha \dot{A}}\epsilon_{\alpha\beta}{(\sigma^{*a})^\beta}_\gamma	\psi^{\gamma\dot{B}}\\
G^{\alpha A}(z) &= \psi^{\alpha \dot{A}}[\pd X]^{\dot{B}A}\epsilon_{\dot{A}\dot{B}}\\
T(z) &= \frac{1}{4}\epsilon_{\dot{A}\dot{B}}\epsilon_{AB}[\pd X]^{\dot{A}A}[\pd X]^{\dot{B}B}
	+ \frac{1}{2}\epsilon_{\alpha\beta}\epsilon_{\dot{A}\dot{B}}
			\psi^{\alpha\dot{A}}\pd\psi^{\beta\dot{B}}.
\end{align}
\end{subequations}
Note that the $SO(4)_I$ of the torus is an outer automorphism and so
while we can make a generator that acts appropriately on the fermions
we cannot make one that also acts appropriately on the bosons. 

% \subsubsection{Relation to Other Papers}
% 
% To relate the susy currents to notations from previous papers, one
% should identify
% \begin{equation}
% G^{11} = G^1\qquad G^{21} = G^2\qquad G^{12} = \tilde{G}_2\qquad G^{22} = -\tilde{G}_1.
% \end{equation}
% and
% \begin{equation}
% \psi^{+\dot{1}} = \tilde{\psi}^2\qquad \psi^{-\dot{1}}= \tilde{\psi}^1\qquad
% \psi^{+\dot{2}} = \psi_1 \qquad \psi^{-\dot{2}} = -\psi_2.
% \end{equation}
 
\subsection{Hermitian conjugation}

Because we work in a Euclidean time formalism, one must address
Hermitian conjugation carefully so that it is consistent with the
physical, real-time formalism.

A quasi-primary field of weight $(\Delta, \bar{\Delta})$ is Hermitian
conjugated as~\cite{difrancesco}
\begin{equation}
[\mathcal{O}(z,\bar{z})]^\dg = \bar{z}^{-2\Delta}z^{-2\bar{\Delta}}
	\mathcal{O}^\dg\big(\tfrac{1}{\bar{z}}, \tfrac{1}{z}\big),
\end{equation}
where $\mathcal{O}^\dg(z, \bar{z})$ has the opposite charges under
$SU(2)_L\times SU(2)_R\times SO(4)_I$.

The fermions Hermitian conjugate as
\begin{equation}
\big(\psi^{\alpha\dot{A}}\big)^\dg(z)
	=
	-\epsilon_{\alpha\beta}\epsilon_{\dot{A}\dot{B}}\psi^{\beta\dot{B}}(z)
	= - \psi_{\alpha\dot{A}}(z).
\end{equation}
This reality condition ensures that there are only four real degrees
of freedom in both the left and right sectors. The specific sign can
be determined from the basic fermion correlator and demanding
a positive-definite norm.

The bosons conjugate as
\begin{equation}
\big(X^i\big)^\dg(z, \bar{z}) = X^i(z, \bar{z})\qquad
\big([X]^{\dot{A}A}\big)^\dg(z,\bar{z}) 
	= -\epsilon_{\dot{A}\dot{B}}\epsilon_{AB}[X]^{\dot{B}B}(z,\bar{z}).
\end{equation}

The stress energy tensor and the $SU(2)_L$ current are both Hermitian
and so conjugate trivially; whereas, the supercurrents conjugate as
\begin{equation}
\big(G^{\alpha A}\big)^\dg(z) = -\epsilon_{\alpha\beta}\epsilon_{AB} G^{\beta B}(z).
\end{equation}
Again, the specific sign is determined by requiring the norm to be
positive-definite.

The Ramond vacua conjugate as
\begin{equation}
\big(\vac_R^\alpha\big)^\dg = {}^R_\alpha\hspace{-4pt}\bvac
\qquad {}_\alpha^R\hspace{-4pt}\braket{\varnothing|\varnothing}_R^\beta = \delta_\alpha^\beta.
\end{equation}

\subsection{OPE}

We normalize the fields so that the basic correlators are
\begin{subequations}
\begin{align}
\vev{X^i(z)X^j(w)} &= -2\delta^{ij}\log |z-w|\label{eq:bos-norm1}\\
\vev{\psi^{\alpha \dot{A}}(z)\psi^{\beta \dot{B}}(w)} &= 
	-\frac{\epsilon^{\alpha\beta}\epsilon^{\dot{A}\dot{B}}}{z-w},
\end{align}
\end{subequations}
where it is also useful to note that Equation~\eqref{eq:bos-norm1}
implies
\begin{equation}
\vev{[X]^{\dot{A}A}(z)[X]^{\dot{B}B}(w)} =4 \epsilon^{\dot{A}\dot{B}}\epsilon^{AB}\log |z-w|.
\end{equation}
From which, the commonly used
\begin{equation}
[\pd X(z)]^{\dot{A}A}[\pd X(w)]^{\dot{B}B}\sim 
	2\frac{\epsilon^{\dot{A}\dot{B}}\epsilon^{AB}}{(z-w)^2},
\end{equation}
immediately follows.

The OPE current algebra for a single copy of the $\Nsc =4$ CFT is
\begin{subequations}\label{eq:currents-OPE}
\begin{align}
J^a(z)J^b(w) &\sim \frac{c}{12}\frac{\delta^{ab}}{(z-w)^2} 
			+ i{\epsilon^{ab}}_c\frac{J^c(w)}{z-w}\\
J^a(z)G^{\alpha A}(w) &\sim \tfrac{1}{2}{(\sigma^{*a})^\alpha}_\beta\frac{G^{\beta A}(w)}{z-w}\\
G^{\alpha A}(z)G^{\beta B}(w) &\sim-\frac{2c}{3}
		\frac{ \epsilon^{AB}\epsilon^{\alpha\beta}}{(z-w)^3}
	+ 2\epsilon^{AB}\epsilon^{\beta\gamma}{(\sigma^{*a})^\alpha}_\gamma
		\left[\frac{2J^a(w)}{(z-w)^2} + \frac{\pd J^a(w)}{z-w}\right]
		-2\epsilon^{AB}\epsilon^{\alpha\beta}\frac{T(w)}{z-w} \\
T(z)J^a(w) &\sim \frac{J^a(w)}{(z-w)^2} + \frac{\pd J^a(w)}{z-w}\\
T(z)G^{\alpha A}(w) &\sim \frac{\tfrac{3}{2}G^{\alpha A}(w)}{(z-w)^2}
			+\frac{\pd G^{\alpha A}(w)}{z-w} \\
T(z)T(w) &\sim \frac{c}{2}\frac{1}{(z-w)^4} + 2\frac{T(w)}{(z-w)^2} + \frac{\pd T(w)}{z-w},
\end{align}
\end{subequations}
which agrees with the above correlators for $c=6$.

For convenient reference, we include the OPEs of the currents with the
basic primary fields, $\pd X$ and $\psi$:
\begin{subequations}\begin{align}
J^a(z)\psi^{\alpha\dot{A}}(w) &\sim \frac{1}{2}{(\sigma^{*a})^\alpha}_\beta 
 	\frac{\psi^{\beta \dot{A}}(w)}{z-w}\\
G^{\alpha A}(z)[\pd X(w)]^{\dot{B}B} &\sim 2\epsilon^{AB}
	\left(\frac{\psi^{\alpha \dot{B}}(w)}{(z-w)^2} 
		+ \frac{\pd \psi^{\alpha \dot{B}}(w)}{z-w}\right)\\
G^{\alpha A}(z)\psi^{\beta \dot{A}}(w) &\sim 
	\epsilon^{\alpha\beta}\frac{[\pd X(w)]^{\dot{A}A}}{z-w}\\
T(z)[\pd X(w)]^{\dot{A}A} &\sim \frac{[\pd X(w)]^{\dot{A}A}}{(z-w)^2} 
	+ \frac{[\pd^2 X(w)]^{\dot{A}A}}{z-w}\\
T(z)\psi^{\alpha \dot{A}}(w) &\sim \frac{\tfrac{1}{2}\psi^{\alpha \dot{A}}(w)}{(z-w)^2}
	+ \frac{\pd\psi^{\alpha \dot{A}}(w)}{z-w}.
\end{align}\end{subequations}

\subsection{Mode algebra}

We define the modes corresponding to the above currents according to
their weight, $\Delta$, by
\begin{equation}\begin{split}
\mathcal{O}_m  &= \oint\frac{\drm z}{2\pi i}\mathcal{O}(z) z^{\Delta + m-1}\\
\mathcal{O}(z) &= \sum_m \mathcal{O}_m\, z^{-(\Delta + m)}.
\end{split}\end{equation}
The weight may be read off from the OPE of the current with the
stress--energy tensor. Fermionic currents have half-integer weight. In
the NS sector, fermions are periodic in the plane and therefore we
need integer powers of $z$. This means the fermionic currents have
modes labeled by half-integer $m$.

Using the OPE current algebra above, one finds that the modes form an
algebra:
\begin{subequations}\label{eq:mode-algebra}
\begin{align}
\com{J^a_m}{J^b_n} 
  &= \tfrac{c}{12}m\delta^{ab}\delta_{m+n, 0} +i{\epsilon^{ab}}_cJ^c_{m+n}\\
\com{J^a_m}{G^{\alpha A}_n}
  &= \frac{1}{2}{(\sigma^{*a})^\alpha}_\beta G^{\beta A}_{m+n}\\
\ac{G^{\alpha A}_m}{G^{\beta B}_n} 
  &=-\tfrac{c}{3}(m^2-\tfrac{1}{4})\epsilon^{AB}\epsilon^{\alpha\beta}\delta_{m+n,0}
 +2(m-n)\epsilon^{AB}\epsilon^{\beta\gamma}{(\sigma^{*a})^\alpha}_\gamma J^a_{m+n}
 -2\epsilon^{AB}\epsilon^{\alpha\beta}L_{m+n}\\
\com{L_m}{J^a_n} &= -n J^a_{m+n}\\
\com{L_m}{G^{\alpha A}_n} &= (\tfrac{m}{2}-n)G^{\alpha A}_{m+n}\\
\com{L_m}{L_n} &= c\tfrac{m^3-m}{12}\delta_{m+n, 0} + (m-n)L_{m+n}.
\end{align}
\end{subequations}

The infinite-dimensional algebra has a finite, anomaly-free subalgebra
which is of primal importance for the AdS--CFT correspondence. The
anomaly-free subalgebra has a basis of $\{J_0^a,\,G^{\alpha
  A}_{\pm\frac{1}{2}},\,L_0,\, L_{\pm 1}\}$. The smaller subalgebra
spanned by $\{J^3_0,\, L_0\}$ is the Cartan subalgebra, which means we
may label states and operators by their charge $m$ and their weight
$h$.

For reference, we provide the mode algebra of the two canonical
primary fields. The $\pd X$'s modes are $\alpha_n$.
\begin{subequations}\begin{align}
\com{\alpha_m^{\dot{A}A}}{\alpha_n^{\dot{B}B}} &= 2m \epsilon^{\dot{A}\dot{B}}
	\epsilon^{A B}\delta_{n+m,0}\\
\ac{\psi^{\alpha\dot{A}}_m}{\psi^{\beta\dot{B}}_n} &= 
	-\epsilon^{\alpha\beta}\epsilon^{\dot{A}\dot{B}}\delta_{m+n,0}\\
\com{J^a_m}{\psi^{\alpha\dot{A}}_n} &= \tfrac{1}{2}{(\sigma^{*a})^\alpha}_\beta
		\psi^{\beta\dot{A}}_{m+n}\\
\com{G^{\alpha A}_m}{\alpha^{\dot{B}B}_n} &= -2n\epsilon^{AB}\psi^{\alpha\dot{B}}_{m+n}\\ 
\ac{G^{\alpha A}_m}{\psi^{\beta\dot{A}}_n} &= \epsilon^{\alpha\beta}\alpha_{m+n}^{\dot{A}A}\\
\com{L_m}{\alpha^{\dot{A}A}_n} &= -n\alpha_{m+n}^{\dot{A}A}\\
\com{L_m}{\psi^{\alpha\dot{A}}_n} 
	&= -(\tfrac{m}{2}+n)\psi^{\alpha\dot{A}}_{m+n}.
\end{align}\end{subequations}

\subsection{Useful identities}

These identities are useful for relating vectors of $SO(4)_I$ to tensors
in $SU(2)_1\times SU(2)_2$:
\begin{subequations}\begin{align}
\epsilon_{\dot{A}\dot{B}}\epsilon_{AB}(\sigma^i)^{\dot{A}A}(\sigma^j)^{\dot{B}B} 
	&= -2\delta^{ij}\\
(\sigma^i)^{\dot{A}A}(\sigma^i)^{\dot{B}B} &= -2\epsilon^{\dot{A}\dot{B}}\epsilon^{AB}.
\end{align}\end{subequations}

It is useful to know how to relate the $(+,-,3)$ basis for the triplet
of $SU(2)$ to the $(1,2,3)$ basis:
\begin{subequations}
\begin{align}
\delta^{++} &= \delta^{--} = \delta_{++} = \delta_{--} = 0\\
\delta^{+-} &= \delta^{-+} = 2\qquad \delta_{+-}= \delta_{-+} = \tfrac{1}{2}\\
\epsilon^{+-3} &= -2i\\
\sigma^+ &= \begin{pmatrix}0 & 2 \\ 0 & 0 \end{pmatrix}\qquad
\sigma^- = \begin{pmatrix}0 & 0 \\ 2 & 0 \end{pmatrix}.
\end{align}
\end{subequations}
One can raise and lower the `$3$' index with impunity.

\subsection{$n$-twisted sector mode algebra}

In the $n$-twisted sector, by which we mean modes whose contour orbits
a twist operator, we can only define the modes by summing over all
$n$-copies of the field. This allows us to define fractional
modes. The modes are defined by~\cite{lm2}
\begin{equation}
\mathcal{O}_\frac{m}{n} = \oint_0\frac{\drm z}{2\pi i}\sum_{k=1}^n\mathcal{O}^{(k)}(z)
	e^{2\pi i\frac{m}{n}(k-1)}z^{\Delta + \frac{m}{n} -1}.
\end{equation}
One can confirm that the integrand is $2\pi$-periodic and therefore
well-defined. If one lifts to a covering space using a map that
locally behaves as
\begin{equation}
z = b t^n,
\end{equation}
then the mode in the base $z$-space can be related to a mode in the
$t$-space:
\begin{equation}\label{eq:frac-mode-cover}
\mathcal{O}_{\frac{m}{n}}^{(z)} = b^\frac{m}{n} n^{1-\Delta}\mathcal{O}_{m}^{(t)},
\end{equation}
where $\Delta$ is the weight of the field.

To compute a correlator in the twisted sector, one may either work in
the base space with the summed-over-copies modes or one may work in
the covering space with the opened-up mode. If one works in the base
space, then one should use the algebra with the \emph{total} central
charge
\begin{equation}
c_\text{tot.} = nc;
\end{equation}
the algebra is otherwise unchanged. If one works in the covering space
then one uses the central charge of a single copy of the CFT, but must
remember to write all of the factors that come in lifting to the
cover. These two methods give identical answers.

\subsection{Spectral flow}\label{sec:spectral-flow}

The $\Nsc = 4$ algebra is a vector space at every point $z$ in the
complex plane, spanned by the local operators $\{J^a(z), G^{\alpha
A}(z), T(z)\}$. This vector space closes under the OPE. It is possible
to make a $z$-dependent change of basis and preserve the algebra.

Making an $SU(2)_L$ transformation in the `3' direction to the local
operators by an angle, $\eta(z)=i\alpha\log z$, at every point $z$
is called `spectral flow' by $\alpha$ units. While this may look like
a nontrivial transformation, the new algebra is isomorphic to the old
algebra~\cite{spectral}.

It is important to remember that $\log z$ has a branch cut, which we
put on the real axis for the following discussion.  Let us suppose we
start in the NS sector, where the local operators are periodic in the
complex plane. Let us spectral flow the local operators by $\alpha$
units. Suppose we start on the (positive imaginary side of the)
positive real axis, where $\eta = 0$ and the new operators are the
same as the old operators. As we make a counter-clockwise circle in
the complex plane, the angle between the old operators and the new
operators increases. Across the branch cut on the real axis, where
before the operators were continuous, now there is a large, finite
$SU(2)_L$ transformation.

More illustratively, consider how the fermions behave under spectral
flow, as described above:
\begin{equation}
\psi^{\pm \dot{A}}(z)\mapsto {\psi^{\pm \dot{A}}}'(z) 
	= e^{\pm\frac{i}{2}\eta(z)}\psi^{\pm\dot{A}}(z)
	= z^{\mp\frac{\alpha}{2}}\psi^{\pm\dot{A}}(z).
\end{equation}
We see that except for even $\alpha$, there is a branch cut. Moreover,
if we spectral flow by an odd number of units, then the new operators
$\psi'(z)$ have the opposite periodicity from $\psi(z)$. In general,
one expects that an operator with charge $m$ under $SU(2)_L$
transforms as
\begin{equation}\label{eq:nice-spec-flow}
\mathcal{O}(z) \mapsto z^{-\alpha m}\mathcal{O}(z);
\end{equation}
however, the superconformal algebra and its $SU(2)_L$ subalgebra, in
particular, is anomalous which leads to nontrivial transformations of
some operators.

Since the spectral flowed algebra and the original algebra are
isomorphic, there is a bijective mapping from states living in the
representations of one algebra to states living in the representation
of its spectral flow. Since the NS sector and the R sector are related
by spectral flow, we can map problems in one sector into problems in
the other.

The operator which maps states into their spectral flow images, we
call $\mathcal{U}_\alpha$,
\begin{equation}
\ket{\psi'} = \mathcal{U}_\alpha\ket{\psi}.
\end{equation} 
Formally, then, we may write the action of spectral flow on operators
as
\begin{equation}
\mathcal{O}'(z) = \mathcal{U}_\alpha\mathcal{O}(z)\mathcal{U}^{-1}_\alpha,
\end{equation}
so that amplitudes are invariant under spectral flow.  The spectral
flow operator, $\mathcal{U}_\alpha$ may be roughly defined as an
`improper gauge transformation'~\cite{spectral, spectral-yu,
  vafa-warner}.  

The spectral flow operator is most naturally defined in the context of
bosonized fermions. We can bosonize the fermions as (conventions
chosen to be consistent with~\cite{lm2})
\begin{equation}
\psi^{+\dot{1}} = e^{-i\phi_6}\qquad
\psi^{+\dot{2}} = e^{i\phi_5} \qquad
\psi^{-\dot{1}} = e^{-i\phi_5}\qquad
\psi^{-\dot{2}} = -e^{i\phi_6},
\end{equation}
which gives the $SU(2)_L$ current in the form\footnote{There are
  implicit cocycles on the exponentials, which make unrelated fermions
  anticommute. Thus, the order of exponentials in expressions
  matters.}
\begin{equation}
J^3(z) = \frac{i}{2}\big(\pd\phi_5(z) - \pd\phi_6(z)\big)\qquad
J^+(z) = e^{-i\phi_6}e^{i\phi_5}(z)\qquad
J^-(z) = e^{-i\phi_5}e^{i\phi_6}(z).
\end{equation}
The fields $\phi_5$ and $\phi_6$ are the (holomorphic half of) real
bosons normalized such that
\begin{equation}
\vev{\phi_i(z)\phi_j(w)} = -\delta_{ij}\log(z-w).
\end{equation} 
They may be expanded as
\begin{equation}
\phi_i = q_i - \frac{i}{2}p_i\,\log z + (\text{modes}),
\end{equation}
where $q_i$ and $p_i$ are the zero-mode position and momentum which
satisfy
\begin{equation}
\com{q_i}{p_j} = i\delta_{ij}.
\end{equation}

With this bosonization, the spectral flow operator can be written as
~\cite{vafa-warner}
\begin{equation}
\mathcal{U}_\alpha = e^{i\alpha(q_5 - q_6)}.
\end{equation}
We see that spectral flow corresponds to increasing and decreasing the
zero mode momentum of the fields $\phi_5$ and $\phi_6$ used to
bosonize the fermions. The Baker--Campbell--Hausdorff formula implies
\begin{equation}
e^{i\alpha q_i}e^{i\beta p_j} = e^{-i\alpha\beta\delta_{ij}}
       e^{i\beta p_j}e^{i\alpha q_i},
\end{equation}
which one can use to confirm that this operator has the correct action
on fermions.

From this perspective, one can see that any operator that is `pure
exponential' in $\phi_5$ and $\phi_6$ transforms as in
Equation~\eqref{eq:nice-spec-flow}. If one considers any of the chiral
primaries of the $\Nsc=4$ orbifold theory, one finds that all of the
chiral primaries are `pure exponential' and therefore transform using
Equation~\eqref{eq:nice-spec-flow}.

% We define the periodicity of the fermions in the complex plane via the
% parameter $\beta_\pm$:
% \begin{equation}
% \psi^{\pm\dot{A}}(ze^{2\pi i}) = e^{i\pi\beta_\pm}\psi^{\pm\dot{A}}(z)\qquad
% \bar{\psi}^{\pm\dot{A}}(\bar{z}e^{2\pi i}) = e^{i\pi\bar{\beta}_\pm}\bar{\psi}^{\pm\dot{A}}(z).
% \end{equation}
% Obviously $\beta_\pm$ is only defined modulo 2 under addition.  We use
% spectral flow to alter the fermion content of the CFT states and to go
% from the NS sector, $\beta_\pm =0$, to the R sector $\beta_\pm =
% 1$. The effect of spectral flow by $\alpha$ units on the left and
% $\bar{\alpha}$ units on the right is
% \begin{equation}
% \beta_\pm \mapsto \beta'_\pm = \beta_\pm \pm \alpha\qquad
% \bar{\beta}_\pm \mapsto \bar{\beta}'_\pm = \bar{\beta}_\pm \pm \bar{\alpha}.
% \end{equation}
 
One finds that the the currents transform under spectral flow as
follows
\begin{equation}\begin{split}
J^3(z) &\mapsto J^3(z) - \frac{c\alpha}{12z}\\
J^\pm(z) &\mapsto z^{\mp\alpha}J^\pm(z)\\
G^{\pm A}(z) &\mapsto z^{\mp\frac{\alpha}{2}}G^{\pm A}(z)\\
T(z) &\mapsto T(z) -\frac{\alpha}{z}J^3(z)+\frac{c\alpha^2}{24z^2},
\end{split}\end{equation}
which gives rise to the transformation of the modes,
\begin{equation}\begin{split}
J^3_m &\mapsto J^3_m - \frac{c\alpha}{12}\delta_{m,0}\\
J^\pm_m &\mapsto J^\pm_{m\mp\alpha}\\
G^{\pm A}_m &\mapsto G^{\pm A}_{m\mp\frac{\alpha}{2}}\\
L_m &\mapsto L_m -\alpha J^3_m+\frac{c\alpha^2}{24}\delta_{m,0}.
\end{split}\end{equation}

Spectral flow also acts on states, changing the weight and charge by
\begin{subequations}\label{eq:spectral-flow-hj}
\begin{align}
h\mapsto h' &= h + \alpha m + \frac{c\alpha^2}{24}\\
m\mapsto m' &= m + \frac{c\alpha}{12},
\end{align}
\end{subequations}
which can be read off from $L_0$ and $J^3_0$.  Frequently, one can
deduce the spectral-flowed state from the weight and charge.

Note that spectral flow by $\alpha_1$ units followed by spectral flow
by $\alpha_2$ units is equivalent to spectral flow by $\alpha_1 +
\alpha_2$ units. Therefore, spectral flow forms an abelian group, and
\begin{equation}
\mathcal{U}_\alpha^{-1} = \mathcal{U}_{-\alpha}.
\end{equation}

\subsection{Ramond sector}

The CFT in the complex plane naturally has periodic fermions, which
corresponds to the NS sector. One can, however, spectral flow by an
odd number of units to the Ramond sector. If one starts with the NS
vacuum and then spectral flows with $\alpha = -1$, then the state in
the R sector has
\begin{equation}
h = \frac{1}{4} \qquad m = \frac{1}{2}.
\end{equation}
From the weight we see that this must be the R ground state. Let us
call this state
\begin{equation}
\vac_R^+.
\end{equation}
Since, we are now in the Ramond ground state we have fermion zero
modes, and therefore may act with $J^-_0$, which gives us the state
\begin{equation}\label{eq:Rminus-expanded}
\vac_R^- = J^-_0\vac_R^+ 
	 = -\tfrac{1}{2}\epsilon_{\dot{A}\dot{B}}\psi^{-\dot{A}}_0\psi^{-\dot{B}}_0\vac_R^+
	 = \psi_0^{-\dot{2}}\psi_0^{-\dot{1}}\vac_R^+.
\end{equation}
The normalization is fixed by the commutation relations of
$J_0^a$. Since $J_0^-$ has zero weight, one can be sure that this
state is also a member of the R vacuum. Acting twice with $J^-_0$
annihilates the state, from which one concludes that these states from
a doublet of $SU(2)_L$,
\begin{equation}
\vac_R^\alpha,
\end{equation}
and one also can determine that
\begin{equation}\label{eq:Rplus-expanded}
\vac_R^+ = J_0^+\vac_R^-
	= \tfrac{1}{2}\epsilon_{\dot{A}\dot{B}}\psi^{+\dot{A}}_0\psi^{+\dot{B}}_0\vac_R^-
	= \psi^{+\dot{1}}_0\psi^{+\dot{2}}_0\vac_R^-.
\end{equation}

What happens if we act on these states not with a pair of fermion zero
modes in the current $J$, but with a single fermion zero mode
directly? Since one cannot raise the charge of the state $\vac_R^+$ or
lower the charge of the state $\vac_R^-$, one must have
\begin{equation}
\psi^{+\dot{A}}_0\vac_R^+ = 0 \qquad \psi^{-\dot{A}}_0\vac_R^- = 0.
\end{equation}
This can also be seen from Equations~\eqref{eq:Rminus-expanded} and
\eqref{eq:Rplus-expanded}. However, one ought to be able to contract the
fermion zero mode index with the R vacuum doublet index to form
\begin{equation}
\vac_R^{\dot{A}} = \frac{1}{\sqrt{2}}\epsilon_{\alpha\beta}\psi_0^{\alpha \dot{A}}\vac_R^\beta,
\end{equation}
where the normalization is determined from the fermion mode
anticommutation relations. Since there are four fermion zero modes (in
the left sector), we expect four Ramond vacuum. We see that those
vacua form a doublet of $SU(2)_L$ and a doublet of $SU(2)_2$.

Of course, the same story holds on the right sector of the theory as
well, which gives a total of 16 Ramond vacua:
\begin{equation}
\vac_R^{\alpha\dot{\alpha}} \qquad \vac_R^{\dot{A}\dot{\alpha}} \qquad
\vac_R^{\alpha\dot{A}} \qquad \vac_R^{\dot{A}\dot{B}}.
\end{equation}
Note that we must be very careful to always write the index
corresponding to the left zero modes first and the index corresponding
to the right zero modes second.

What are the images of the Ramond vacua in the NS sector? From the
action of spectral flow on the charge and weight of a state, one can
conclude that the Ramond vacua are one unit of spectral flow from
chiral primary ($h=m$) states in the NS sector; or equivalently,
negative one units of spectral flow from anti-chiral primary ($h=-m$)
states. Therefore, there is a one-to-one correspondence between the R
vacua and the NS chiral primary states. There are four left chiral
primary states in the NS sector,
\begin{equation}
\vac_{NS} \qquad
\psi^{+ \dot{A}}_{-\frac{1}{2}}\vac_{NS}\qquad 
\epsilon_{\dot{A}\dot{B}}\psi^{+\dot{A}}_{-\frac{1}{2}}\psi^{+\dot{B}}_{-\frac{1}{2}}\vac_{NS},
\end{equation}
and there are also four states in the right sector. Thus a total of 16
chiral primary states in the NS sector that get mapped onto the 16
Ramond vacua via spectral flow.  These are all of the chiral primary
states for a single strand of the CFT. In the twisted sector, there
are more chiral primary states which correspond to Ramond `vacua' in the
twisted sector.

\subsection{Chiral primaries}

Consider the supercurrents in the anomaly free subalgebra. In
particular we have
\begin{equation}
\ac{G^{-A}_{+\frac{1}{2}}}{G^{+B}_{-\frac{1}{2}}} = 2\epsilon^{AB}(J_0^3 - L_0).
\end{equation}
Since
\begin{equation}
\big(G^{+ B}_{-\frac{1}{2}}\big)^\dg = -\epsilon_{BC}G^{-C}_{+\frac{1}{2}},
\end{equation}
one finds that for a normalized state $\ket{\psi}$ of weight $h$ and
charge $m$
\begin{calc}
\bra{\psi} \big(G^{+B}_{-\frac{1}{2}}\big)^\dg G^{+B}_{-\frac{1}{2}}\ket{\psi}
	&= -\epsilon_{BC}\bra{\psi}G^{-C}_{+\frac{1}{2}}G^{+B}_{-\frac{1}{2}}\ket{\psi}\\
	&= \epsilon_{BC}\bra{\psi}G^{+B}_{-\frac{1}{2}}G^{-C}_{+\frac{1}{2}}\ket{\psi}
	   -2\epsilon_{BC}\epsilon^{CB}\bra{\psi}(J_0^3 - L_0)\ket{\psi}\\
	&= -\bra{\psi}\big(G_{+\frac{1}{2}}^{-C}\big)^\dg G^{-C}_{+\frac{1}{2}}\ket{\psi}
	   -2(m-h)\\
\Longrightarrow 
&\bra{\psi} \big(G^{+B}_{-\frac{1}{2}}\big)^\dg G^{+B}_{-\frac{1}{2}}\ket{\psi}
	= h-m.
\end{calc}
This and a similar calculation from the anticommutation relation
\begin{equation}
\ac{G^{+A}_{+\frac{1}{2}}}{G^{-B}_{-\frac{1}{2}}} = 2\epsilon^{AB}(J_0^3 + L_0),
\end{equation}
imply that for our theory to be unitary, we need all physical states
to have weight greater than (the absolute value of) charge:
\begin{equation}
h \geq |m|.
\end{equation}

A \emph{chiral} state is a state that is annihilated by
$G_{-\frac{1}{2}}^{+A}$ for $A=1,2$. A \emph{primary} state is killed
by all the positive modes of the theory. Thus, a state or operator
that is both chiral and primary must saturate the above bound, having
$h=m$.  In fact, the converse is also true: any state or operator that
saturates the bound is a chiral primary.

Since an $h=m$ state $\ket{\chi}$ saturates the bound, its weight
cannot be lowered without lowering the charge, and thus
\begin{equation}
L_m\ket{\chi} = J^3_m \ket{\chi} = 0\qquad m>0.
\end{equation}
From the bound, one can immediately conclude that
\begin{equation}
G^{+A}_{+\frac{1}{2}}\ket{\chi} = 0.
\end{equation}
On the other hand, it is also possible to show
\begin{equation}
G^{-A}_{+\frac{1}{2}}\ket{\chi} = 0\qquad
G^{+A}_{-\frac{1}{2}}\ket{\chi} = 0.
\end{equation}
To demonstrate this, one must use the algebra to show that the sum of
the norms of the above two states must vanish. Then, the above follows
from the positive-definiteness of the norm. Therefore, the states
saturating the bound are annihilated by all of the $G$'s in the
anomaly-free subalgebra except for $G_{-1/2}^{-A}$.  Since all of the
positive modes kill the state, the state must also be primary. That
is, $h=m$ implies chiral primary.

The importance of the chiral primary operators is that they correspond
to the top member\footnote{Frequently, this is called the `highest
weight state', even though it actually has the lowest conformal mass
dimension of the multiplet.} of the superconformal multiplets \emph{in
the NS sector}. Therefore, it suffices to find the chiral primary
operators to catalog the representations of the superconformal
algebra.

\section{Cartesian to spherical Clebsch--Gordan coefficients}
\label{ap:spherical}

In this section, we outline our conventions for relating irreducible
spherical tensors to ordinary cartesian tensors in flat space. This
fixes the factors in going from Equation~\eqref{eq:general-S-int} to
Equation~\eqref{eq:general-decay-rate} for the D1D5 case, and explains
how we define the correctly normalized differential operator, so that
Equation~\eqref{eq:diff-op-norm} is satisfied.

Our goal is to show how to construct the coefficients, $Y_{l,m_\psi,
m_\phi}^{j_1\dots j_l}$, that define the differential operator in
Equation~\eqref{eq:diff-op} such that it satisfies
Equation~\eqref{eq:diff-op-norm},
\[
Y^{k_1k_2\cdots k_l}_{l,m_\psi,m_\phi}\pd_{k_1}\pd_{k_2}\cdots\pd_{k_l}
		\big[r^{l'} Y_{l', m_\psi', m_\phi'}(\Omega_3)\big]
	= \delta_{ll'}\delta_{m_\psi m'_\psi}\delta_{m_\phi m'_\phi}.
\]
We take the spherical harmonics as a starting point. The cartesian
coordinates for the noncompact space are related to the angular
coordinates via
\begin{equation}\begin{split}
x^1 &= r \cos\theta\cos\psi\\
x^2 &= r \cos\theta\sin\psi\\
x^3 &= r \sin\theta\cos\phi\\
x^4 &= r \sin\theta\sin\phi,
\end{split}\end{equation}
where the $(\theta, \psi, \phi)$ are restricted to
\begin{equation}
\theta\in \big[0,\tfrac{\pi}{2}\big)\qquad
\psi,\phi \in [0,2\pi).
\end{equation}

Spherical harmonics can be written as a homogeneous polynomial of the
cartesian unit vector components of the form
\begin{equation}
Y_{l,m_\psi,m_\phi}(\Omega_3) = \frac{1}{r^l}
     \mathcal{Y}^{l,m_\psi, m_\phi}_{j_1\cdots j_l}
     x^{j_1}\cdots x^{j_l} 
  = \mathcal{Y}^{l,m_\psi, m_\phi}_{j_1\cdots j_l}n^{j_1}\cdots n^{j_l}.
\end{equation}
The $\mathcal{Y}$ must be pairwise symmetric and traceless. One can
compute these tensors by the usual methods of breaking up a tensor
into irreducible components, or by inspection of the explicit form of
the spherical harmonics in angular components.

Given the spherical harmonic normalization
\begin{equation}\label{eq:sphere-harm-norm}
\int Y_{l,m_\psi, m_\phi}^* Y_{l',m_\psi', m_\phi'} \drm \Omega 
   = \delta_{ll'}\delta_{m_\psi m_\psi'}\delta_{m_\phi m_\phi'},
\end{equation}
one can determine an orthogonality relation for the $\mathcal{Y}$'s:
\begin{equation}\label{eq:ortho-1}
\big(\mathcal{Y}^{l,m_\psi, m_\phi}_{j_1\cdots j_l}\big)^*
\mathcal{Y}^{l',m'_\psi,m'_\phi}_{k_1\cdots k_{l'}}
\int (n^{j_1}\cdots n^{j_l})(n^{k_1}\cdots n^{k_{l'}})\drm \Omega
= \delta_{ll'}\delta_{m_\psi m_\psi'}\delta_{m_\phi m_\phi'}.
\end{equation}
The integral over $l+l'$ unit vectors defines a natural inner product
on the Clebsch--Gordan coefficients $\mathcal{Y}$.

We label the integral
\begin{equation}
I^{j_1\cdots j_lk_1\cdots k_{l'}},
\end{equation}
and note that $I$ must be symmetric in all of its indices.
Furthermore, from the symmetry of the integral one must conclude that
$I$ vanishes unless every index appears an even number of times. For
instance,
\begin{equation}
I^{j} = \int n^j \drm\Omega = 0.
\end{equation}
As a corollary, $I$ vanishes unless it has an even number of indices.
Having picked off the easiest properties, let's without further
comment give the general form. Let $a_i$ be the total number of times
the index $i$ appears in the collection of indices on $I$, then
\begin{calc}
I^{[a_1,a_2,a_3,a_4]} &=\int (n^1)^{a_1}(n^2)^{a_2}(n^3)^{a_3}(n_4)^{a_4}
                           \drm\Omega_3\\
	&= 
\left[\int_0^{\frac{\pi}{2}}\cos^{a_1 + a_2+1}\theta\sin^{a_3+a_4+1}\theta\drm\theta\right]
	\left[\int_0^{2\pi}\cos^{a_1}\psi\sin^{a_2}\psi\drm\psi\right]
	\left[\int_0^{2\pi}\cos^{a_3}\phi\sin^{a_4}\phi\drm\phi\right].\\
\end{calc}
We recognize the above definite integrals as different representations
of the beta function:
\begin{subequations}
\begin{align}
\int_0^\frac{\pi}{2}\cos^\alpha\theta\sin^\beta\theta\,\drm\theta
	&= \frac{1}{2}B\big(\tfrac{\alpha+1}{2},\tfrac{\beta+1}{2}\big)\\
\int_0^{2\pi}\cos^\alpha\theta\sin^\beta\theta\,\drm\theta
	&= \frac{1}{2}[1 + (-1)^\alpha + (-1)^{\alpha + \beta} + (-1)^{\beta}]
	 B\big(\tfrac{\alpha+1}{2},\tfrac{\beta+1}{2}\big),
\end{align}
\end{subequations}
where the second equation follows from the first. Therefore, one sees
that \emph{provided all the $a_i$ are even}
\begin{calc}
I^{[a_1, a_2, a_3, a_4]} &=2  B\big(\tfrac{a_1+a_2+2}{2},\tfrac{a_3+a_4+2}{2}\big)
	 B\big(\tfrac{a_1+1}{2},\tfrac{a_2+1}{2}\big)
	 B\big(\tfrac{a_3+1}{2},\tfrac{a_4+1}{2}\big)\\
&= 2\pi^2\left(\frac{1}{\pi^2}
\frac{\Gamma\big(\tfrac{a_1+1}{2}\big)\Gamma\big(\tfrac{a_2+1}{2}\big)
      \Gamma\big(\tfrac{a_3+1}{2}\big)\Gamma\big(\tfrac{a_4+1}{2}\big)}
	{\Gamma\big(\tfrac{a_1+a_2+a_3+a_4+4}{2}\big)}\right)\\
&= \frac{2\pi^2(a_1-1)!(a_2-1)!(a_3-1)!(a_4-1)!}
         {2^{a_1+a_2+a_3+a_4-4}(\frac{a_1}{2}-1)!(\frac{a_2}{2}-1)!
           (\frac{a_3}{2}-1)!(\frac{a_4}{2}-1)!(\frac{a_1+a_2+a_3+a_4}{2}+1)!}.
\end{calc}
Note that in the last line one must use the limit
\begin{equation}
\lim_{x\to 0} \frac{(x-1)!}{(\frac{x}{2}-1)!} = \frac{1}{2},
\end{equation}
in the event that some of the $a_i$ are zero.

The orthogonality condition from Equation~\eqref{eq:ortho-1} is
\begin{equation}
I^{j_1\cdots j_lk_1\cdots k_{l'}}
 \big(\mathcal{Y}^{l, m_\psi, m_\phi}_{j_1\cdots j_l}\big)^*
 \mathcal{Y}^{l', m_\psi', m_\phi'}_{k_1\cdots k_{l'}}
 =  \delta_{ll'}\delta_{m_\psi m_\psi'}\delta_{m_\phi m_\phi'},
\end{equation}
which motivates the choice of
\begin{equation}
Y_{l, m_\psi, m_\phi}^{j_1\cdots j_l} 
  \propto I^{j_1\cdots j_lk_1\cdots k_l}
    \big(\mathcal{Y}^{l, m_\psi, m_\phi}_{k_1\cdots k_l}\big)^*.
\end{equation}
We can think of the $2l$-index $I$ as defining an inner product on the
space of symmetric traceless $l$-index tensors, spanned by the
$\mathcal{Y}^{l,m_\psi, m_\phi}_{j_1\dots j_l}$. Then, we can think of
$Y_{l,m_\psi, m_\phi}$ as (proportional to) the dual of
$\mathcal{Y}^{l,m_\psi,m_\phi}$.

Since the $l$ derivatives are symmetrized and the spherical harmonic's
cartesian form is also symmetrized, we get a factor of $l!$. One finds
that
\begin{equation}
I^{j_1\cdots j_lk_1\cdots k_l}
    \big(\mathcal{Y}^{l, m_\psi, m_\phi}_{k_1\cdots k_l}\big)^*
	\pd_{j_1}\cdots \pd_{j_l} 
    r^l Y_{l, m_\psi, m\phi}(\theta, \psi, \phi)
	\bigg|_{r\to 0}
 = l!
\end{equation}
and therefore we define
\begin{equation}
Y_{l, m_\psi, m_\phi}^{j_1\cdots j_l} 
  = \tfrac{1}{l!} I^{j_1\cdots j_lk_1\cdots k_l}
    \big(\mathcal{Y}^{l, m_\psi, m_\phi}_{k_1\cdots k_l}\big)^*.
\end{equation}

\bibliography{fuzzball-decays}

%%%%%%%%% End Document %%%%%%%%
\end{document}